\documentclass[]{jfm_arxiv}

\usepackage{subcaption}
\usepackage{graphicx}
\usepackage{epstopdf,epsfig}
\usepackage{newtxtext}
\usepackage{newtxmath}
\usepackage{natbib}
\usepackage{hyperref}
\usepackage{soul}

\hypersetup{
    colorlinks = true,
    urlcolor   = blue,
    citecolor  = black,
}

\newcommand{\RomanNumeralCaps}[1]
\linenumbers

\newcommand{\avepsilonxt}{\langle\varepsilon\rangle}

\title{Turbulent diffusion-cascade interaction}

\author{E. Fuentes Noriega\aff{1} 
	\and J.C. Vassilicos\aff{1}
	\corresp{\email{john-christos.vassilicos@cnrs.fr}}}

\affiliation{\aff{1}Univ. Lille, CNRS, ONERA, Arts et Metiers
  Institute of Technology, Centrale Lille, UMR 9014, LMFL -
  Laboratoire de Mécanique des Fluides de Lille, Kampé de Fériet,
  F-59000 Lille, France}

\begin{document}

\maketitle

\begin{abstract}
In the decay region around the centreline of three qualitatively
different turbulent plane wakes, the turbulence is non-homogeneous and
two-point turbulent diffusion counteracts the turbulence cascade all
the way down to scales smaller than the Taylor
length. It is found that the sum of the inter-space
transfer rate and the horizontal part of the inter-scale transfer rate
of horizontal two-point turbulent kinetic energy is approximately
proportional to the turbulence dissipation rate in the inertial range
with a constant of proportionality between $-0.6$ and $-1$ depending
on wake and location within the wake, except at the near-field edge of
the decay region.

  \end{abstract}

\begin{keywords}
wakes, turbulence theory
\end{keywords}

\section{Introduction}
\label{sec:Introduction}

Inter-scale turbulence transfers and the turbulence cascade are
pivotal in turbulent flows. In statistically stationary homogeneous
turbulence, the average inter-scale transfer rate balances the
turbulence dissipation rate over an inertial range of length scales
which widens as the Reynolds number increases. This scale-by-scale
equilibrium is a prediction of the Kolmogorov theory which is
specifically designed for statistically stationary homogeneous
turbulence \cite[see][]{frisch1995turbulence}. However, turbulent
flows are typically non-homogeneous and inter-space turbulence
transfers (two-point turbulent diffusion) do not necessarily average
to zero. In this case they contribute to the average scale-by-scale
turbulent kinetic energy budget and must therefore be taken into
account. In fact, concurrent inter-scale and 
inter-space transfers were identified by \citet{marati2004energy} 
  more than twenty years ago in direct numerical
  simulations of fully developed turbulent channel flow. \cite{cimarelli2013paths,cimarelli2016cascades,cimarelli2021spatially,cimarelli2024spatially} studied how turbulent energy evolves through
  both physical and scale spaces and identified paths for this
  evolution in both wall-bounded turbulent flows and planar
  jets. Direct effects of spatial non-homogeneity and coherent
  structures on inter-scale transfer rates were also reported in
  spatially evolving wakes \citep[see][]{thiesset2020illusion}.  
  Even in statistically homogeneous turbulence where
inter-space transfer rates average to zero at all length-scales, 
one cannot fully describe inter-scale transfer rate
  fluctuations without taking inter-space transfer rates into
  account. Indeed, the fluctuations of the solenoidal part of the
  inter-space transfer rates have recently been shown to be
anti-correlated with the fluctuations of the solenoidal part of the
inter-scale turbulence transfer rates \citep{larssen2023spatio}. Are
there such clear simple relations between parts of average inter-scale
turbulence transfers and average inter-space turbulence transfers in
statistically non-homogeneous turbulence? How do these average
transfer rates depend on two-point separation and how do they compare
with the turbulence dissipation rate?  Is it necessary that there
should be a tendency towards local homogeneity at small enough
inertial length scales as commonly believed?

We answer these questions for three turbulent wakes of two
side-by-side parallel identical square prisms by analysing
Two-Dimensional Two-Component Particle Image Velocimetry (2D2C PIV)
data obtained by \cite{chen2021turbulence}. These three planar
turbulent wakes are qualitatively different, thereby allowing us to
test the generality of our results. We concentrate attention on the
decaying wake, i.e. the region around the centreline which is far
enough from the square prisms for the local Reynolds number to decay
with streamwise distance.

Given that the 2D2C PIV data at our disposal provide information about
the turbulent fluctuating velocities $u'_{1}$ and $u'_{2}$ in the
streamwise and cross-stream directions only, both of which are in the
horizontal plane normal to the spanwise direction of the vertical
prisms, this paper's focus is on the inter-scale transfer of
horizontal turbulent kinetic energy.

In the following section we present the paper's theoretical
framework. In section 3 we describe the 2D2C PIV data of
\cite{chen2021turbulence} and in section 4 we present our analysis of
these data with particular focus on the two-point turbulent kinetic
energy transfer processes. Conclusions are drawn in section 5 where we
highlight some key open questions regarding the intimate link between
the dissipation rate, the cascade and the turbulent diffusion in
non-homogeneous situations.

\section{Two-point horizontal turbulent kinetic energy budget}
\label{sec:2-point}

We are interested in the budget of the horizontal two-point turbulent
kinetic energy, which is defined as $\delta K_{h}\equiv {1\over 2}
[(\delta u'_{1})^{2} + (\delta u'_{2})^{2}]$ where $\delta u'_{i}
\equiv {1\over 2}[u'_{i} ({\bf X} + {\bf r},t) - u'_{i} ({\bf X} -
  {\bf r},t)]$ (for $i=1,2,3$) is a half-difference (high-pass
filtered) fluctuating velocity component \citep{germano2012simplest}. 
Given the horizontal 2D nature of the data used in
  this paper, we limit ourselves to horizontal separation vectors
  ${\bf r} = (r_{1}, r_{2}, 0)$.
We consider statistically stationary turbulent velocity fields so that
the budget is averaged over time and is for $\overline{\delta
  K_h}\equiv {1\over 2} \overline{(\delta u'_{1})^{2} + (\delta
  u'_{2})^{2}}$ where the overbar signifies time-averaging.
This two-point average energy defines the following
  scale-by-scale decompositions of the one-point horizontal energy
  $K_h(\mathbf{X}) = \vert \frac{1}{2} {\bf u}_{h}'\vert^{2}$ (with
  ${\bf u}_{h}' \equiv (u'_1 , u'_2 , 0))$ for $i=1$ and $i=2$:
  \begin{equation}
\int_0^{L_{i0}} \frac{d}{dr_i}\overline{\delta K_{h}}dr_i =
\overline{K_h({\bf X} + {\bf R}_{i})} + \overline{K_h({\bf X} - {\bf
    R}_{i})}
  \end{equation}
  where ${\bf R}_{1} = (L_{10}, 0 , 0)$ and ${\bf
    R}_{2} = (0,L_{20}, 0)$ are the separation vectors with the
  smallest separations $L_{i0}$ for which the two-point correlations
  $\overline{{\bf u}_{h}' ({\bf X}+{\bf R}_{i})\cdot {\bf u}_{h}'
    ({\bf X}-{\bf R}_{i})}$ vanish. These separations $L_{10}$ and
  $L_{20}$ can be thought of as characteristic streamwise and
  cross-stream sizes of the largest ``eddy'' centered at ${\bf X}$,
  and $\frac{d}{dr_i}\overline{\delta K_{h}}$ are energy densities per
  unit separation distance for each $i=1,2$. These energy densities
  decompose the sum of the horizontal turbulent kinetic energies at
  ${\bf X} + {\bf R}_{i}$ and ${\bf X} - {\bf R}_{i}$. Of course,
  these decompositions can be generalised to any other quantity
  (e.g. the full turbulent kinetic energy) and makes sense only when
  finite separations $L_{i0}$ exist; we checked that this is indeed
  the case for all the data used here.

For the budget of $\delta K_{h}$ we also need to define the half-sum
(low-pass filtered) fluctuating velocity vector ${\bf u'_{X}} \equiv
{1\over 2}[{\bf u'} ({\bf X} + {\bf r},t) + {\bf u'} ({\bf X} - {\bf
    r},t)]$ \citep{germano2012simplest} and the half-difference
fluctuating pressure $\delta p' \equiv {1\over 2\rho}[p'({\bf X} +
  {\bf r},t) - p' ({\bf X} - {\bf r},t)]$ (in terms of the fluid
density $\rho$ and the fluctuating pressure $p'$) as well as the half-sum and half-difference mean velocity fields
${\overline {\bf u_{X}}} \equiv {1\over 2}[\overline{{\bf u}} ({\bf X}
  + {\bf r}) + \overline{{\bf u}} ({\bf X} - {\bf r},t)]$ and $\delta
\overline{{\bf u}} \equiv {1\over 2}[\overline{{\bf u}} ({\bf X} +
  {\bf r}) - \overline{{\bf u}} ({\bf X} - {\bf r})]$ in terms of the
mean flow field $\overline{{\bf u}}$. At any position ${\bf X}$ in
physical space and for any two-point separation vector $2{\bf r}$ this
budget can be written as
\citep[see][]{hill2001equations,hill2002approach,chen2022scalings,beaumard2024scale}

\begin{equation}
L_T - P + T_X + \Pi_h= T_p +D -\tilde{\varepsilon_{1}}
-\tilde{\varepsilon_{2}},
	\label{eq:KHMH}
\end{equation}
where the linear transport rate $L_{T}$, the two-point turbulence
production rate $P$, the inter-space transport rate $T_X$, the
inter-scale transfer rate $\Pi_h$, the two-point pressure-velocity
term $T_p$ and the viscous diffusion rate $D$ are defined below.

\begin{equation}
L_T \equiv \overline{{\bf u}_{{\bf X}}}\cdot \nabla_{{\bf X}}
\overline{\delta K_h} + \delta \overline{\bf u}\cdot \nabla_{{\bf
    r}}\overline{\delta K_h}, \label{eq:LT}
\end{equation}  
where $\nabla_{{\bf X}}$ and $\nabla_{{\bf r}}$ are the gradients with
respect to ${\bf X}$ and ${\bf r}$ respectively. This
  term represents transport of two-point turbulent kinetic energy by
  the mean flow and is therefore refered to as linear to distinguish
  it from the non-linear terms $T_X$ and $\Pi_h$ which represent
  transport of two-point turbulent kinetic energy by fluctuating
  turbulent velocities.

\begin{equation}
  P \equiv -\overline{\delta u'_{1} {\bf u'_{X}}}\cdot \nabla_{{\bf X}}
  \delta \overline{u_{1}}-\overline{\delta u'_{2} {\bf u'_{X}}}\cdot
  \nabla_{{\bf X}} \delta \overline{u_{2}}- \overline{\delta u'_{1}
    \delta {\bf u'}}\cdot \nabla_{{\bf r}} \delta
  \overline{u_{1}}-\overline{\delta u'_{2} \delta {\bf u'}}\cdot
  \nabla_{{\bf r}} \delta \overline{u_{2}}, \label{eq:Prod}
\end{equation}

\begin{equation}
  T_{X} \equiv \nabla_{{\bf X}}\cdot \overline{{\bf u'_{X}} \delta K_h}, \label{eq:TX}
\end{equation}

\begin{equation}
  \Pi_h \equiv \nabla_{{\bf r}}\cdot \overline{\delta{\bf u'}
    \delta K_h}, \label{eq:Pi}
\end{equation}

\begin{equation}
  T_{p} \equiv - \overline{\delta u'_{1} {\partial \over \partial
      X_{1}}\delta p'} - \overline{\delta u'_{2} {\partial \over
      \partial X_{2}}\delta p'}, \label{eq:Tp}
\end{equation}
where $X_1$ and $X_2$ are the horizontal streamwise and cross-stream coordinates of $\bf X$ and

\begin{equation}
  D \equiv {\nu\over 2}(\nabla_{{\bf X}}^{2} + \nabla_{{\bf r}}^{2})
  \overline{\delta K_h},\label{eq:Diff}
\end{equation}
where $\nu$ is the fluid's kinematic viscosity. The two-point
turbulence dissipation rates $\tilde{\varepsilon_1}$ and
$\tilde{\varepsilon_2}$ in (\ref{eq:KHMH}) are defined as
$\tilde{\varepsilon_{i}} \equiv {1\over 2}(\varepsilon_{i} ({\bf
  X}+{\bf r}) + \varepsilon_{i} ({\bf X}-{\bf r}))$ where
$\varepsilon_{i} ({\bf X}) \equiv \nu \overline{({\partial
    u'_{i}/\partial x_{j}})^{2}}$ with a sum over $j=1,2,3$ for any
$i=1,2$. This definition also holds for $i=3$.

If the turbulence is statistically homogeneous this budget reduces to
\begin{equation}
\Pi_h= T_p +D -\varepsilon_{1} -\varepsilon_{2},
	\label{eq:KHMH-SSHT}
\end{equation}
which means that the inter-scale transfer rate $\Pi_h$ of horizontal
two-point turbulent kinetic energy is balanced by the
pressure-redistribution rate $T_p$ between $\overline{\delta K_h}$ and
${1\over 2} \overline{(\delta u'_{3})^{2}}$, the rate of molecular
diffusion of $\overline{\delta K_h}$ which is known to be negligible
at scales larger than the Taylor length $\lambda$
\citep[see][]{laizet2013interscale,valente2015energy} and the
dissipation rate of $\overline{\delta K_h}$. At scales larger than the
Taylor length where we can neglect molecular diffusion, the
inter-scale transfer rate $\Pi_{3}$ of the vertical/spanwise two-point
turbulent kinetic energy ${1\over 2} \overline{(\delta u'_{3})^{2}}$
obeys $\Pi_{3} \approx-T_{p} - \varepsilon_{3}$ in statistically
stationary homogeneous turbulence so that the total inter-scale
transfer rate $\Pi_h + \Pi_3$ of the entire two-point turbulent
kinetic energy balances the total turbulence dissipation rate,
i.e. $\Pi_h + \Pi_3 \approx -\varepsilon$ (where $\varepsilon =
\varepsilon_{1} + \varepsilon_{2} + \varepsilon_{3}$). This is the
scale-by-scale Kolmogorov equilibrium which is a prediction
  specifically designed for statistically stationary and homogeneous
  turbulence
\citep[see][]{kolmogorov1941dissipation,frisch1995turbulence}.

The situation is different if the statistically stationary turbulent
flow is not statistically homogeneous, in which case at least one of
the three terms $L_T$, $P$ and $T_X$ on the left hand side of equation
(\ref{eq:KHMH}) is not negligible. In this paper, we study the budget
equation (\ref{eq:KHMH}) in three qualitatively different turbulent
wakes of two parallel square prisms placed normal to an incoming
uniform flow. These flows are non-homogeneous in the plane normal to
the square prisms but homogeneous and symmetric in the prisms'
spanwise direction, thereby allowing the three terms $L_T$, $P$ and
$T_X$ to be fully evaluated from velocity data obtained by 2D2C PIV in
the plane normal to the spanwise direction. The Taylor length Reynolds
numbers $Re_{\lambda}$ in the regions of the three wakes that we
analyse range from about $150$ to about $500$.

In the following section we briefly describe the three turbulent wakes
and the experimental data of \cite{chen2021turbulence} that we use.
These experiments were conducted in the low speed closed circuit wind
tunnel of the \emph{Laboratoire de Mécanique de Fluides de Lille}
(LMFL) in 2020. Its test section is 2m wide by 1m high and 20m long
and is transparent on all four sides for maximal use of optical
techniques. A comprehensive description of these experiments can of
course be found in \cite{chen2021turbulence}.

\section{The wake experiments of \cite{chen2021turbulence} and flow characteristics}

\begin{figure}
	\centering
        \includegraphics[width=0.99\textwidth]{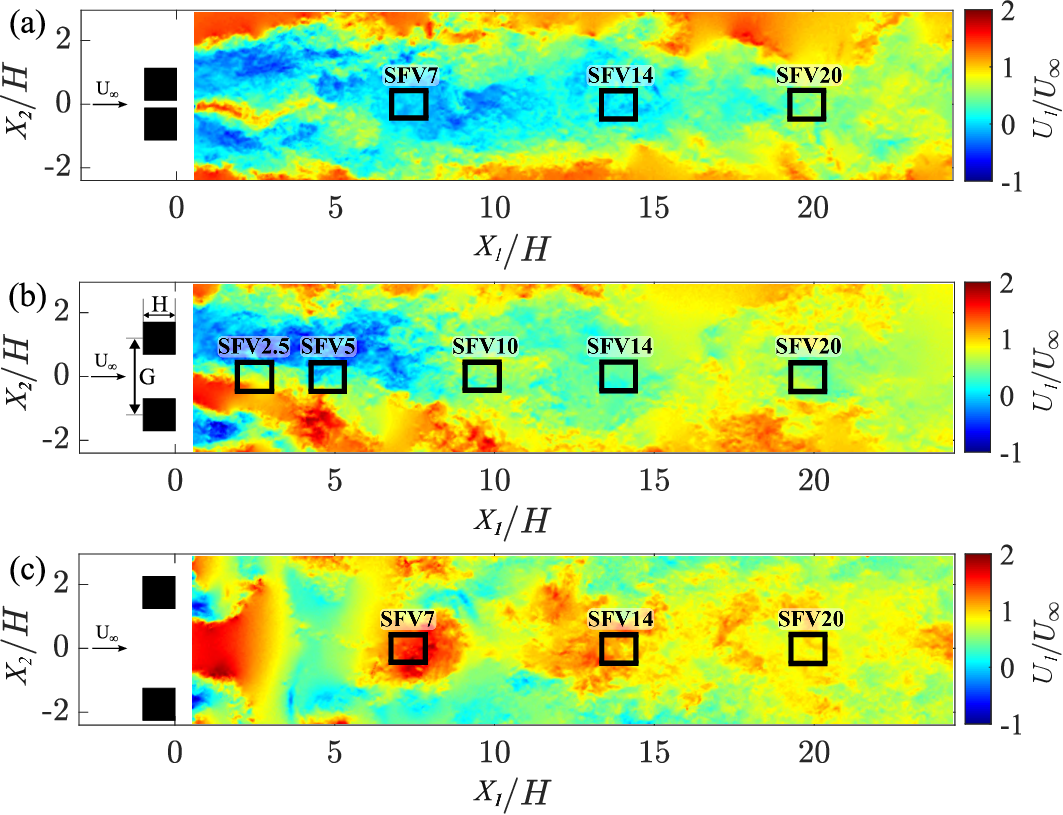}
	\caption{Example of normalised instantaneous streamwise
          velocity realisations $U_1/U_\infty$ in the wakes generated by
          two square prisms ($\blacksquare$) of side $H$ separated by
          a gap $G$ at $Re_H=1.0\times 10^4$. Flow patterns are shown
          for gap ratios (a) $G/H=1.25$, (b) $G/H=2.4$ and (c)
          $G/H=3.5$. 2D2C PIV data courtesy of \cite{chen2021turbulence}.}
	\label{fig:SFVs_visu}
\end{figure}

\cite{chen2021turbulence} collected 2D2C PIV data from three
qualitatively different wakes generated with a simple single-parameter
set-up. They measured the wake of two side-by-side identical square
prisms of side-width $H=0.03$m in small fields of view
(SFV$\mathcal{N}$) of size similar to the horizontal size of the
prisms with a high magnification factor at different streamwise
distances $X_{1}=\mathcal{N} H$ from the middle point between the two
square prisms. The three different wake regimes are obtained by
varying the gap distance $G$ between the middle points of each prism
in the cross-stream direction (measured by spatial coordinate
$X_2$). The three different gap ratios $G/H$ chosen by
\cite{chen2021turbulence} correspond to three qualitatively different
flows in terms of dynamics, bistability, large-scale features and
non-homogeneity as explained in \cite{chen2021turbulence} and
references therein. The resulting three wakes are illustrated in
Fig. \ref{fig:SFVs_visu}. The velocity fields in
Fig. \ref{fig:SFVs_visu} come from a large field of view PIV also
performed by \cite{chen2021turbulence} mainly for integral length
scale measurements and shown here for illustrative purposes only. We
make use of some of their integral length scale measurements as
reference length scales but we do not use their large field of view
data. We only use some of their small field of view PIV data as they
are spatially well resolved. The small fields of view SFV$\mathcal{N}$
with $\mathcal{N}$ from 2.5 to 20 are indicated in the figure. Their
size is 1H in the streamwise direction by 0.9H in the cross-stream
direction and their centre coincide with the geometric centreline
($X_2=0$).

A dual-camera PIV set-up was used by \cite{chen2021turbulence} for
each one of the small fields of view. Two sCMOS cameras, one over the
top and one under the bottom of the test section were aimed at the
same small field of view so as to obtain two independent measurements
of the same velocity field for PIV noise reduction.  As
comprehensively explained in \cite{chen2021turbulence} and in
\cite{beaumard2024scale}, this noise reduction method is a key step to
obtain accurate dissipation rate estimates. The acquisition frequency
was 5 Hz and 20000 velocity fields were captured for each measurement
corresponding to about 67 min. The PIV analysis' final interrogation
window size was $24\times24$ pixels with about 58\% overlap which
corresponds to a 312$\mu m$ interrogation window which ranged from
4.5 to 2.5 times the Kolmogorov length scale $\eta$ from nearest to
farthest SFV$\mathcal{N}$ (i.e. with increasing $\mathcal{N}$). For
all the small fields of view data used in the present paper, the
interrogation window size is below $3.2\eta$.

\begin{table}
	\begin{center}
		\def~{\hphantom{0}}
		\begin{tabular}{llccccc}
			Case $G/H=1.25$ &                                 & SFV7 & \multicolumn{2}{c}{SFV14} & \multicolumn{2}{c}{SFV20} \\
			& $Re_H$ ($\times 10^4$) & 1    & 1          & 1.2          & 1          & 1.2          \\
			& $Re_\lambda$           & 142  & 321        & 381          & 430        & 476            \\
			& $\avepsilonxt$ (m$^2$.s$^{-3}$) & 18.6  & 23.5  & 34.9   & 9.6   & 16.4    \\
			& $\eta$ (m)       & 120 $\times 10^{-6}$  & 113 $\times 10^{-6}$  & 102 $\times 10^{-6}$  & 141 $\times 10^{-6}$  & 123 $\times 10^{-6}$  \\
			& $\lambda$ (m)  & 2.8 $\times 10^{-3}$    & 4.0 $\times 10^{-3}$  & 3.9 $\times 10^{-3}$  & 5.8 $\times 10^{-3}$  & 5.3 $\times 10^{-3}$  \\
			&  $\langle \mathcal{L}_v \rangle /H$ & 0.98 & 1.4 & 1.4 & 1.6  &  1.6 \\
			& $\langle \mathcal{L}_v \rangle /\lambda$ & 10.5 & 10.8 & 11.0 & 8.4 & 9.1\\
			& $  L_{I} /\lambda $ & 2.1 & 1.5 & 1.2 & 2.6 & 2.7
		\end{tabular}
		\caption{Flow characteristics for the available wake
                  regions and global Reynolds numbers for case
                  $G/H=1.25$.}
		\label{tab:G_H_1.25_charac}
	\end{center}
\end{table}

\begin{table}
	\begin{center}
		\def~{\hphantom{0}}
		\begin{tabular}{llcccccc}
			Case $G/H=2.4$ &                                 & SFV2.5 & SFV5 & SFV10 & SFV14 & \multicolumn{2}{c}{SFV20} \\
			& $Re_H$ ($\times 10^4$) & 1      & 1    & 1     & 1.2   & 1          & 1.2          \\
			& $Re_\lambda$    & 260  & 266    & 185     & 186     & 148          & 164            \\
			& $\avepsilonxt$ (m$^2$.s$^{-3}$)     & 94.6  & 72.2    & 41.2     & 29.6     & 8.9          & 15.0            \\
			& $\eta$ (m)  & 80  $\times 10^{-6}$    & 85  $\times 10^{-6}$  & 98 $\times 10^{-6}$  & 106 $\times 10^{-6}$ & 144 $\times 10^{-6}$    & 126 $\times 10^{-6}$            \\
			& $\lambda$ (m)   & 2.5 $\times 10^{-3}$     & 2.7  $\times 10^{-3}$  & 2.6 $\times 10^{-3}$     & 2.9 $\times 10^{-3}$     & 3.4 $\times 10^{-3}$    & 3.2 $\times 10^{-3}$ \\
			& $\langle \mathcal{L}_v \rangle /H$ & 0.42 $\pm$ 0.1 & 0.79 $\pm$ 0.1 & 0.92 & 0.94  & 0.94  & 0.96 \\
			& $\langle \mathcal{L}_v \rangle /\lambda$ & 5.0 & 8.7 & 10.5 & 9.9 & 8.2 & 9.0\\
			& $  L_{I} /\lambda $ & 2.6 & 2.3 & 1.3 & 0.9 & 1.1 & 1.6       
		\end{tabular}
		\caption{Flow characteristics for the available wake
                  regions and global Reynolds numbers for case
                  $G/H=2.4$. The uncertainties are shown for the cases
                  where the integral length scale varies more than
                  20\% across the SFV.}
		\label{tab:G_H_2.4_charac}
	\end{center}
\end{table}

\begin{table}
	\begin{center}
		\def~{\hphantom{0}}
		\begin{tabular}{llcccccc}
			Case $G/H=3.5$ &                                 & SFV7 & \multicolumn{2}{c}{SFV14} & \multicolumn{3}{c}{SFV20} \\
			& $Re_H$ ($\times 10^4$) & 1    & 1          & 1.2          & 1     & 1.2     & 1.5     \\
			& $Re_\lambda$           & 183    & 129      & 167         & 118   & 135     & 160      \\
			& $\avepsilonxt$ (m$^2$.s$^{-3}$) & 41.9    & 26.8    & 33.2            & 12.5     & 21.0      & 34.7       \\
			& $\eta$ (m)    & 98 $\times 10^{-6}$    & 109 $\times 10^{-6}$ & 103 $\times 10^{-6}$ & 132 $\times 10^{-6}$ & 116 $\times 10^{-6}$ & 102 $\times 10^{-6}$  \\
			& $\lambda$ (m) & 2.6 $\times 10^{-3}$ & 2.4 $\times 10^{-3}$ & 2.6 $\times 10^{-3}$ & 2.8 $\times 10^{-3}$     & 2.6 $\times 10^{-3}$       & 2.6 $\times 10^{-3}$ \\
			& $\langle \mathcal{L}_v \rangle/H$ & 0.74  & 0.64  & 0.67 & 0.58 & 0.60   & 0.62 \\
			& $\mathcal{L}_v/\lambda$ & 8.5 & 7.9 & 7.7 & 6.2   & 6.8  & 7.3\\
			& $  L_{I} /\lambda $ & 1.4 & 1.7 & 0.9 & 1.1 & 1.2 & 1.3  
		\end{tabular}
		\caption{Flow characteristics for the available wake
                  regions and global Reynolds numbers for case
                  $G/H=3.5$.}
		\label{tab:G_H_3.5_charac}
	\end{center}
\end{table}

\cite{chen2021turbulence} acquired data for three incoming free stream
velocities $U_{\infty}=5.0, 6.0, 7.35$ m/s corresponding to three
values of the global Reynolds number $Re_{H} \equiv U_{\infty} H/\nu$
$=1.0 \times 10^4, 1.2 \times 10^4$ and $1.5 \times 10^4$,
respectively. The characteristics of the turbulence in each
SFV$\mathcal{N}$ are reported in tables \ref{tab:G_H_1.25_charac},
\ref{tab:G_H_2.4_charac} and \ref{tab:G_H_3.5_charac} for all
available global Reynolds numbers for all cases $G/H=1.25$, $G/H=2.4$
and $G/H=3.5$. The Kolmogorov length $\eta\equiv
(\nu^3/\avepsilonxt)^{1/4}$ and the Taylor length estimate $\lambda =
(10\nu K_{h}/\avepsilonxt)^{1/2}$ have been computed using our
estimation of the space-time average turbulent dissipation rate
$\avepsilonxt$ (see following paragraph) and the space-time average
horizontal one-point turbulent kinetic energy $K_{h}
\equiv {1\over 2}\langle \overline{u'^2_{1} + u'^2_{2}} \rangle$. The
angular brackets $\langle \rangle$ represent a space average over the
entire small field of view. The Taylor length-based Reynolds number
has been computed as $Re_{\lambda} =
\sqrt{(2/3)K_{h}}\lambda/\nu$. Both $Re_\lambda$ and $\lambda = (10\nu
K_{h}/\avepsilonxt)^{1/2}$ are under-estimated because they have been
computed with $K_{h}$ rather than ${1\over 2}<\overline{u'^2_{1} +
  u'^2_{2} + u'^2_{3}}>$ which is not accessible using the 2D2C PIV
data of \cite{chen2021turbulence}.

To obtain the time-averaged turbulence dissipation rate $\varepsilon$
we make use of the assumption of local axisymmetry around the
streamwise ($X_1$) axis which implies $\varepsilon_{3} = \varepsilon_{2}$
and therefore $\varepsilon = \varepsilon_{1}+2\varepsilon_{2}$
\citep{george1991locally}. This local axisymmetry has been found to be
a very good approximation for the calculation of $\varepsilon$ both
experimentally and numerically in numerous flows. In particular,
\cite{lefeuvre2014statistics} found that this approximation yields
accurate results of $\varepsilon$ in the wake of a square
cylinder. More importantly, \cite{AlvesPortela2022} achieved the same
conclusions using Direct Numerical Simulations (DNS) for the same
flows studied in this paper. We are therefore able to estimate the
full turbulence dissipation rate $\varepsilon$ from the 2D2C PIV data of
\citep{chen2021turbulence} as they give access to $\varepsilon_{1}$
and $\varepsilon_{2}$. This axisymmetric estimate has also been
denoised using a technique based on a dual camera set-up
\citep{chen2021turbulence,foucaut2021optimization}.

We also include the space-averaged integral length scale $\langle
\mathcal{L}_v \rangle$ as a fraction of $H$ and $\lambda$ in tables
\ref{tab:G_H_1.25_charac}, \ref{tab:G_H_2.4_charac} and
\ref{tab:G_H_3.5_charac}. The length scale $\mathcal{L}_v$ was
estimated by \cite{chen2021turbulence} from the cross-stream velocity
auto-correlation function in the streamwise direction as the
integration of the streamwise velocity autocorrelation function in
that same direction does not always converge
\citep{chen2021turbulence}. Although \cite{chen2021turbulence}
reported some variations in the very near fields, $\mathcal{L}_v$
remains fairly constant in each SFV$\mathcal{N}$ except for SFV2.5 and
SFV5 of configuration $G/H=2.4$. In terms of multiples of $\langle
\mathcal{L}_v \rangle$, the SFV$\mathcal{N}$ locations are much closer
to the prisms in the $G/H=1.25$ than in the other two cases. This is
reflected in table $\ref{tab:G_H_1.25_charac}$ where $\langle
\mathcal{L}_v \rangle$ reaches values up to $1.6H$ at SFV20 for
instance.

  Not only do these wakes exhibit distinct flow
  characteristics arising from their respective inlet conditions
  ($G/H$ parameter), one should also expect the physics to be
  different in the near field where the turbulence increases with the
  streamwise distance compared to further downstream where the
  turbulence decreases with streamwise distance. The present work
  concentrates on the further downstream decaying wake regions, hence
  on SFV7, SFV14 and SFV20 for $G/H = 3.5$, SFV10, SFV14 and SFV20 for
  $G/H=2.4$ and SFV20 for $G/H=1.25$ (see
  Fig. \ref{fig:SFVs_visu}). One can see in tables
  \ref{tab:G_H_1.25_charac}, \ref{tab:G_H_2.4_charac} and
  \ref{tab:G_H_3.5_charac} that the local Reynolds number
  $Re_{\lambda}$ decreases from SFV7 to SFV14 to SFV20 for $G/H=3.5$
  and from SFV10 to SFV14 to SFV20 for $G/H=2.4$. The following
  one-point energy analysis confirms that one-point production is
  small in these SFV stations and that they are therefore in the
  downstream decaying region of the $G/H=3.5$ and $G/H=2.4$ flow
  cases. For flow case $G/H=1.25$, $Re_{\lambda}$ actually increases
  from SFV7 to SFV14 to SFV20. Furthermore, the one-point energy
  analysis below suggests that SFV20 may not be in the decaying region
  of the $G/H=1.25$ wake. We nevertheless keep $G/H=1.25$ SFV20 in our
  study for comparison.

\begin{figure}
	\centering
	\begin{subfigure}{0.48\textwidth}
		\includegraphics[width=\textwidth]{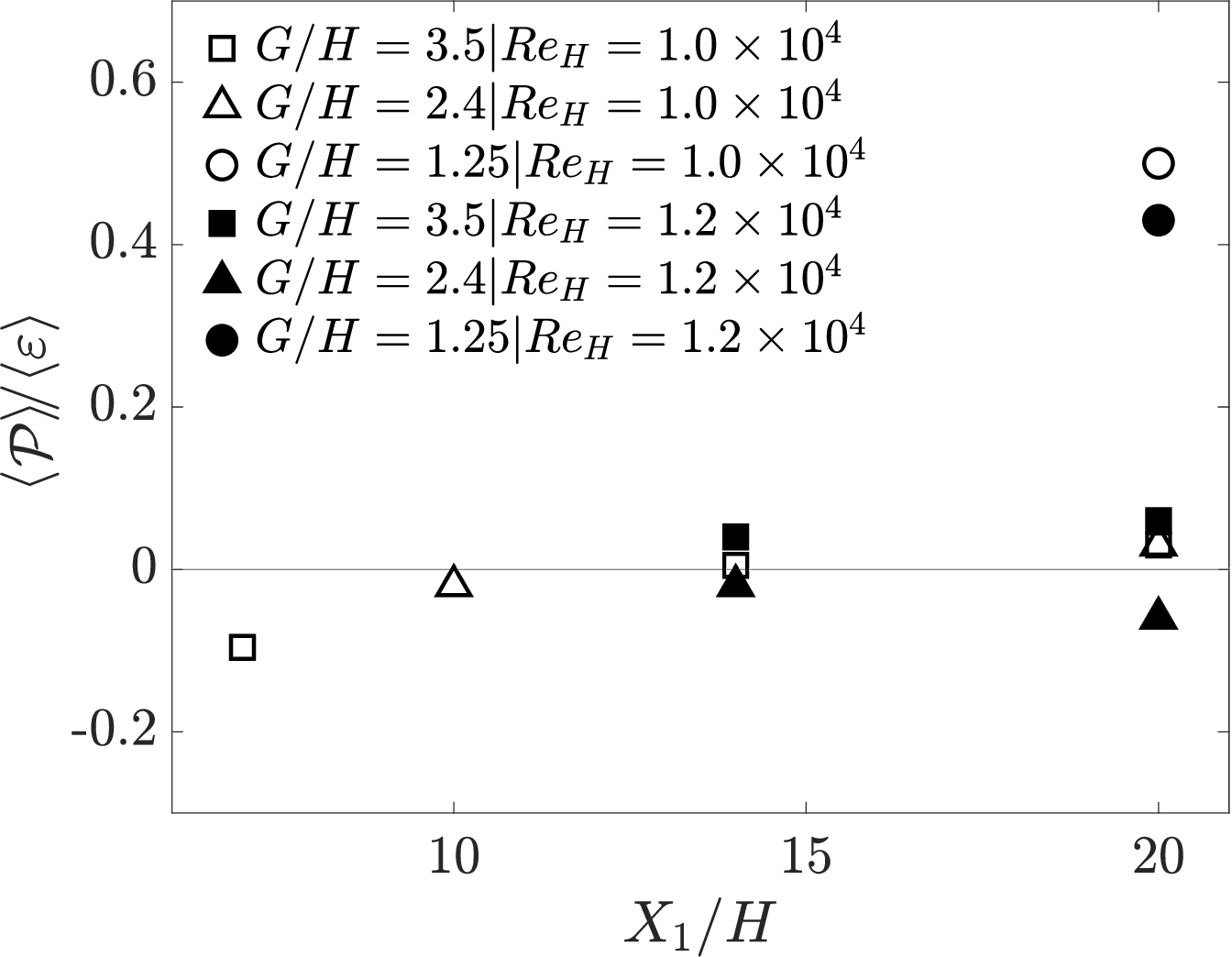}
		\caption{}
	\end{subfigure}
	\begin{subfigure}{0.48\textwidth}
		\includegraphics[width=\textwidth]{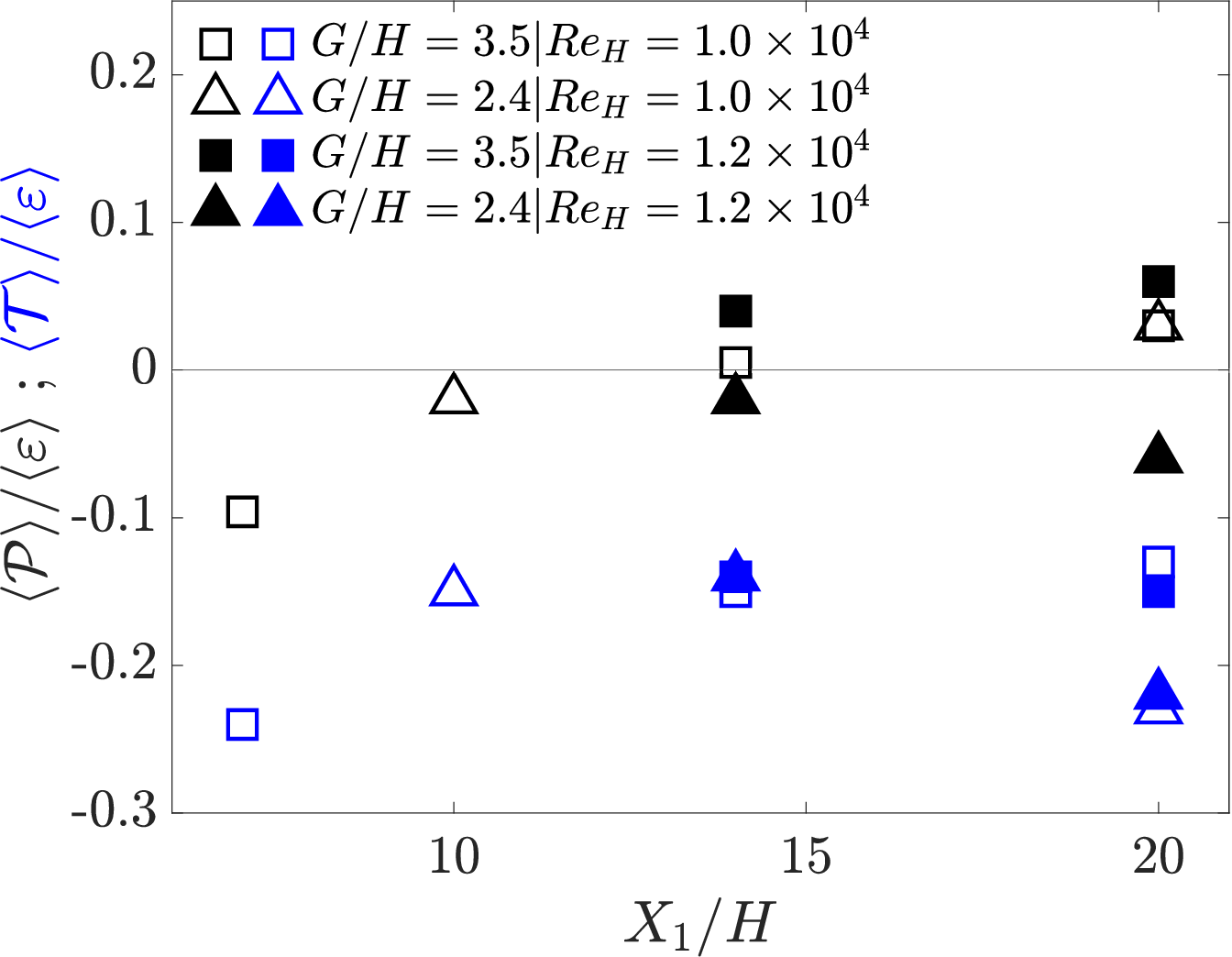}
		\caption{}
	\end{subfigure}
	\caption{Average values of one-point production $\langle
          \mathcal{P} \rangle$ (a) and turbulent diffusion $\langle
          \mathcal{T} \rangle$ (b) normalised by the average
          dissipation rate $\langle \varepsilon \rangle$ across the
          studied SFVs for two $Re_H$ values.}
	\label{fig:1pt_TKE}
\end{figure}

  To further characterise the turbulence at the SFV
  locations of interest before starting our two-point energy analysis,
  we look at the one-point horizontal turbulent kinetic energy
  transport equation. Taking advantage of the up-down symmetry and
  statistical homogeneity in the spanwise direction (normal to the
  $(X_{1}, X_{2})$ plane), this equation reads
\begin{equation}
		\overline{u_1}\frac{\partial \overline{K_h}}{\partial
                  X_1} + \overline{u_2}\frac{\partial
                  \overline{K_h}}{\partial X_2} - \mathcal{P} +
                \mathcal{T} = \mathcal{T}_p + \mathcal{D} -
                \varepsilon,
	\label{eq:1pt_TKE_eq}
\end{equation}
    where $\mathcal{P}=-\overline{u'_{1}
    u'_{1}}\frac{\partial \overline{u_1}}{\partial X_1} -
  \overline{u'_{1} u'_{2}}\frac{\partial \overline{u_1}}{\partial X_2}
  - \overline{u'_{1} u'_{2}}\frac{\partial \overline{u_2}}{\partial
    X_1} - \overline{u'_{2} u'_{2}}\frac{\partial
    \overline{u_2}}{\partial X_2}$ is the one-point turbulence
  production rate, $\mathcal{T} = \frac{\partial }{\partial
    X_1}(\overline{u'_{1}K_h}) + \frac{\partial }{\partial
    X_2}(\overline{u'_{2}K_h})$ is the one-point turbulent diffusion
  rate and where $\mathcal{T}_p$ and $\mathcal{D}$ are the one-point
  pressure-velocity correlation and molecular diffusion terms
  respectively. Note that our 2D2C PIV enables full evaluations of
  $\mathcal{P}$ and $\mathcal{T}$ because of the spanwise homegenity.
  We focus on three key terms, $\mathcal{P}$, $\mathcal{T}$ and
  $\varepsilon$, with particular emphasis on how the first two compare
  in magnitude to $\varepsilon$. Figure \ref{fig:1pt_TKE}a shows
  $\langle \mathcal{P}\rangle / \langle \varepsilon \rangle$ (the
  averaging being over the entire SFV) for all flow cases. At all SFV
  locations considered in the $G/H=3.5$ and $G/H=2.4$ wakes, $\langle
  \mathcal{P}\rangle$ generally remains between 1-5\% of $\langle
  \varepsilon\rangle$ except for our closest station (SFV7) in the
  $G/H=3.5$ flow case where $\langle \mathcal{P}\rangle$ approaches
  10\% of $\langle \varepsilon \rangle$ and has a negative sign. This
  is in contrast with $G/H=1.25$ SFV20 where $\langle
  \mathcal{P}\rangle$ is no longer negligible, reaching values up to
  50\% of $\avepsilonxt$ for $Re_H=1.0\times10^4$ (but positive). This
  sharp difference suggests that at our farthest measurement station
  (SFV20) the $G/H=1.25$ wake flow may not have yet entered the
  downstream decaying regime. In the following section we report that
  there are also significant differences in the two-point turbulent
  kinetic energy budget between $G/H=1.25$ $SFV20$ and the SFV
  stations that we consider in the decay regions of the $G/H=3.5$ and
  $G/H=2.4$ flow cases.

  Figure \ref{fig:1pt_TKE}b displays normalised
  space-averaged one-point turbulent diffusion rates $\langle
  \mathcal{T}\rangle / \avepsilonxt$ and compares them with $\langle
  \mathcal{P}\rangle / \avepsilonxt$ for the $G/H=3.5$ and $G/H=2.4$
  wakes. We find that $\langle \mathcal{T} \rangle$ ranges from
  approximately 15\% to 25\% of $\avepsilonxt$, consistently exceeding
  $\langle \mathcal{P}\rangle$ by a factor close to three or more
  across all SFV stations in the decay regions of these two
  wakes. Whereas average production is small, $\langle \mathcal{T}
  \rangle$ is significant, which is a clear sign of non-homogeneity, a
  non-homogeneity which may qualify as ``non-producing
  non-homogeneity''. Notably, $\langle \mathcal{T} \rangle$ has a
  negative sign, indicating that, on average, turbulence is
  transporting turbulent kinetic energy into these SFVs.

  The SFV20 station in the $G/H=1.25$ wake is very
  different as $\langle \mathcal{T} \rangle$ is positive there,
  specifically $\langle \mathcal{T} \rangle/\avepsilonxt = 1.75$ and
  $1.67$ for $Re_H = 1.0\times 10^4$ and $1.2\times 10^4$
  respectively. This is a station with a non-homogeneity caused by
  both turbulent production and turbulent diffusion where turbulence
  is transporting turbulent kinetic energy outside SFV20.

  To quantify the non-homogeneity characteristic of
  turbulent diffusion, a length scale is introduced and computed
  within the SFVs. To the authors' knowledge, no prior studies have
  characterised the degree of such non-homogeneity through a dedicated
  length scale. Although the Corrsin length, as recently used in
  \citep{kaneda2020linear,chen2021turbulence} for example,
  distinguishes between scales influenced or not influenced by mean
  flow gradients (and therefore also potentially by turbulence
  production), it does not capture the degree of non-homogeneity
  related to turbulent diffusion in the near-absence of
  production. Length scales such as $L_{IK_1} \equiv
  \frac{K_h}{|\frac{\partial k_h}{\partial X_1}|}$, $L_{IK_2} \equiv
  \frac{K_h}{|\frac{\partial k_h}{\partial X_2}|}$ and $L_{IK} =
  \frac{k_h}{|{\bf \nabla k_h}|}$ characterise the non-homogeneity of
  the one-point turbulent kinetic energy. In the seven regions in total
  considered here within the three turbulent wakes $G/H=1.25, 2.4,
  3.5$ we find $L_{IK_j}\gg L_{IK}\gg H$ for $j=1,2$ with values that
  are one to two orders of magnitude larger than $H$ for $L_{IK_j}$
  across each SFV and between 6 to 15 times $H$ for $\langle
  L_{IK}\rangle$. This suggests that the turbulent kinetic energy is
  not strongly non-homogeneous in our SFVs. We therefore propose a
  length scale based on the one-point turbulent diffusion of kinetic
  energy, defined as follows:
\begin{equation}
	L_{I} = \frac{\langle \overline{|u'_2K_h|}\rangle}{\langle
          |\frac{\partial}{\partial x_j}\overline{u'_jK_h}|\rangle}
	\label{eq:L_inhomog}
	\end{equation}
  with an implicit sum over $j=1,2$. The space-average
  in this definition is over an SFV area in the present paper, but can
  be generalised (or perhaps even lifted) for other studies. A
  turbulence with little turbulence production but significant
  turbulent diffusion may be considered to be locally homogeneous over
  local regions of size smaller than $L_I$. In the extreme case of
  homogeneous turbulence in an infinite or periodic domain, $L_{I}$ is
  infinite everywhere.

  Values of $L_{I}$ are reported in tables
  \ref{tab:G_H_1.25_charac}, \ref{tab:G_H_2.4_charac} and
  \ref{tab:G_H_3.5_charac} and are compared to the corresponding SFV's
  Taylor length. For all SFV$\mathcal{N}$ locations, irrespective of
  type of wake, position in the wake and even global or local Reynolds
  number, the values of $L_{I}$ are comparable to those of $\lambda$
  and well below the integral length scale. These findings confirm
  that non-homogeneity related to turbulent diffusion is present
  throughout all wake regions down to scales of the order of the
  Taylor length $\lambda$. Having established one-point
  characteristics of the various wake regions, the following section
  addresses the budget (\ref{eq:KHMH}) with particular attention to
  the inter-scale and inter-space transfer rates of horizontal
  two-point turbulent kinetic energy which constitute the core subject
  of this study.

\section{Two-point statistics results}\label{sec:results}

Given the up-down symmetry in the spanwise direction (i.e. along the
direction normal to the horizontal $(X_1,X_2)$ plane, the spanwise
components of $\overline{{\bf u}_{{\bf X}}}$ and $\delta \overline{\bf
  u}$ are zero. Hence, the linear transport rate (\ref{eq:LT}) can be
fully determined from 2D2C measurements in the $(X_1,X_2)$ plane if we
limit ourselves to ${\bf r} = (r_1 , r_2 , 0)$. Statistical
homogeneity in the spanwise direction implies that the gradients
$\nabla_{{\bf X}} \delta \overline{u_{i}}$ and $\nabla_{{\bf r}}
\delta \overline{u_{i}}$ have zero spanwise components for any index
$i$, thereby implying that the two-point turbulence production rate
(\ref{eq:Prod}) can also be fully determined from 2D2C measurements in
the $(X_1, X_2)$ plane. Spanwise statistical homogeneity also implies
${\partial \over \partial X_{3}} \overline{u'_{X_3} \delta K_h}=0$ so that
(\ref{eq:TX}) reduces to

\begin{equation}
  T_{X} \equiv {\partial \over \partial X_{1}} \overline{u'_{X1}
    \delta K_h} + {\partial \over \partial X_{2}} \overline{u'_{X2}
    \delta K_h}, \label{eq:TX2D2C}
\end{equation}
where ${\bf X} = (X_{1}, X_{2}, X_{3})$ and ${\bf u'_{X}} = (u'_{X1},
u'_{X2}, u'_{X3})$.

The only terms in the scale-by-scale turbulent kinetic energy budget
(\ref{eq:KHMH}) which cannot be fully determined from 2D2C PIV
measurements in the $(X_1, X_2)$ plane even for $r_3=0$ are the inter-scale
transfer rate $\Pi_h$ defined in equation (\ref{eq:Pi}), the
two-point pressure-velocity term $T_p$ defined in equation
(\ref{eq:Tp}) and the viscous diffusion rate $D$ defined in equation
(\ref{eq:Diff}). Whilst $T_p$ is not at all accessible by such
measurements, the part $\Pi_r \equiv {\partial \over \partial r_{1}}
\overline{\delta u'_{1} \delta K_h} + {\partial \over \partial r_{2}}
\overline{\delta u'_{2} \delta K_h}$ of $\Pi_h$ is accessible whereas
the part $\Pi_z \equiv {\partial \over \partial r_{3}}
\overline{\delta u'_{3} \delta K_h}$ is not. Concerning $D$, most of it is
accessible for $r_3 =0$ except $D_{z}\equiv {\partial^{2}\over
  \partial r_{3}^{2}} \overline{\delta K_h}$, see equation
(\ref{eq:Diff}).

In summary, the scale-by-scale budget of the horizontal two-point
turbulent kinetic energy can be expressed as
\begin{equation}
L_T - P + T_X + \Pi_{r} + \Pi_{z}= T_p +D_{r} + D_{z}
-\tilde{\varepsilon_{1}} -\tilde{\varepsilon_{2}} ,
	\label{eq:KHMH2}
\end{equation}
where $D_{r}\equiv D-D_{z}$ and $\Pi_r \equiv
  \Pi_h-\Pi_z$. Every term in (Eq. \ref{eq:KHMH2}) can be fully
obtained from the small field of view 2D2C PIV measurements of
\cite{chen2021turbulence} for ${\bf r}= (r_{1}, r_{2}, 0)$ except
$\Pi_z$, $T_p$ and $D_z$. In this paper we primarily study the
space-time average turbulence transfer rates $\langle T_X \rangle$ and
$\langle \Pi_{r}\rangle$. We verified that the
  spatial average does not impact our results by replacing the average $\langle
  \rangle$ over the entire small field of view by an average over any
  straight line $X_1 = const$ or $X_2 = const$ within the small field
  of view and checking that the same conclusions presented here are reached, albeit with less statistical convergence
  (see examples of such checks for $\langle T_X \rangle$ and $\langle \Pi_r \rangle$ in Appendix
  \ref{appA}).

Following \cite{larssen2023spatio}, we ask whether the accessible
part, $\langle \Pi_r \rangle$, of the average inter-scale turbulence
transfer rate counteracts or cooperates with the average inter-space
turbulence transfer rate $\langle T_X \rangle$, and how they both
compare with the average turbulence dissipation rate $\avepsilonxt$ in
terms of magnitude.  The $\Pi_r$ part of the inter-scale turbulence
transfer rate $\Pi_h$ is the part which is fully determined by the
horizontal velocity field without any direct influence from the
spanwise (out-of-plane) velocity field, very much like $T_X$.  Note
that space-time averages are used to achieve statistical convergence
of the third order statistics involved in these transfer rates (over
20,000 uncorrelated samples and over the entire field of view).  How
do these average transfer rates depend on $r_1$ for $r_2 = r_3 =0$ and
on $r_2$ for $r_1 = r_3 =0$ and over what scale-ranges? We answer
these questions and also calculate $\langle L_T \rangle$ and $\langle
P \rangle$ to complement the analysis which is carried out in the
decay region around the centreline of the three qualitatively
different turbulent wakes described in the previous
section. Note that in the SFV14 and SFV20 locations of the
  $G/H=2.4$ and $G/H=3.5$ wakes, \cite{chen2022scalings} have shown,
  using the exact same data used here, that the longitudinal and
  transverse second order structure functions vary with two-point
  separation $r_1$ as $r_{1}^{2/3}$ in an inertial subrange of scales
  $r_1$. They also found that these two structures functions depart
  from this scaling and evolve faster than $r_{1}^{2/3}$ at the SFV20
  location of the $G/H=1.25$ wake.

  The horizontal two-point turbulent kinetic energy
  $\overline{\delta K_h}$ is the sum of these structure functions and
  thus scales in the same manner for $r_1$ and $r_2$ as shown in
  Appendix \ref{appA:deltaKh_r_profiles} where we also show that there
  is a departure from the $2/3$ power law at the SFV7 location of the
  $G/H=3.5$ wake. We now present our results on the average
  inter-scale and inter-space transfer rates, starting below with an
  overview of the main results detailed in the following
  sub-sections. This overview is intended to help the reader's focus
  when reading through the discussion of our results one wake at a
  time in the following sub-sections. We also advance some hypotheses
  on which we base a tentative qualitative explanation of some of our
  results.

As can be seen in the subsequent figures, in all the small fields of
view in the decaying wake region of both $G/H=3.5$ and $G/H=2.4$
turbulent wakes as well as in $SFV20$ of the $G/H=1.25$ turbulent
wake, we find that $\langle T_X \rangle$ is positive whilst $\langle
\Pi_r \rangle$ is negative for all accessible length-scales $r_{1}\not
= 0$ and $r_{2}\not = 0$ at the very least smaller or equal to
$\langle \mathcal{L}_v \rangle$. This means that, on average, scales
smaller than $r_1$ or $r_2$ within the small field of view gain
horizontal two-point turbulent kinetic energy via the interscale
transfers but also lose it to the neighbouring physical space outside
the small field of view by turbulent diffusion. The subsequent figures
also show that, in all these cases, the fully horizontal inter-scale
transfer rate $\langle \Pi_r \rangle$ is significantly larger or
sometimes approximately equal in magnitude to $\langle \varepsilon
\rangle$ over all accessible length scales, or sometimes a significant
range of them. Furthermore, the subsequent figures
make it clear that the two-point turbulent diffusion is not at all
negligible compared to $\avepsilonxt$ for all length scales $r_1$ or
$r_2$ larger than a fraction of the Taylor length and, at the very
least, smaller than $\langle \mathcal{L}_v \rangle$. 
(Note that $\mathcal{L}_v$ varies by at most 8\% of
  $\langle \mathcal{L}_v \rangle$, and typically much less, within
  each one of the SFVs we consider.) Non-homogeneity is therefore
present at all inertial length scales all the way down to
viscosity-affected length scales for all of our local Reynolds numbers
$Re_{\lambda}$ which range up to nearly $500$, in agreement with the
theory of \cite{chen2022scalings} and of \cite{beaumard2024scale}
which predicts that non-homogeneity can be present over the entire
inertial range even in the limit of infinite Reynolds number.

  Non-homogeneity all the way down to the smallest
  scales is not inconceivable in the presence of inter-scale turbulent
  energy transfers. The argument runs as follows. An increase in
  inter-space turbulence transfer can remove energy from the
  inter-scale transfer process to smaller scales. If we hypothesise,
  for simplicity of argument, that the turbulence dissipation rate is
  somehow independently set by some mechanism in the flow, then the
  rate of inter-scale transfer of the remaining energy may accelerate
  to ensure the turbulence dissipation rate is met. If this leads to
  an increase of the inter-scale turbulence transfer, the energy
  available for turbulent diffusion at a given scale may reduce which
  could bring the inter-space turbulence transfer rate down at that
  scale. (This is a two-point analogue of the observation made by
  \cite{alexakis2023far} that the turbulence dissipation (hence the
  turbulent cascade) can reduce, even inhibit, one-point turbulent
  diffusion.) Going back to the start of our argument, a reduction in
  inter-space turbulence transfer may have the inverse effect of an
  increase and may decelerate the rate of inter-scale turbulence
  transfer to ensure the turbulence dissipation rate is met. In turn,
  this may increase the energy available for turbulent diffusion and
  bring the inter-space turbulence transfer rate back up. A balance
  between the two transfers may consequently be achieved so that none
  of them vanishes and non-homogeneity persists at all scales
  irrespective of Reynolds number. Of course, the mechanism setting
  the turbulence dissipation rate is likely to interact with the
  interplay between inter-scale and inter-space transfers in which
  case a balance may somehow be dynamically reached between these two
  transfer mechanisms and the turbulence dissipation. We stress that
  this is an argument for plausibility not a definitive explanation of
  the small-scale non-homogeneity reported in the following
  sub-sections. We leave this explanation and the important question
  of what sets the local turbulence dissipation rate in
  non-homogeneous turbulence \citep[see][]{lumley1992some} for future
  investigation.

  A particular aspect of the small-scale
  non-homogeneity observed in the subsequent figures is that $\langle
  T_X \rangle$ is uniformly positive at all length scales $r_1$ and
  $r_2$ equal to or smaller than $\langle \mathcal{L}_v\rangle$ and in
  most cases even above $\langle \mathcal{L}_v\rangle$. On the other
  hand, the one point turbulent diffusion rate $\mathcal{T}$ is
  negative in all SFV stations except SFV20 in the $G/H=1.25$ where it
  is positive. The two-point inter-space transfer rate can be
  decomposed as $\langle T_X \rangle = \langle \mathcal{T}^{+}\rangle
  +\langle \mathcal{T}^{-}\rangle + Corr \approx 2 \langle
  \mathcal{T}\rangle + Corr$ where $\mathcal{T}^{\pm} \equiv
  \mathcal{T} ({\bf X} \pm {\bf r})$ and $Corr$ is the space-average
over the small field of view of the sum of all two-point correlation
terms making up $\langle T_X \rangle$. At the SFV20 station of the
$G/H=1.25$ wake, $\langle T_X \rangle$ and $\mathcal{T}$ are both
positive and, as shown in Appendix \ref{sec:appC_corr}, $Corr$ is negative and
decreasing in magnitude with increasing $r_1$ and $r_2$. At all the
other six stations considered here, $Corr$ is positive and generally
non-increasing in magnitude with increasing $r_1$ and $r_2$ above
$\lambda$. This is a striking illustration of the difference that we
observe in terms of qualitatively different two-point correlations
between the scale-by-scale non-homogeneity for SFV20 $G/H=1.25$ and
the scale-by-scale non-homogeneity at the other six stations
considered in the $G/H=2.4$ and $G/H=3.5$ wakes. The explanation of
this difference requires a more comprehensive analysis of our
turbulent wakes which goes beyond the present study.

We now give a detailed presentation of our results in the following
subsections, one gap ratio $G/H$ value at a time.
  (An even more detailed presentation of the
  inter-space energy transfer rate $\langle T_X \rangle$ in terms of
  its decomposition into a streamwise transfer rate $\langle {\partial \over
    \partial X_{1}} \overline{u'_{X1} \delta K_h} \rangle$ and a cross-stream
  transfer rate $\langle {\partial \over \partial X_{2}} \overline{u'_{X2}
    \delta K_h}\rangle$ is presented in Appendix \ref{appA} where it is shown
  that both are positive across a wide range of length-scales all the
  way down to the Taylor length.)

\subsection{Horizontal two-point turbulent energy transfer rates in the $G/H=3.5$ wake}
\begin{figure}
	\begin{subfigure}{0.97\textwidth}
	
		\includegraphics[width=0.96\textwidth]{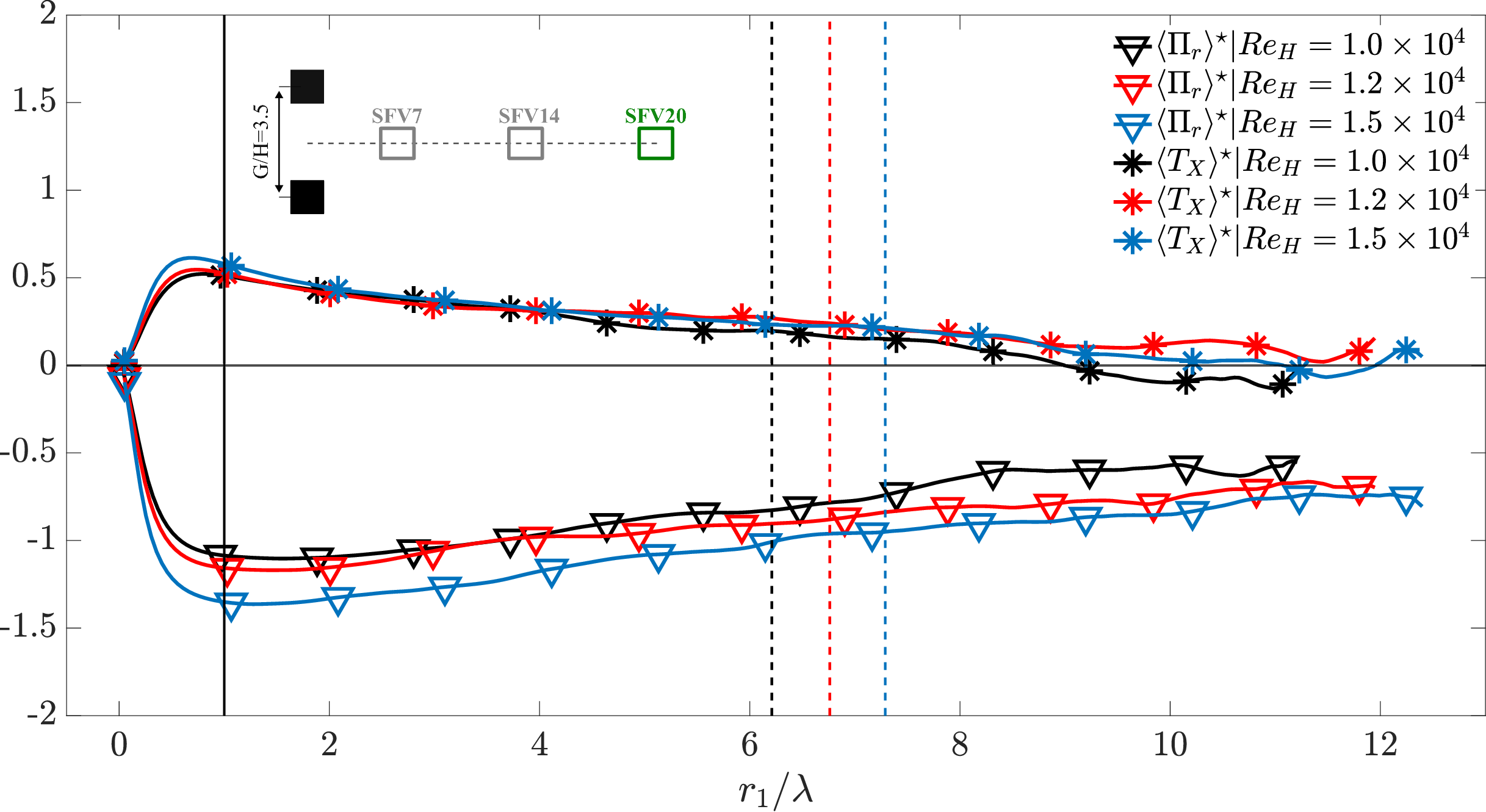}
		\caption{}
	\end{subfigure}
	\begin{subfigure}{0.97\textwidth}
	
		\includegraphics[width=0.96\textwidth]{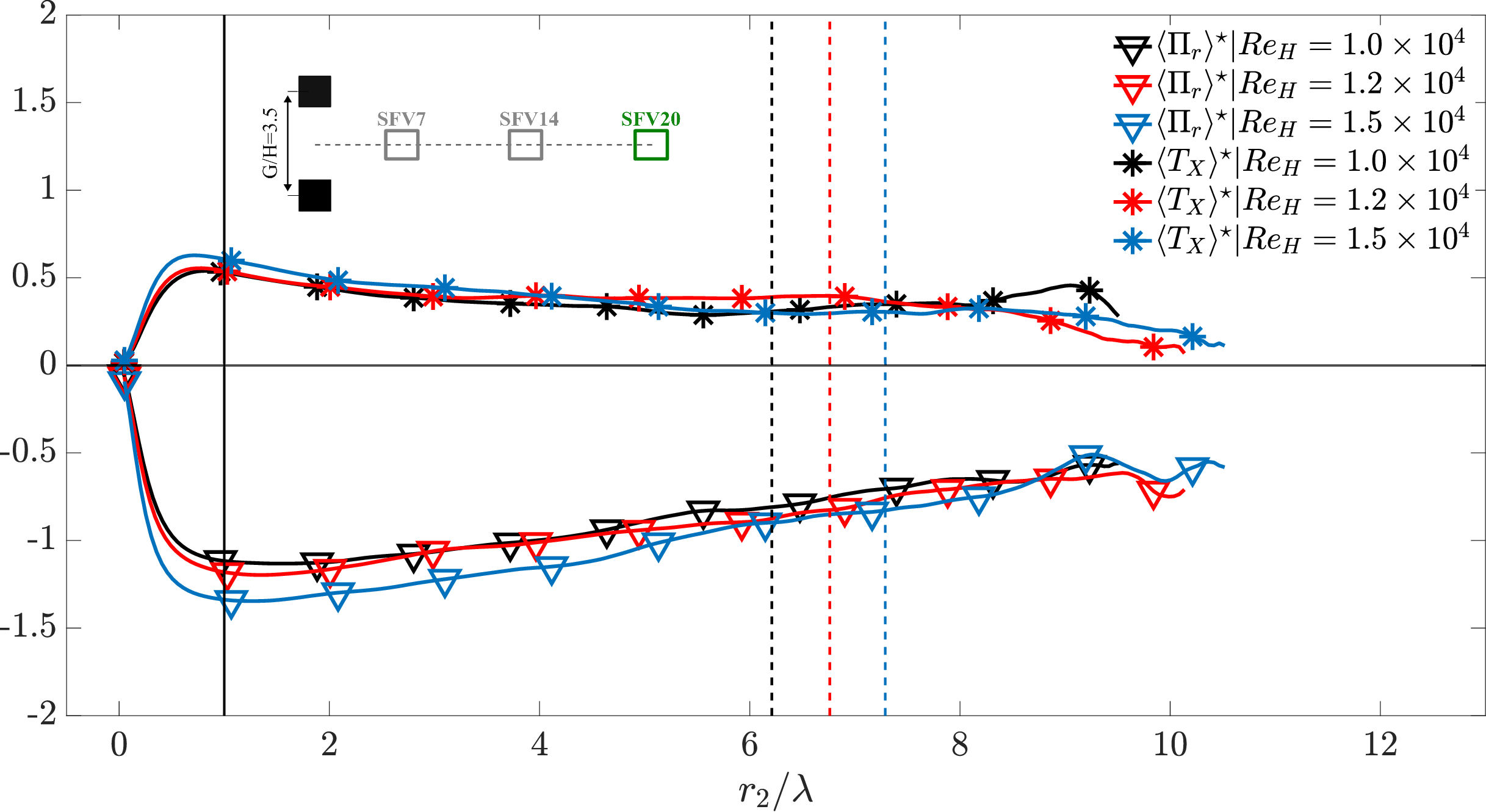}
		\caption{}
	\end{subfigure}
	\caption{Normalised average scale-by-scale energy transfer
		rates in scale $\langle \Pi_r \rangle^\star$ and in physical
		$\langle T_X \rangle^\star$ space for configuration
		$G/H=3.5$ SFV20. Plots are shown for separations scales in the
		streamwise (a) and in the cross-stream (b)
		direction. Vertical dashed lines: $\langle\mathcal{L}_v\rangle/\lambda$. Only 1/20
		of the markers are shown for clarity and a sketch is displayed to depict the $G/H$ and SFV (green box) probed here. The same is done in the following figures.}
	\label{fig:7.35mps_G_H_3.5_SFV20_transfer_rates}
\end{figure}

In Fig. \ref{fig:7.35mps_G_H_3.5_SFV20_transfer_rates} we plot
normalised inter-scale and inter-space energy transfer rates, denoted
as $\langle \Pi_r \rangle^\star \equiv \langle \Pi_r
\rangle/\avepsilonxt$ and $\langle T_X \rangle^\star\equiv \langle T_X
\rangle/\avepsilonxt$ respectively, for $SFV20$ in the $G/H=3.5$
turbulent wake for the three available global Reynolds numbers
$Re_H$. For all accessible values of $r_1$ and $r_2$ up to
significantly above $\langle \mathcal{L}_v \rangle$, $\langle \Pi_r
\rangle^\star <0$ and $\langle T_X \rangle^\star >0$, and $\langle T_X
\rangle^\star$ increases as either $r_1$ or $r_2$ decrease towards
$\lambda$ or $\lambda/2$ where it reaches its maximum at a value of
about $0.5$ or $0.6$ depending on $Re_H$. $\langle \Pi_r
\rangle^\star$ also increases in magnitude as either $r_1$ or $r_2$
decrease towards $\lambda$ where it reaches its minimum at a value of
about $-1.1$ for $Re_H = 1.0\times 10^4$ and $1.2\times 10^4$ and of
about $-1.4$ for $Re_H = 1.5\times 10^4$. Whilst $\langle T_X
\rangle^\star$ does not vary significantly with $Re_H$ in the small
range of $Re_H$ values considered, $-\langle \Pi_r \rangle^\star$ is
significantly larger for $Re_H = 1.5\times 10^4$ than for $Re_H =
1.0\times 10^4$ and $1.2\times 10^4$ over all $r_1$ scales and most
$r_2$ scales under $\langle \mathcal{L}_v \rangle$. Note that
$-\langle \Pi_r \rangle^\star$ is larger than 1 for either $r_1$ or
$r_2$ between $\lambda /2$ and $\langle \mathcal{L}_v \rangle /3$,
i.e. over the inertial range.

\begin{figure}
	\centering
	\begin{subfigure}{0.49\textwidth}
		\includegraphics[width=\textwidth]{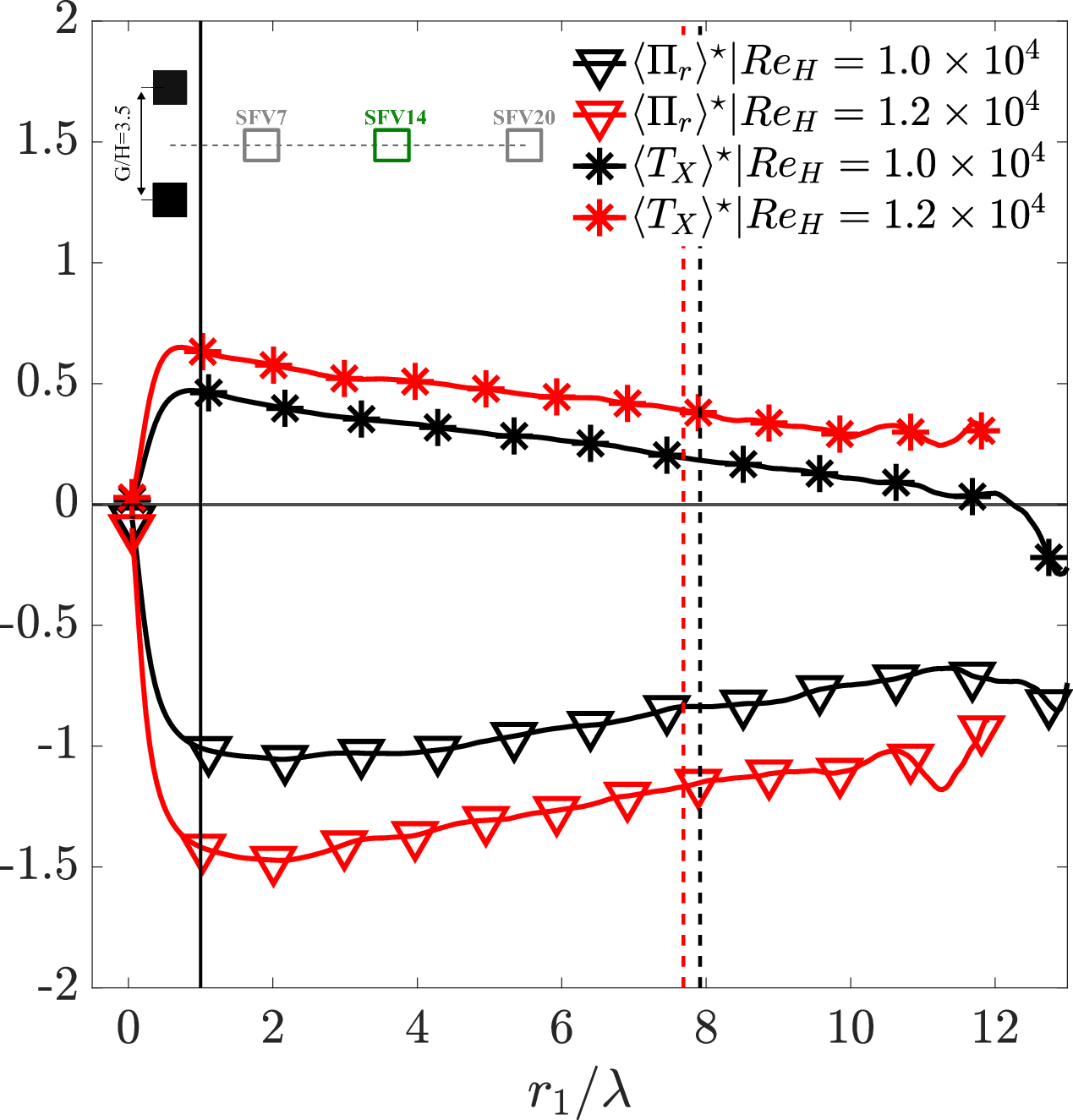}
		\caption{}
	\end{subfigure}
	\begin{subfigure}{0.49\textwidth}
		\includegraphics[width=\textwidth]{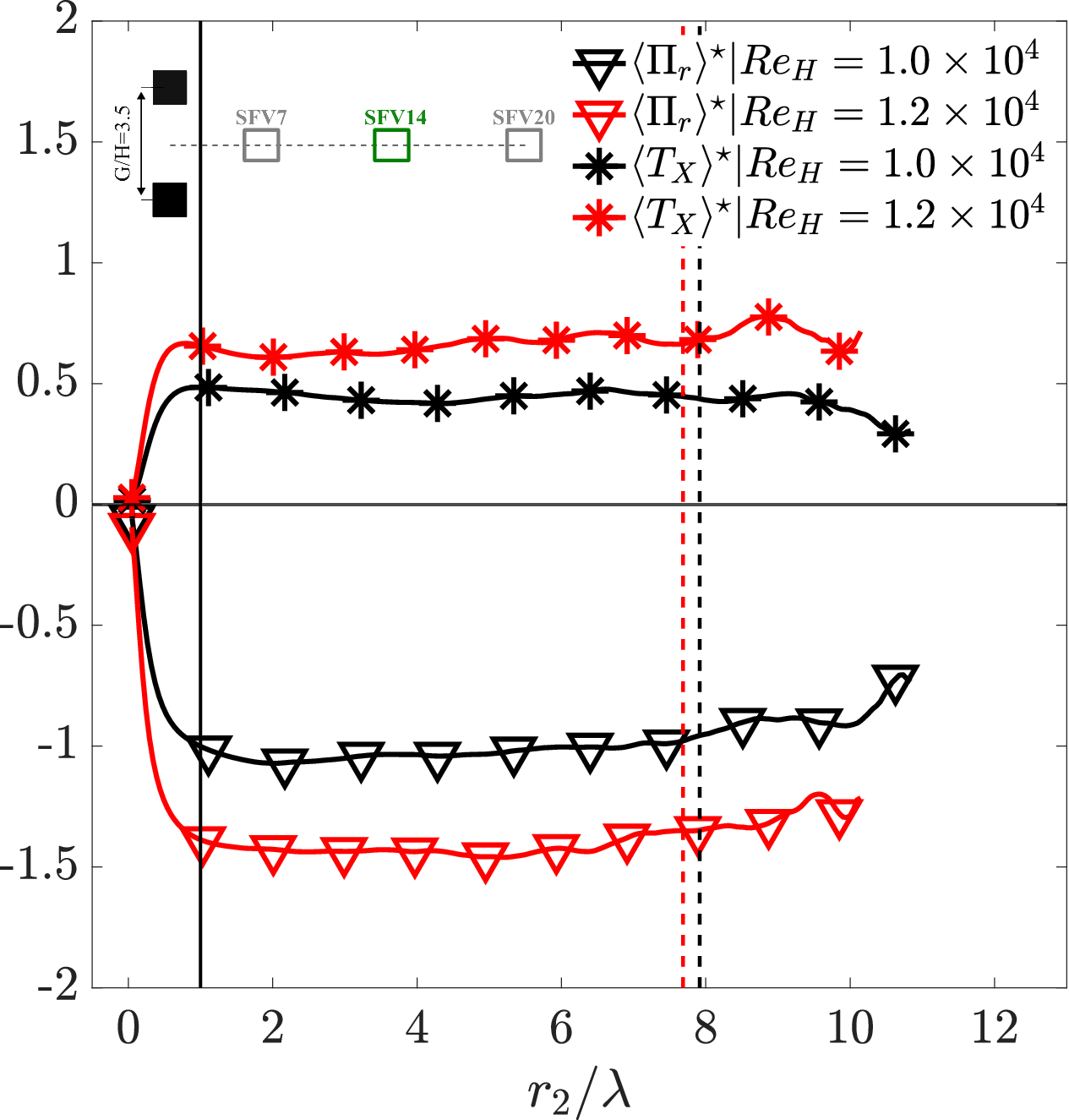}
		\caption{}
	\end{subfigure}

	\caption{Normalised average scale-by-scale energy transfer
		rates in scale $\langle \Pi_r \rangle^\star$ and in physical
		$\langle T_X \rangle^\star$ space for configuration
		$G/H=3.5$ SFV14 at $Re_H=1.0$ and $1.2 \times 10^4$. Plots
		are shown for separations scales in the streamwise (a)
		and in the cross-stream (b) direction.}
	\label{fig:G_H_3.5_SFV14_transfer_rates}
\end{figure}

\begin{figure}
	\centering
	\begin{subfigure}{0.49\textwidth}
		\includegraphics[width=\textwidth]{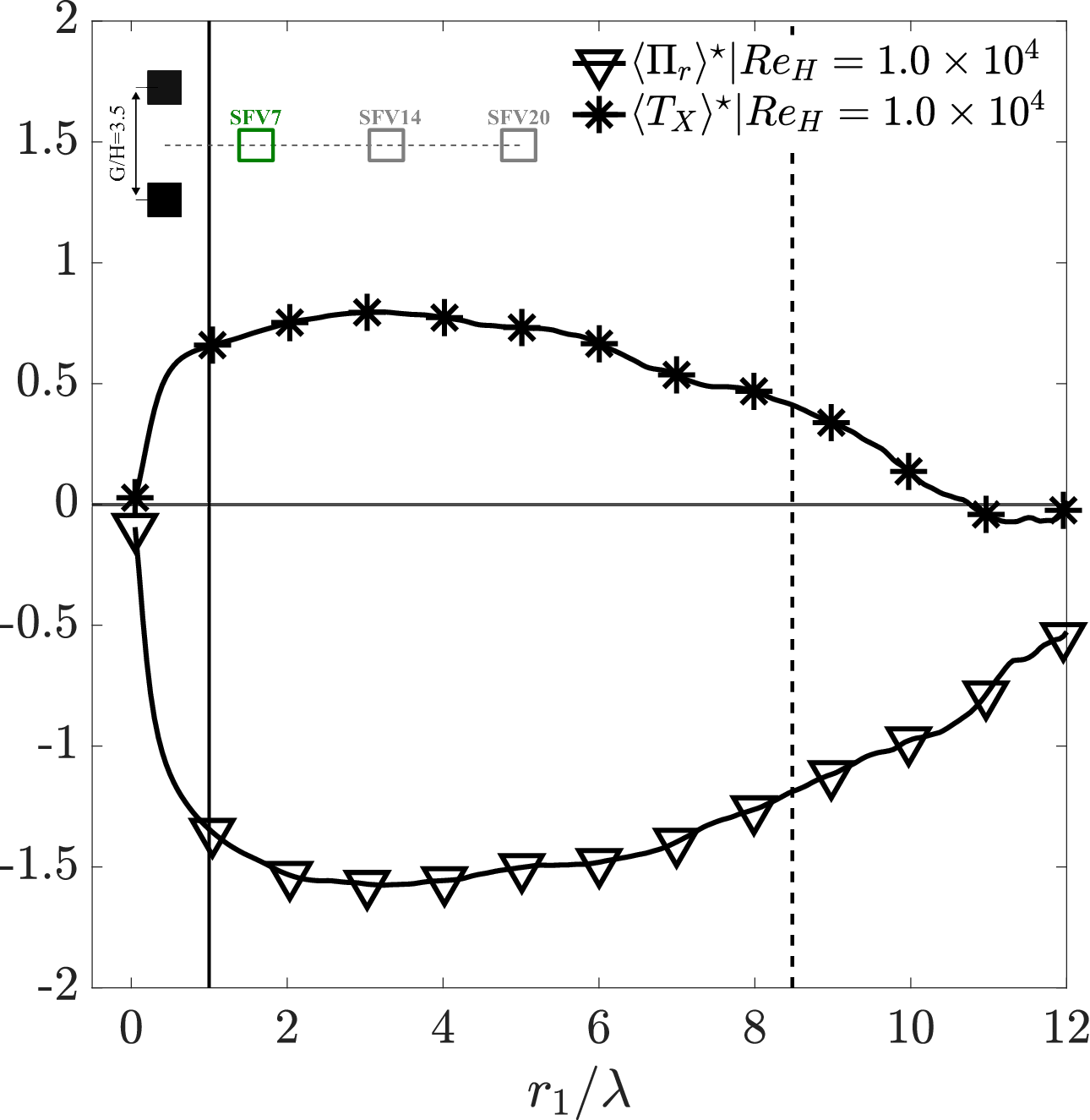}
		\caption{}
	\end{subfigure}
	\begin{subfigure}{0.49\textwidth}
		\includegraphics[width=\textwidth]{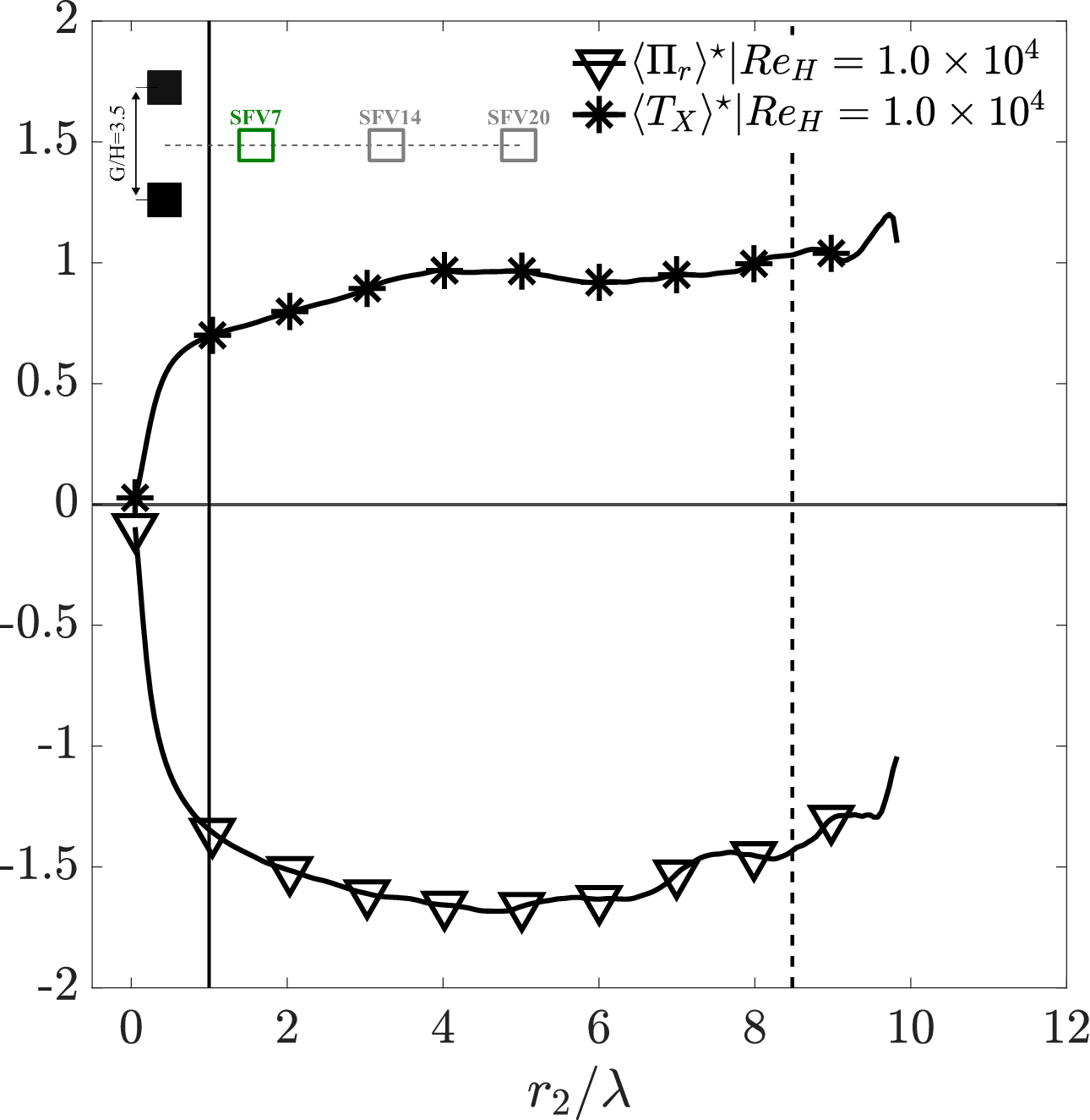}
		\caption{}
	\end{subfigure}
	
	\caption{Normalised average scale-by-scale energy transfer
		rates in scale $\langle \Pi_r \rangle^\star$ and in physical
		$\langle T_X \rangle^\star$ space for configuration
		$G/H=3.5$ SFV7 at $Re_H=1.0 \times 10^4$. Plots are shown
		for separations scales in the streamwise (a) and in the
		cross-stream (b) direction.}
	\label{fig:G_H_3.5_SFV7_transfer_rates}
\end{figure}

The same plots are shown in
Fig. \ref{fig:G_H_3.5_SFV14_transfer_rates} and
Fig. \ref{fig:G_H_3.5_SFV7_transfer_rates} but for SFV14 and SFV7,
respectively and the same observations are made. In particular,
$\langle T_X \rangle^\star$ is equal to or larger than $0.4$ at $r_1
=\lambda/2$ and $r_2 = \lambda/2$ and equal to or larger than $0.5$ at
$r_1 =\lambda/2$ and $r_2 = \lambda/2$ in all three small fields of
view for all available global Reynolds numbers. The turbulence is
therefore invariably very significantly non-homogeneous at the
smallest turbulent length-scales. However, one also observes the
following quantitative differences.

(i) There is a much greater sensitivity on $Re_H$ at SFV14 than SFV20
which are the two fields of view for which two values of $Re_H$ are
available. The normalised average two-point inter-scale and
inter-space transfer rates are approximately the same at SFV20 for
$Re_H = 1.0\times 10^4$ and $Re_H = 1.2\times 10^4$ but they both
increase very significantly in magnitude with this 20\% jump in $Re_H$
at SFV14.

(ii) At constant $Re_H$, $\langle \Pi_r \rangle^\star$ and $\langle
T_X \rangle^\star$ increase in magnitude as the small field of view
gets closer to the square prisms along the centreline and the local
$Re_{\lambda}$ increases appreciably,
i.e. from SFV20 ($Re_{\lambda} = 135$) to SFV14 ($Re_{\lambda}= 167$)
for $Re_H =1.2\times 10^4$ and from SFV14 ($Re_{\lambda} = 129$) to
SFV7 ($Re_{\lambda} = 183$) for $Re_H = 1.0\times 10^4$. (There is no
significant increase in $Re_{\lambda}$ from SFV20 to SFV14 for $Re_H =
1.0\times 10^4$ and no significant change in the $r_1$ and $r_2$
profiles of $\langle \Pi_r \rangle^\star$ and $\langle T_X
\rangle^\star$.)

(iii) From SFV20 ($Re_{\lambda} = 135$) to SFV14 ($Re_{\lambda}= 167$)
for $Re_H =1.2\times 10^4$, there is a tendency for $\langle \Pi_r
\rangle^\star$ and $\langle T_X \rangle^\star$ to get closer to
constant (near $-1.5$ for $\langle \Pi_r \rangle^\star$ and near $0.6$
for $\langle T_X \rangle^\star$) with either $r_1$ or $r_2$ in the
inertial range between $\lambda$ and $\langle \mathcal{L}_v \rangle$.
There may be a similar but less well-defined tendency from SFV14
($Re_{\lambda} = 129$) to SFV7 ($Re_{\lambda} = 183$) for $Re_H =
1.0\times 10^4$.

\begin{figure}
	\centering
	\begin{subfigure}{0.49\textwidth}
		\includegraphics[width=\textwidth]{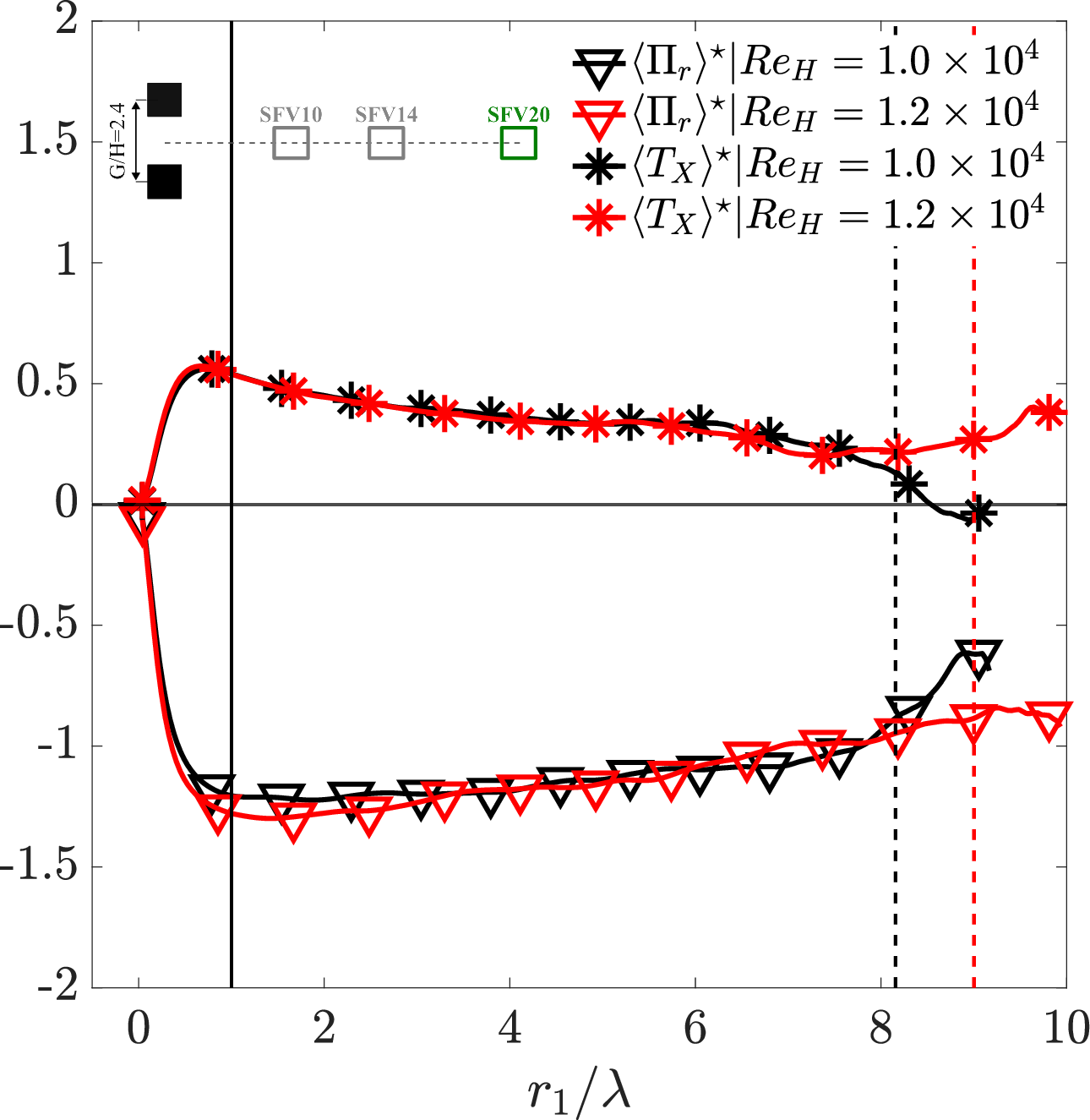}
		\caption{}
	\end{subfigure}
	\begin{subfigure}{0.49\textwidth}
		\includegraphics[width=\textwidth]{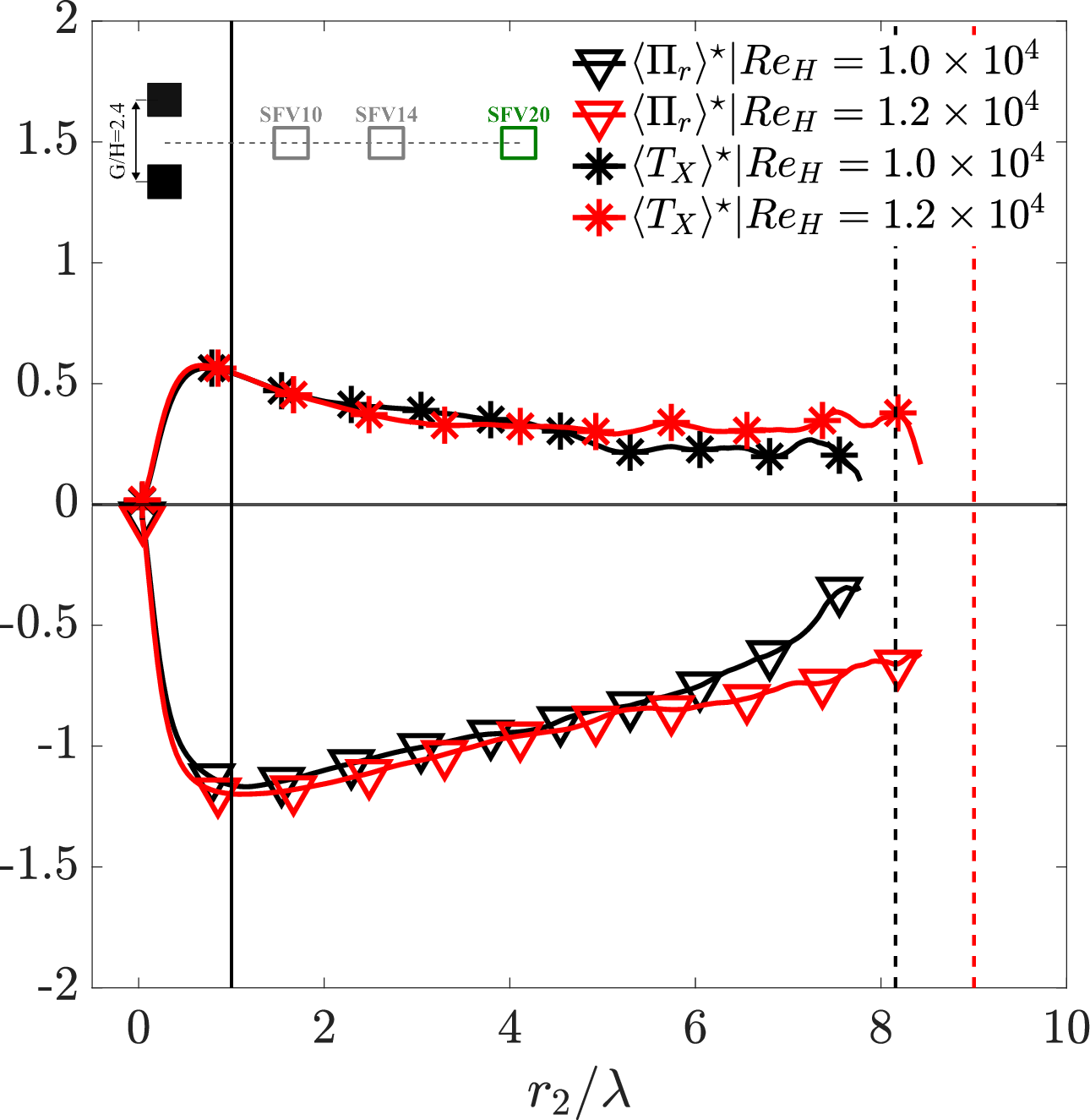}
		\caption{}
	\end{subfigure}
	\caption{Normalised average scale-by-scale energy transfer
		rates in scale $\langle \Pi_r \rangle^\star$ and in physical
		$\langle T_X \rangle^\star$ space for configuration
		$G/H=2.4$ SFV20 at $Re_H=1.0$ and $1.2 \times 10^4$. Plots
		are shown for separations scales in the streamwise (a)
		and in the cross-stream (b) direction.}
	\label{fig:6mps_G_H_2.4_SFV20_transfer_rates}
\end{figure}

Let us now consider a different turbulent wake by decreasing the gap
ratio from $G/H=3.5$ to $G/H=2.4$.

\subsection{Horizontal two-point turbulent energy transfer rates in the $G/H=2.4$ wake}

If one compares Figs.  \ref{fig:7.35mps_G_H_3.5_SFV20_transfer_rates}
and \ref{fig:6mps_G_H_2.4_SFV20_transfer_rates} (same SFV, different
wake), the following remarks can be made. The $r_1$ and $r_2$ profiles
of $\langle \Pi_r \rangle^\star$ and $\langle T_X \rangle^\star$ at
SFV20 are very similar in the $G/H=2.4$ and $G/H=3.5$ turbulent wakes
for the two global Reynolds numbers $Re_H = 1.0\times 10^4$ and $Re_H
= 1.2\times 10^4$, except that the local $Re_{\lambda}$ values and
$\langle\mathcal{L}_v\rangle/\lambda$ ratios are higher for $G/H=2.4$
at the same $Re_H$. Our observations and conclusions for SFV20
$G/H=2.4$ and SFV20 $G/H=3.5$ are therefore the same (even
quantitatively bearing in mind the $Re_{\lambda}$ difference) for
these two global Reynolds numbers.

\begin{figure}
		\centering
		\begin{subfigure}{0.49\textwidth}
			\includegraphics[width=\textwidth]{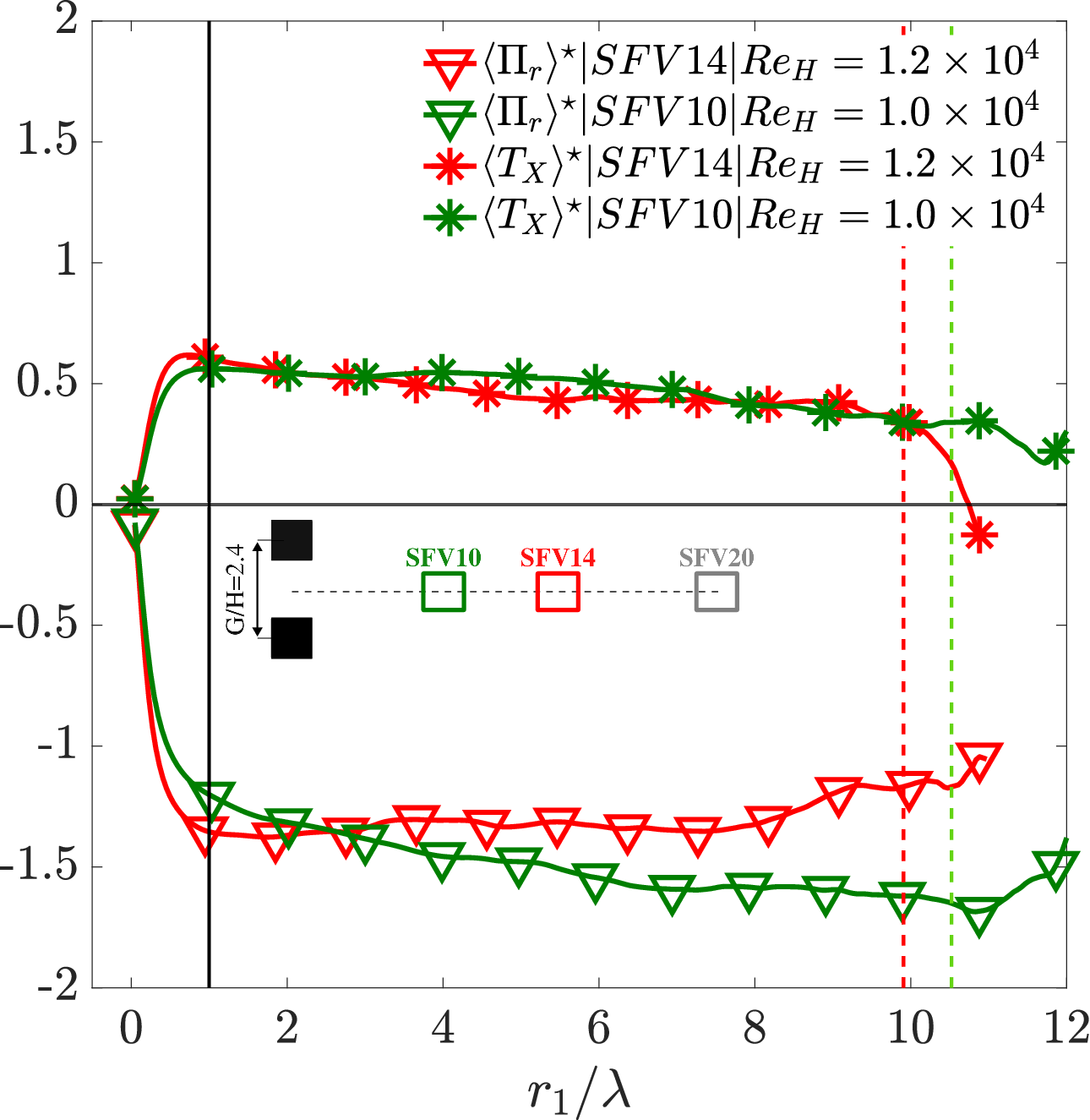}
			\caption{}
		\end{subfigure}
		\begin{subfigure}{0.49\textwidth}
			\includegraphics[width=\textwidth]{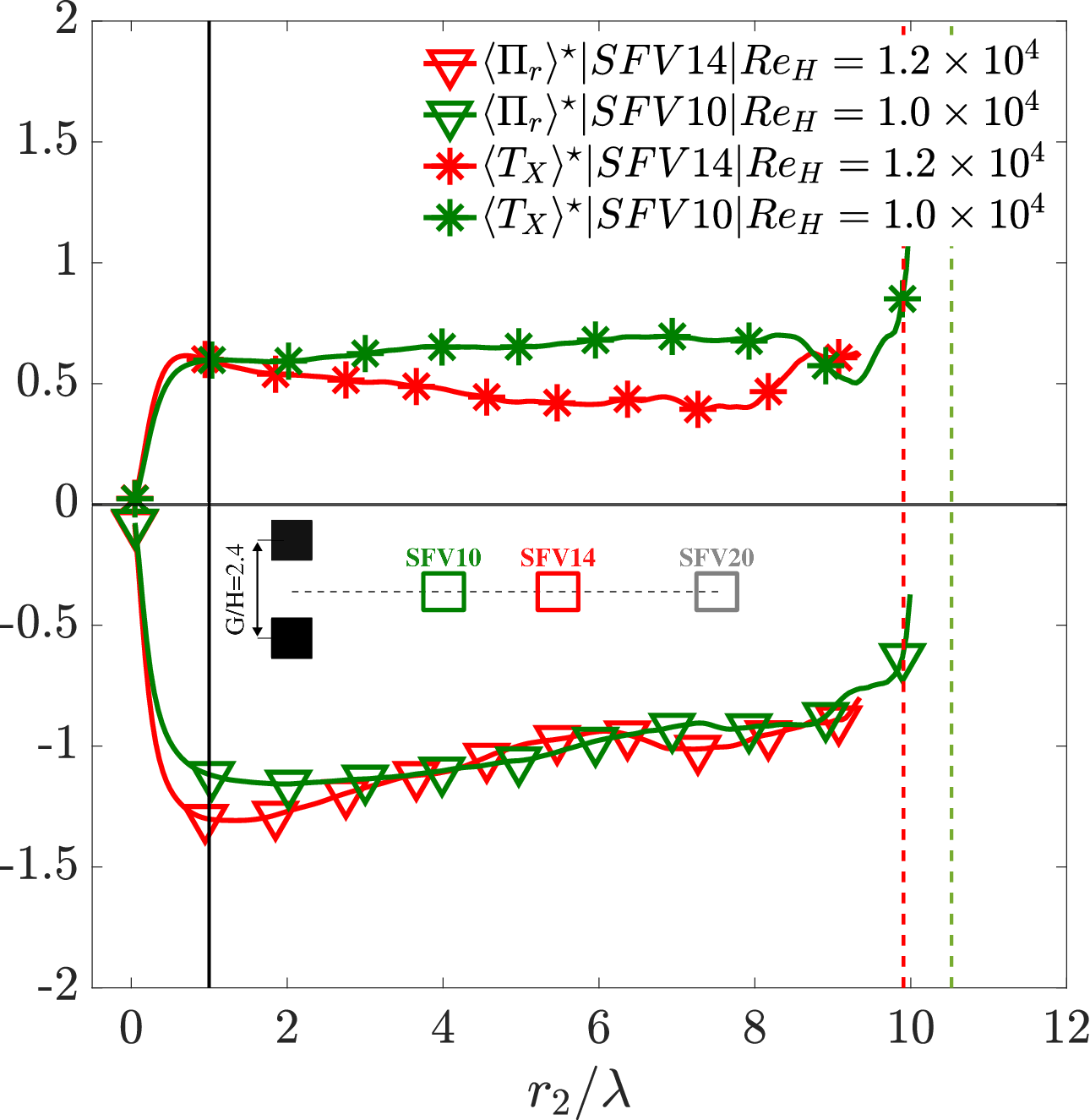}
			\caption{}
		\end{subfigure}
		
		\caption{Normalised average scale-by-scale energy
			transfer rates in scale $\langle \Pi_r
			\rangle^\star$ and in physical $\langle T_X
			\rangle^\star$ space for configuration $G/H=2.4$
			SFV14 at $Re_H=1.2 \times 10^4$ and SFV10 at
			$Re_H=1.0 \times 10^4$. Plots are shown for
			separations scales in the streamwise (a) and in
			the cross-stream (b) direction.}
		\label{fig:6mps_G_H_2.4_SFV14_transfer_rates}
	\end{figure}

In Fig. \ref{fig:6mps_G_H_2.4_SFV14_transfer_rates} we plot the $r_1$
and $r_2$ profiles of $\langle \Pi_r \rangle^\star$ and $\langle T_X
\rangle^\star$ closer to the square prisms along the centreline, at
SFV14 and SFV10. There is only one global Reynolds number available
for each one of these small fields of view, $Re_H = 1.2\times 10^4$
for SFV14 and $Re_H = 1.0\times 10^4$ for SFV7 but
$Re_{\lambda}\approx 185$ for both. Whilst the inter-scale and
inter-space transfer rates behave at SFV14 more or less as they do at
SFV20 for $G/H=2.4$ (see Fig.
\ref{fig:6mps_G_H_2.4_SFV20_transfer_rates}), a substantial difference
appears at the SFV10 station of the $G/H=2.4$ wake compared to all
other stations examined in this paper in both the $G/H=2.4$ and
$G/H=3.5$ wakes. It is the only one of these stations where $\langle
\Pi_r \rangle^\star$ and $\langle T_X \rangle^\star$ decrease
monotonically together with increasing $r_1$ and where $\langle \Pi_r
\rangle^\star$ and $\langle T_X \rangle^\star$ increase monotonically
together with increasing $r_2$. This difference is particularly
striking by comparison to the SFV14 station of the same wake at a 20\%
higher global Reynolds number even though they both have the same
$Re_{\lambda}$ (see Fig.
\ref{fig:6mps_G_H_2.4_SFV14_transfer_rates}). Note, however, that in
line with the other stations in the $G/H=3.5$ and $G/H=2.4$ wakes,
$\langle \Pi_r \rangle^\star$ is particularly high in magnitude,
between $-1$ to $-1.6$ in the inertial range $\lambda$ to
$\langle\mathcal{L}_v\rangle$. Furthermore, $\langle T_X
\rangle^\star$ is consistently above $0.3$ and it is equal to or
larger than $0.5$ at $r_1 =\lambda/2$ and $r_2 = \lambda/2$. Once
again, the turbulence is very significantly non-homogeneous at the
smallest, and in fact viscosity-affected, turbulent
length-scales. There is clearly no tendency towards local homogeneity
at small enough turbulent length scales in any of the small fields of
view in the centreline decay region of both $G/H=3.5$ and $G/H=2.4$
turbulent wakes and for all available global Reynolds numbers.

\subsection{The other terms in the scale-by-scale budget}

\begin{figure}
	\centering
	\begin{subfigure}{0.98\textwidth}
		\includegraphics[width=\textwidth]{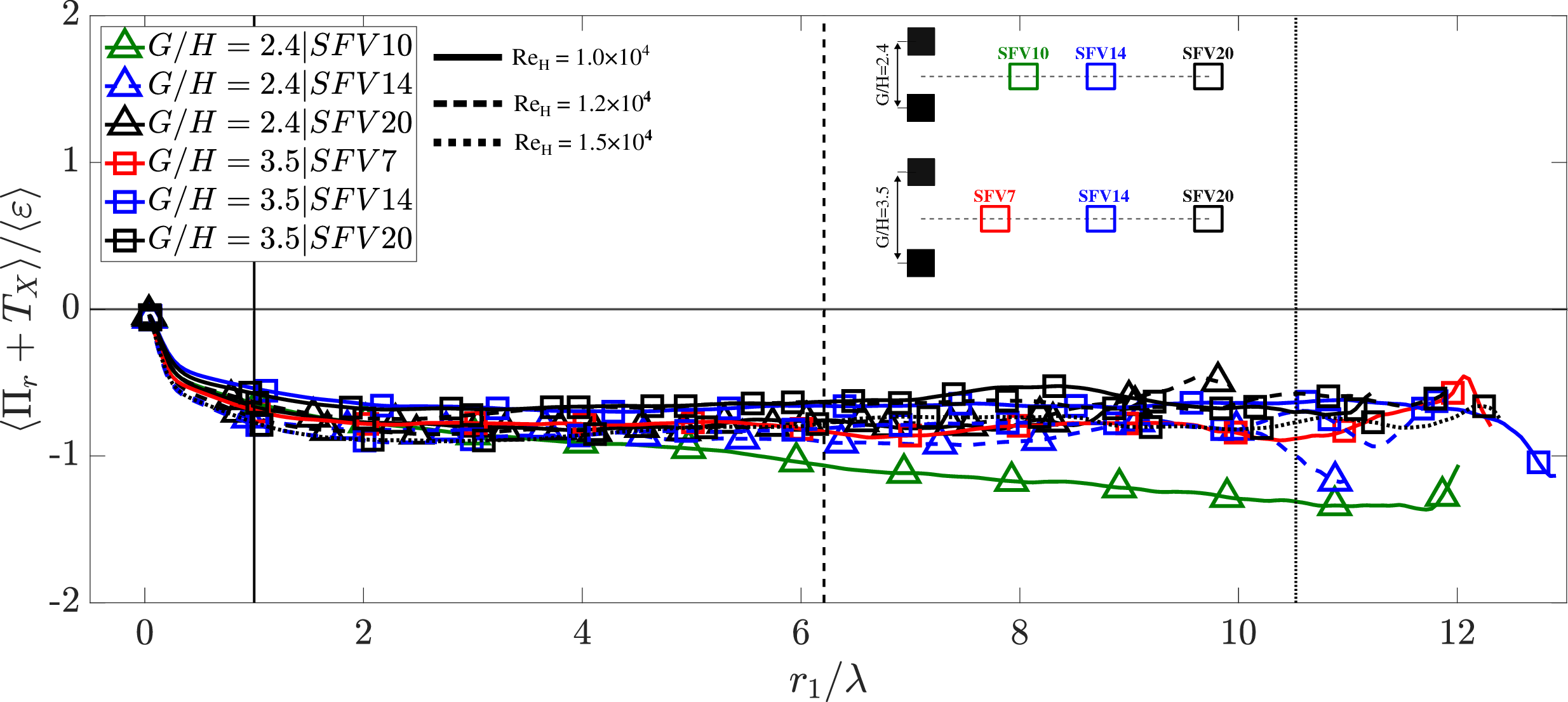}
		\caption{}
	\end{subfigure}
	\begin{subfigure}{0.98\textwidth}
		\includegraphics[width=\textwidth]{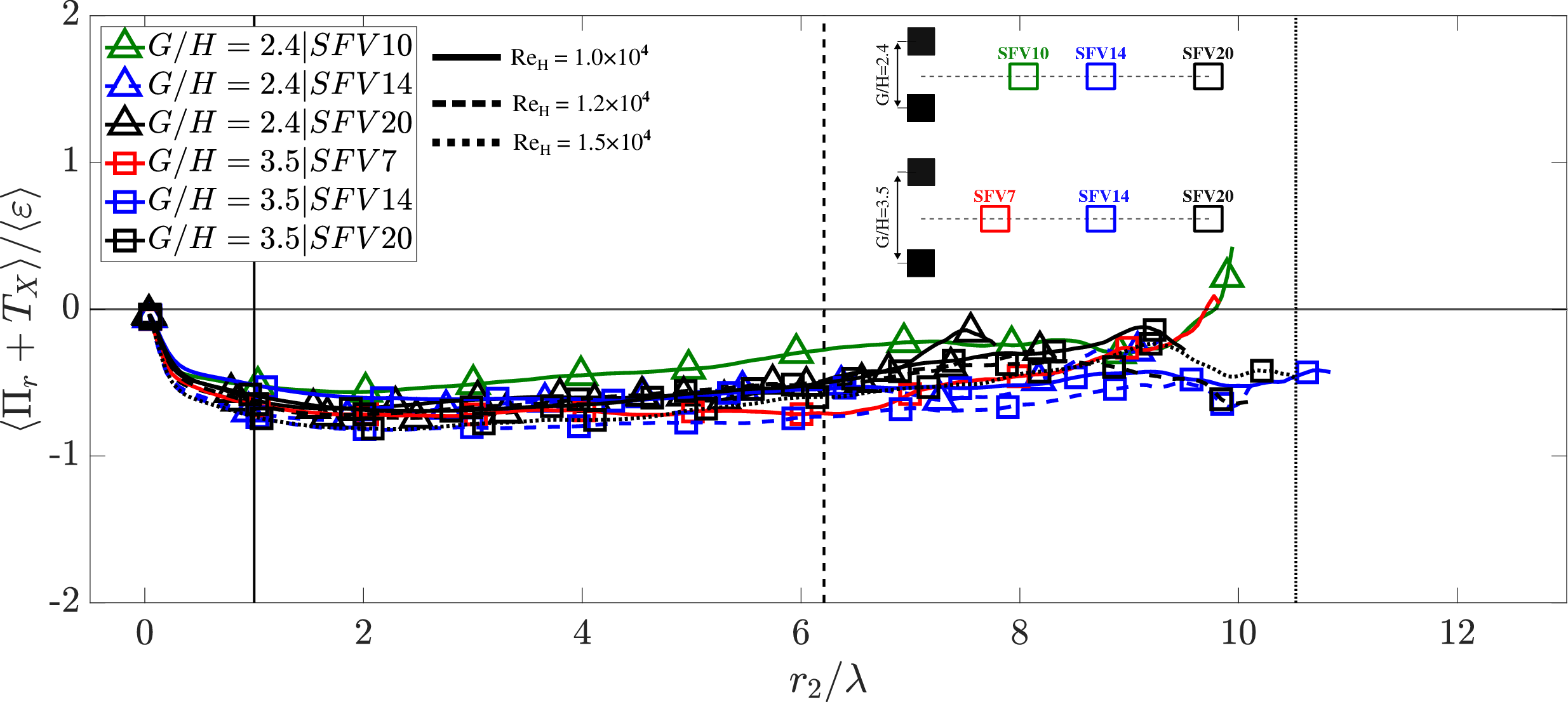}
		\caption{}
	\end{subfigure}
	\caption{Normalised scale-by-scale plots of the sum of
		interscale and interspace small scale energy transfer rate
		$\langle \Pi_r+T_X
		\rangle$/$\avepsilonxt$. Results are shown for separations scales in the streamwise (a)
		and in the cross-stream (b) directions. The vertical
		dashed and dotted lines represent the smallest and largest
		integral length scales for the SFVs, respectively.}
	\label{fig:sum_transfers_Far_fields}
\end{figure}

In all the SFV$\mathcal{N}$ stations of the $G/H=2.4$ and $G/H=3.5$
turbulent wakes that we examined except SFV10 for $G/H=2.4$, $\langle
\Pi_r \rangle^\star$ ranges from around -1 to -1.5 and remains
constant or increases with increasing $r_1$ or $r_2$ whereas $\langle
T_X \rangle^\star$ is close to $0.5$ and remains constant or decreases
with increasing $r_1$ or $r_2$. Consequently, $\langle \Pi_r
\rangle^\star + \langle T_X \rangle^\star$ remains constant over the
inertial range $\lambda$ to $\langle\mathcal{L}_v\rangle$ as shown in
Fig. \ref{fig:sum_transfers_Far_fields} except for SFV10,
$G/H=2.4$. This constant is between $-0.6$ and $-1.0$ depending on
wake and location in the decaying wake.

\begin{figure}
	\centering
	\begin{subfigure}{0.98\textwidth}
		\includegraphics[width=\textwidth]{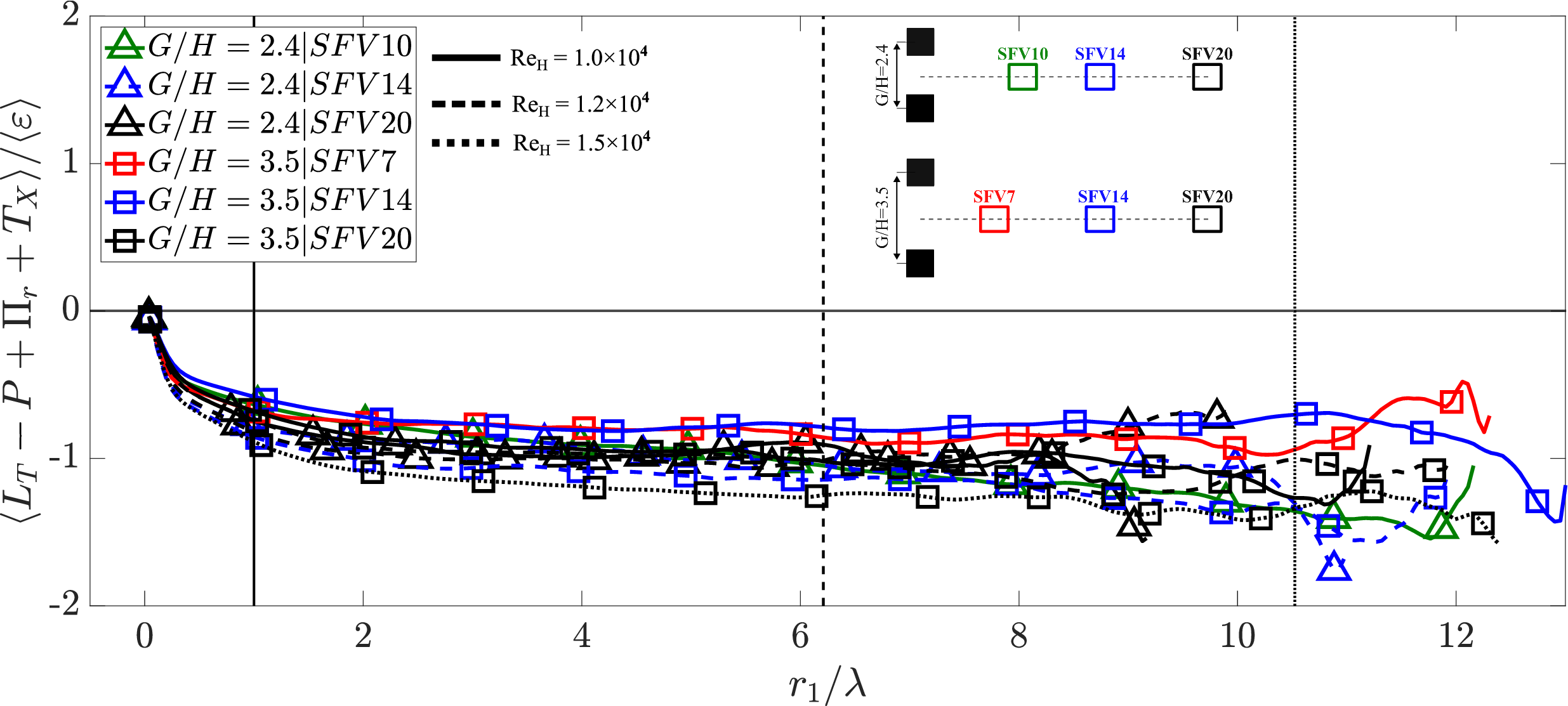}
		\caption{}
	\end{subfigure}
	\begin{subfigure}{0.98\textwidth}
		\includegraphics[width=\textwidth]{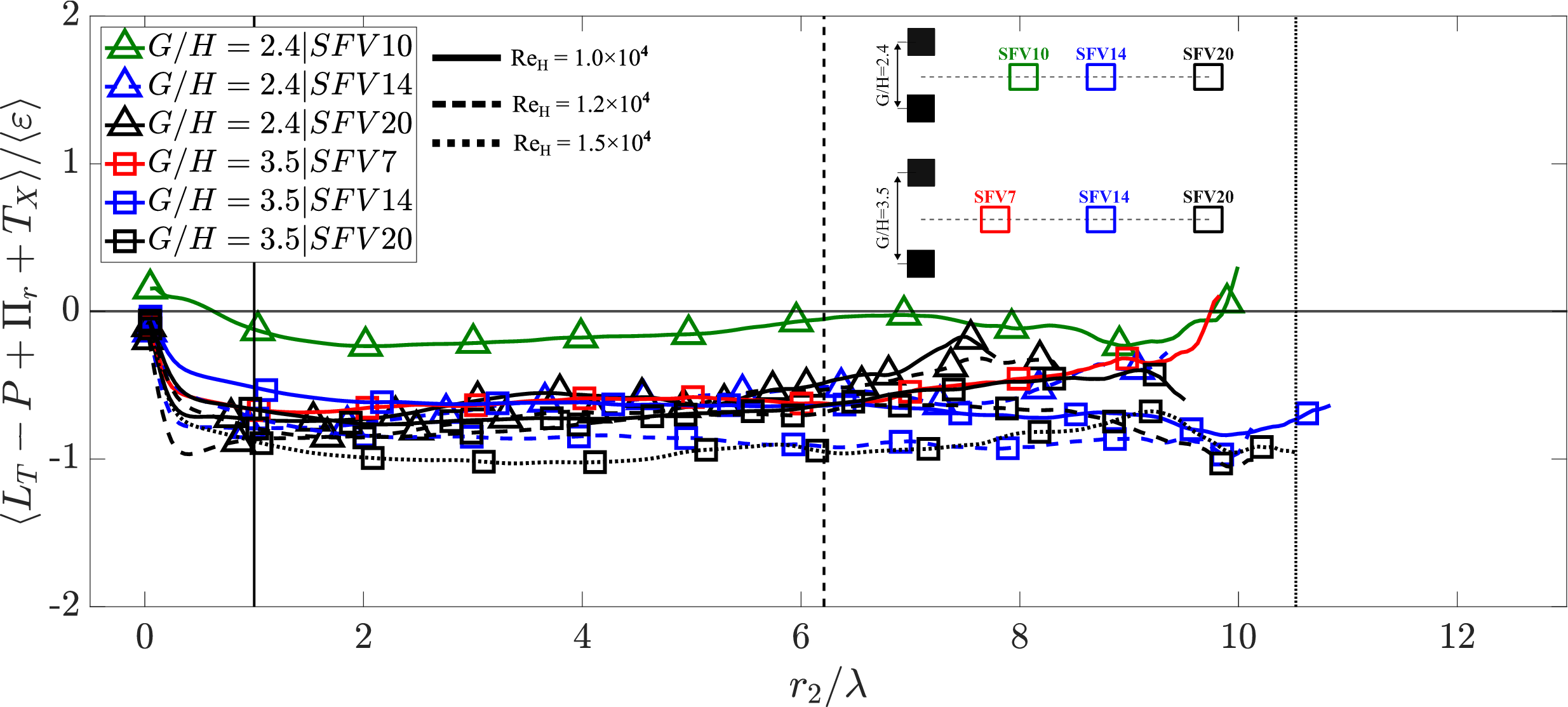}
		\caption{}
	\end{subfigure}
	\caption{Normalised scale-by-scale plots of the sum of
		interscale and interspace transfer rates, two-point production and linear transport rates. Results are shown for separations scales in the streamwise (a)
		and in the cross-stream (b) directions. The vertical
		dashed and dotted lines represent the smallest and largest
		integral length scales for the SFVs, respectively.}
	\label{fig:sum_KHMH_Far_fields}
\end{figure}

As already mentioned, the linear transport rate $L_T$ and the
two-point turbulence production rate $P$ are fully accessible from the
2D2C PIV data in our disposal. We find that they make a small
contribution to the scale-by-scale energy budget (\ref{eq:KHMH2}) in
all the SFV$\mathcal{N}$ stations of the $G/H=2.4$ and $G/H=3.5$
turbulent wakes examined in the present work except SFV10 for
$G/H=2.4$. This can be seen in Fig. \ref{fig:sum_KHMH_Far_fields}
which shows that $\langle L_T - P + T_X + \Pi_{r}\rangle/\langle
\varepsilon \rangle$ is approximately constant in the inertial range
$\lambda$ to $\langle\mathcal{L}_v\rangle$ with a constant which is
between $-0.7$ and $-1.3$ in the $r_1$ inertial range and between
$-0.5$ and $-1.0$ in the $r_2$ inertial range, with $SFV10$, $G/H=2.4$
being the one exception. It is worth pointing out that the
scale-by-scale energy budget (\ref{eq:KHMH2}) implies $\langle L_T - P
+ T_X + \Pi_{r}\rangle \approx \langle T_p \rangle - \langle
\Pi_z\rangle - \langle \varepsilon_{1}\rangle -\langle
\varepsilon_{2}\rangle$ for $r_1 , r_2 > \lambda$, suggesting the
perhaps remarkable balance
\begin{equation}
 \langle T_p \rangle - \langle \Pi_z\rangle \sim \langle \varepsilon
 \rangle
 	\label{eq:TpPi}
\end{equation}  
in the inertial $r_1$ and $r_2$ ranges, again with the one SFV10,
$G/H=2.4$ exception.

\subsection{The SFV20 station in the $G/H=1.25$ wake}

 \begin{figure}
	\centering
	\begin{subfigure}{0.49\textwidth}
		\includegraphics[width=\textwidth]{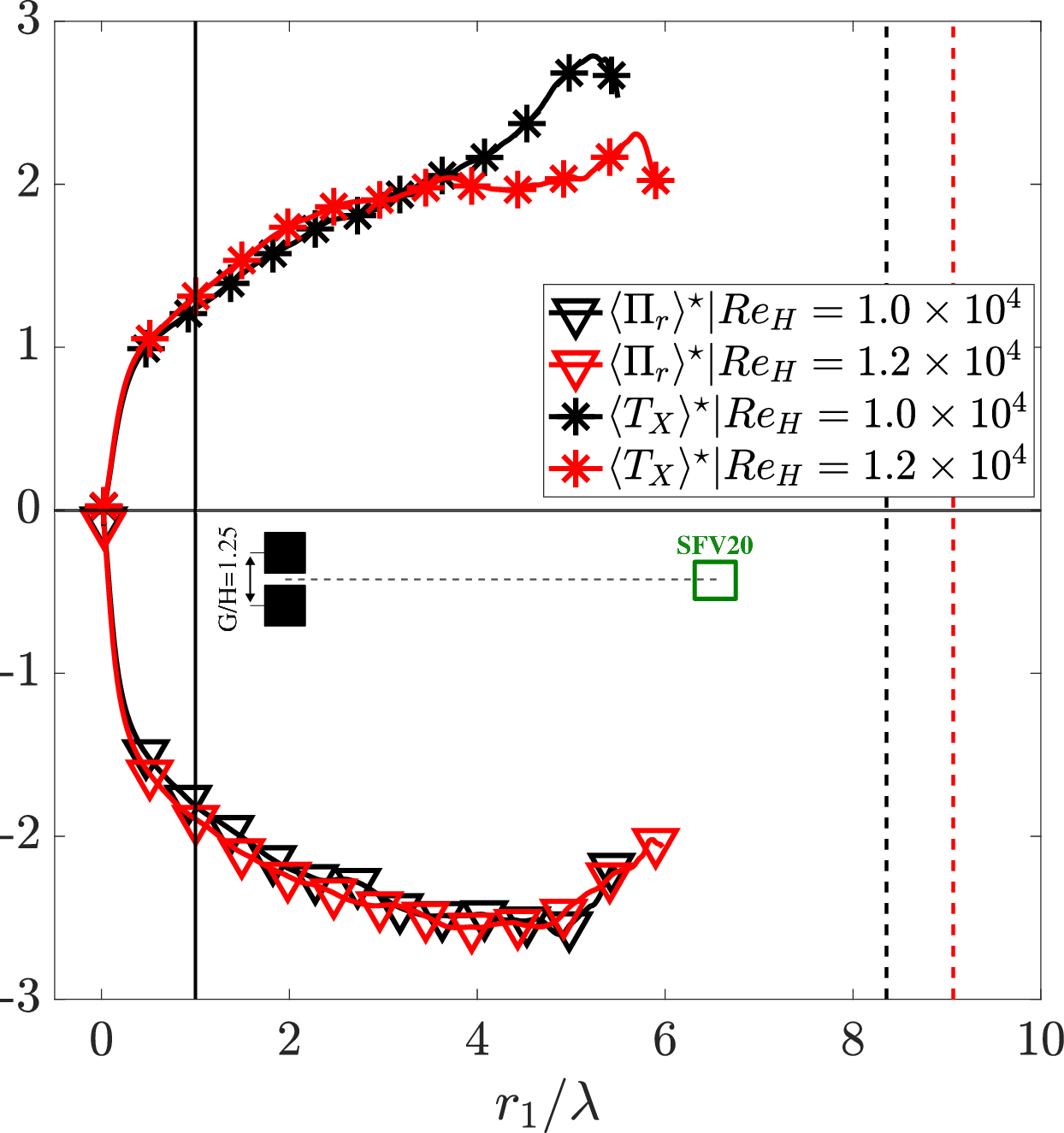}
		\caption{}
	\end{subfigure}
	\begin{subfigure}{0.49\textwidth}
		\includegraphics[width=\textwidth]{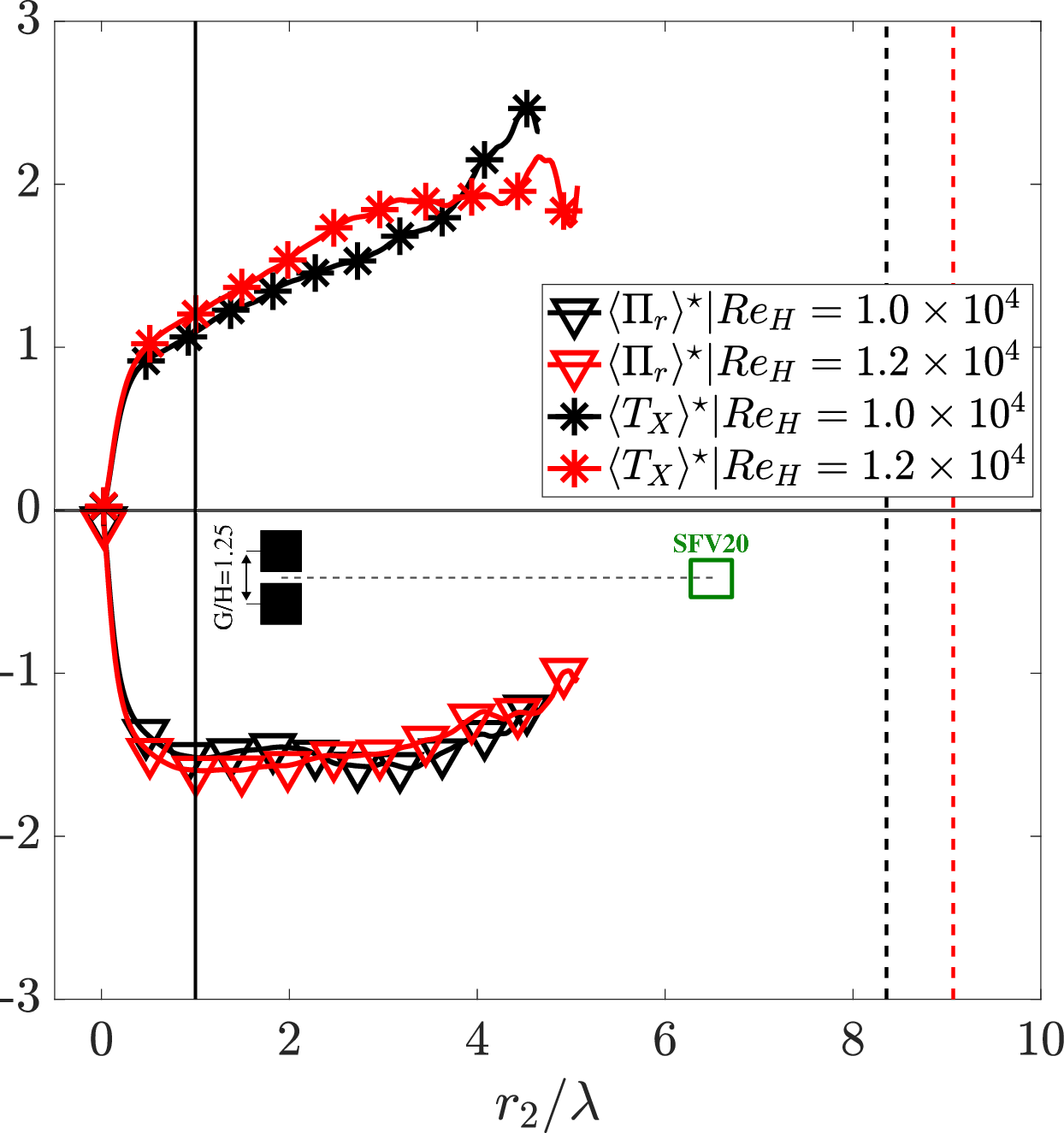}
		\caption{}
	\end{subfigure}
	\caption{Normalised average scale-by-scale energy transfer
		rates in scale $\langle \Pi_r \rangle^\star$ and in physical
		$\langle T_X \rangle^\star$ space for configuration
		$G/H=1.25$ SFV20 at $Re_H=1.0$ and $1.2 \times 10^4$. Plots
		are shown for separations scales in the streamwise (a)
		and in the cross-stream (b) direction.}
	\label{fig:G_H_1.25_SFV20_transfer_rates}
\end{figure}

The exceptional SFV10 station in the $G/H=2.4$ turbulent wake is at or
close to the border between the near field where the turbulence and
$Re_{\lambda}$ increase and the subsequent field where the turbulence
and $Re_{\lambda}$ decrease with streamwise distance (see table
\ref{tab:G_H_2.4_charac}). It is not clear how close to the streamwise
border between increasing and decreasing $Re_{\lambda}$ the SFV20
station is in the $G/H=1.25$ wake. Table \ref{tab:G_H_1.25_charac}
shows that $Re_{\lambda}$ increases from SFV7 to SFV20 in this wake
and we can therefore expect SFV20 in the $G/H=1.25$ wake to be
exceptional too. Nevertheless, $\langle T_X \rangle$ and $\langle
\Pi_{r}\rangle$ have opposite signs, specifically $\langle T_X \rangle
>0$ and $\langle \Pi_{r}\rangle <0$, and non-homogeneity is very
significant down to the smallest turbulent scales (see
Fig. \ref{fig:G_H_1.25_SFV20_transfer_rates}) as for all the stations
studied here in the other two wakes. There is no tendency towards
local homogeneity at small enough turbulent length scales even at the
$SFV20$ station of the $G/H=1.25$ wake where $Re_{\lambda}$ nears 500.

The SFV20 station of the $G/H=1.25$ wake is actually unique among all
the other stations studied in this paper in that $\langle T_X
\rangle^\star$ grows very sharply with both $r_1$ and $r_2$, in fact
reaching values between 2 and 3 at $r_1$, $r_2$ between $4\lambda$ and
$6\lambda$, see Fig. \ref{fig:G_H_1.25_SFV20_transfer_rates}. The
$G/H=1.25$ wake seems to be the one of the three wakes with
recirculations reaching the furthest downstream as illustrated in
Fig. \ref{fig:SFVs_visu} suggesting stronger non-homogeneity over a
longer streamwise distance in multiples of $H$. In fact
non-homogeneity is not only stronger at larger but also at the
smallest scales in this wake's SFV20, with $\langle T_X \rangle^\star$
between $0.6$ and $0.7$ at both $r_1 =\lambda/4$ and $r_2 =\lambda/4$.

\begin{figure}
	\centering
	\begin{subfigure}{0.49\textwidth}
		\includegraphics[width=\textwidth]{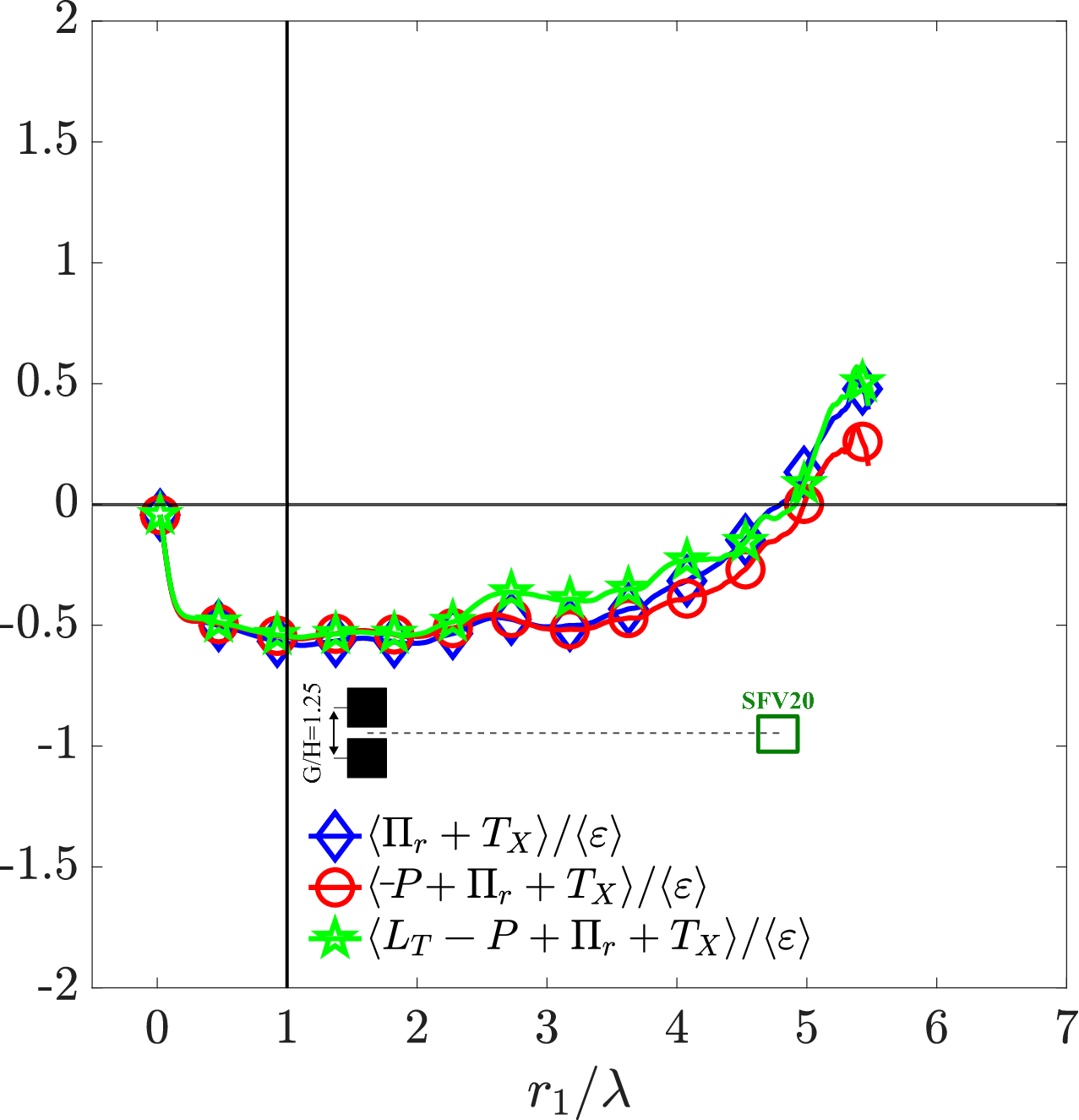}
		\caption{}
	\end{subfigure}
	\begin{subfigure}{0.49\textwidth}
		\includegraphics[width=\textwidth]{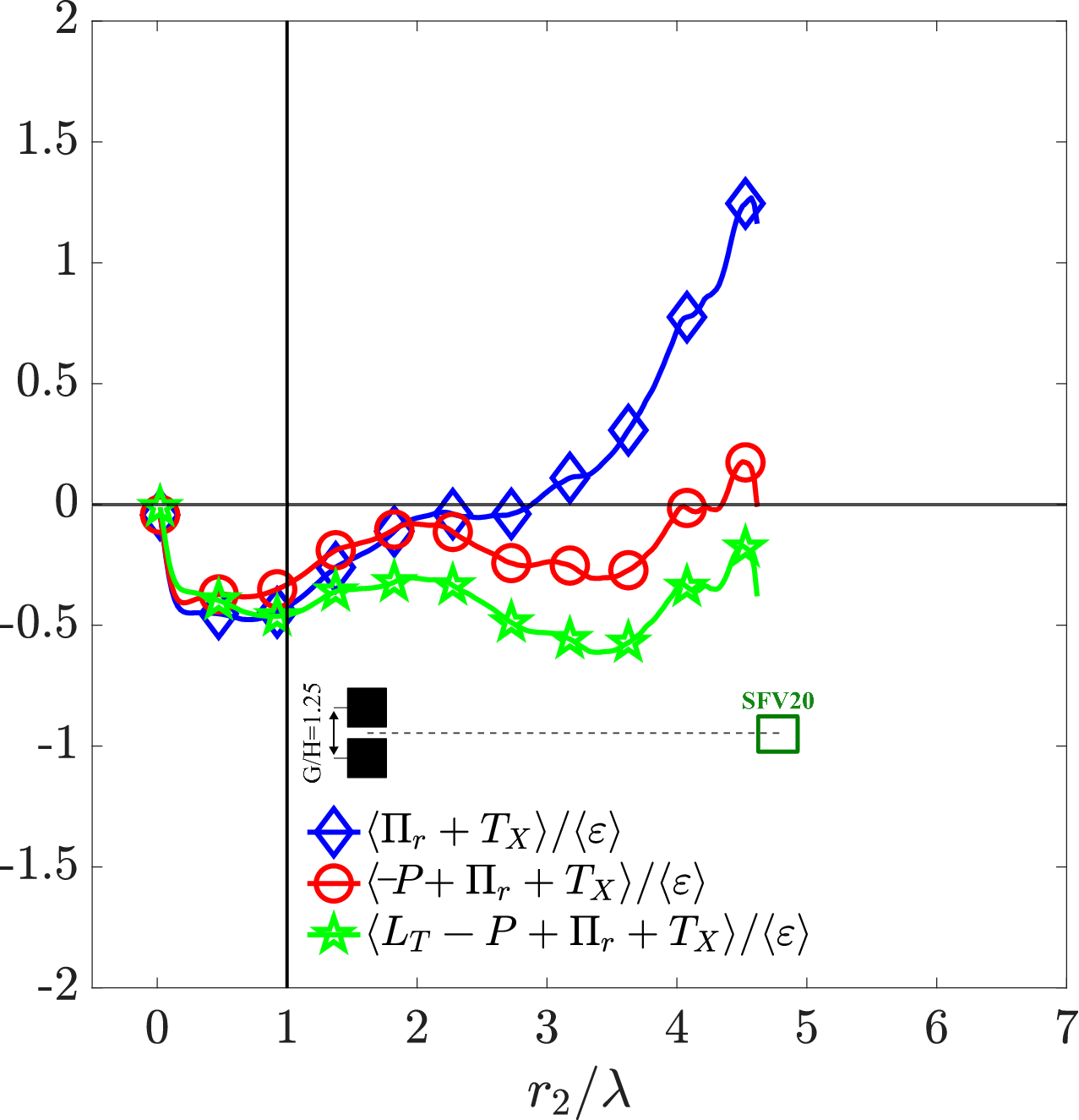}
		\caption{}
	\end{subfigure}
	\caption{Normalised scale-by-scale plots of the energy
		transfer rate budget taking into account two-point
		production and linear transport for configuration $G/H=1.25$
		SFV20 at $Re_H=1.0 \times 10^4$. Plots are shown for
		separations scales in the streamwise (a) and in the
		cross-stream (b) direction.}
	\label{fig:G_H_1.25_SFV20_sum_transfer_rates}
\end{figure}

\begin{figure}
	\centering
	\begin{subfigure}{0.49\textwidth}
		\includegraphics[width=\textwidth]{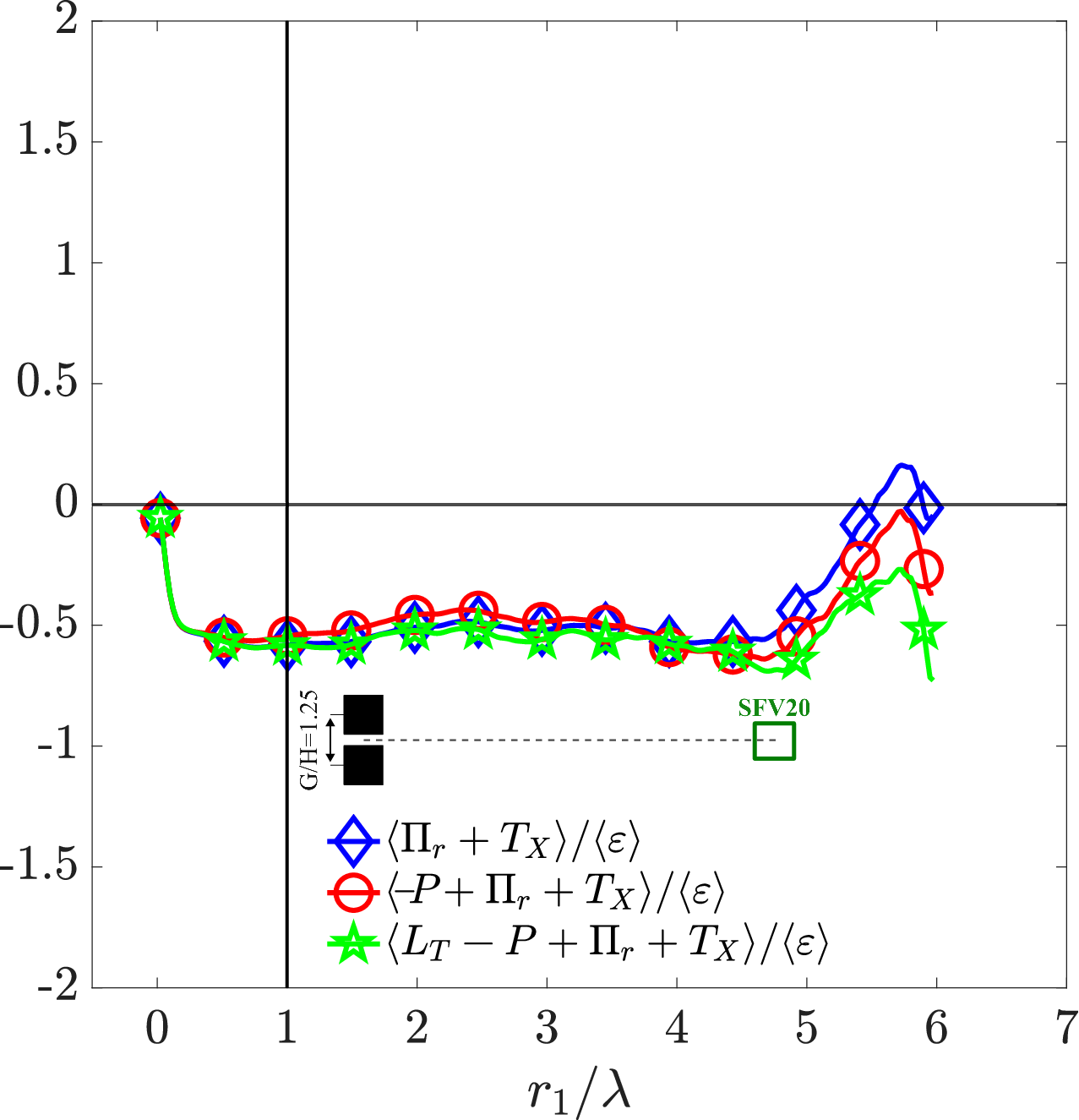}
		\caption{}
	\end{subfigure}
	\begin{subfigure}{0.49\textwidth}
		\includegraphics[width=\textwidth]{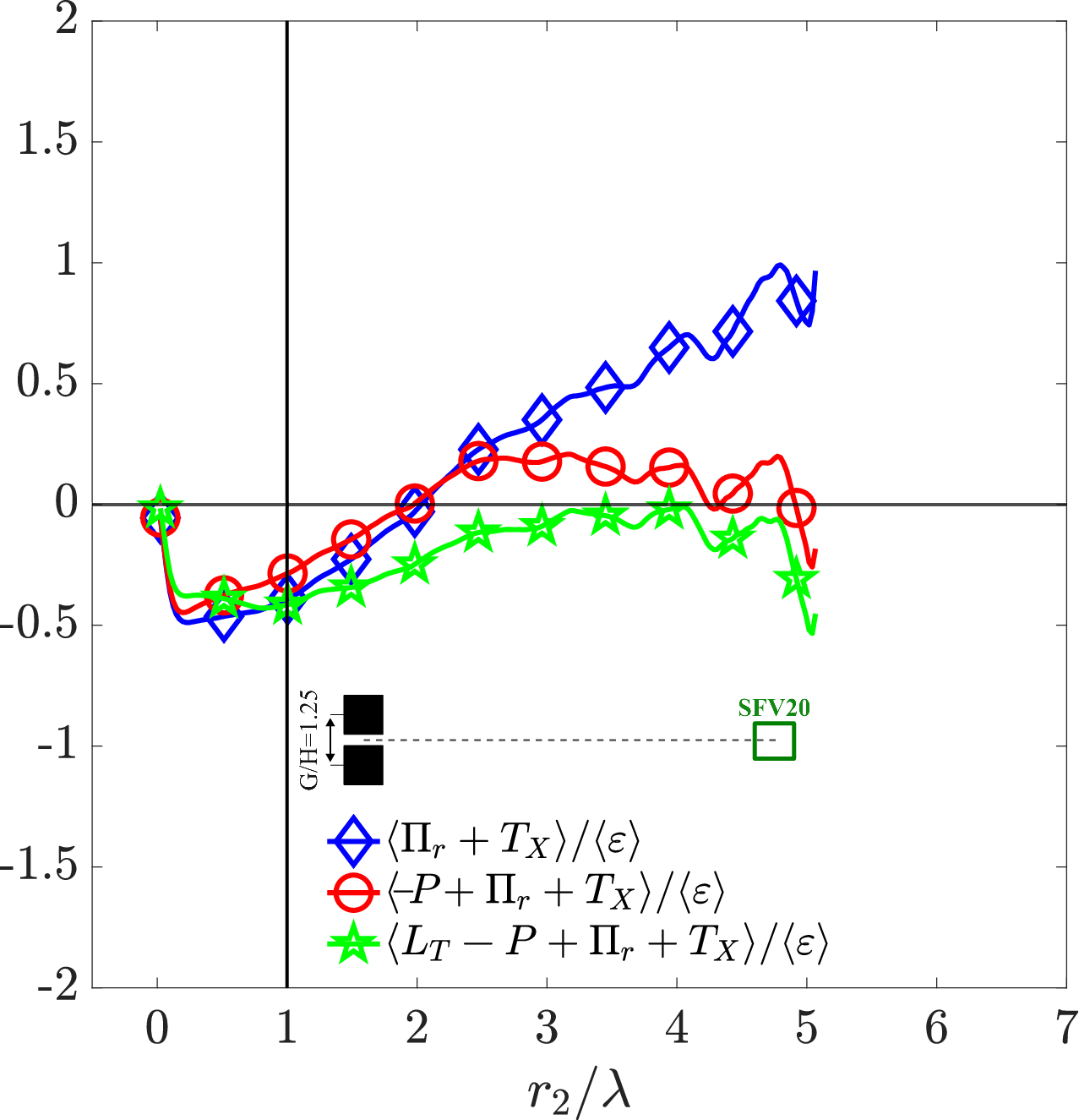}
		\caption{}
	\end{subfigure}
	\caption{Normalised scale-by-scale plots of the energy
		transfer rate budget taking into account two-point
		production and linear transport for configuration $G/H=1.25$
		SFV20 at $Re_H=1.2 \times 10^4$. Plots are shown for
		separations scales in the streamwise (a) and in the
		cross-stream (b) direction.}
	\label{fig:6mps_G_H_1.25_SFV20_sum_transfer_rates}
\end{figure}

For $r_2 = 0$, the $r_1$ profile of $\langle \Pi_r \rangle^\star$ is
as sharp as the $r_1$ profile of $\langle T_X \rangle^\star$ but
opposite (see Fig. \ref{fig:G_H_1.25_SFV20_transfer_rates}). Hence,
similarly to all stations in the other two wakes except $SFV10$ for
$G/H=2.4$ which is not quite in the decaying wake, $\langle \Pi_r
\rangle^\star + \langle T_X \rangle^\star$ remains about constant at
$-0.5$ over a $r_1$ range from $0.2\lambda$ to
$\langle\mathcal{L}_v\rangle/2$, see
Fig. \ref{fig:G_H_1.25_SFV20_sum_transfer_rates} for $Re_H = 1.0\times
10^4$ ($\langle\mathcal{L}_v\rangle/\lambda = 8.4$) and
Fig. \ref{fig:6mps_G_H_1.25_SFV20_sum_transfer_rates} for $Re_H =
1.2\times 10^4$ ($\langle\mathcal{L}_v\rangle/\lambda = 9.1$). These
two figures also show that the very strong non-homogeneity in $r_1$ is
limited to two-point turbulent diffusion as two-point linear transport
and turbulence production are negligible compared to the other terms
in the scale-by-scale horizontal two-point turbulent kinetic energy
budget.

For $r_1 =0$, the $r_2$ profiles of $\langle \Pi_r \rangle^\star$ and
$\langle T_X \rangle^\star$ in
Fig. \ref{fig:G_H_1.25_SFV20_transfer_rates} are qualitatively closer
to the exception SFV10 of $G/H=2.4$: whereas $\langle T_X
\rangle^\star$ increases, $\langle \Pi_r \rangle^\star$ does not
decrease with $r_1$. In fact, $\langle \Pi_r \rangle^\star$ remains
remarkably constant at about $-1.5$ for $r_2$ from $\approx \lambda$
to $\approx 4\lambda$. Similarly to SFV10 of $G/H=2.4$,
Fig. \ref{fig:G_H_1.25_SFV20_sum_transfer_rates} and
Fig. \ref{fig:6mps_G_H_1.25_SFV20_sum_transfer_rates} show that
$\langle \Pi_r \rangle^\star + \langle T_X \rangle^\star$ increases
with $r_2$ and that two-point linear transport and/or turbulence
production are not at all negligible except at length scales below
$\lambda$ at SFV20 of $G/H=1.25$. (In fact we checked that it is
two-point turbulence production which is not negligible, two-point
linear transport remains small in magnitude.)

We conclude from this and the previous subsections that

\begin{equation}
\langle \Pi_r \rangle + \langle T_X \rangle \approx -C \langle
\varepsilon \rangle
	\label{eq:space-scaleEQ}  
\end{equation}
holds for either $r_1$ or $r_2$ in the inertial range (bounded from
below by $\lambda$ or a fraction of $\lambda$ and from above by an
outer scale between $\langle\mathcal{L}_v\rangle$ and
$\langle\mathcal{L}_v\rangle/2$), with the dimensionless constant
coefficient $C$ between $0.6$ and $1.0$ depending on wake
and location within the wake, unless two-point production rate is not
negligible (as is the case for the $r_2$ dependencies at SFV10,
$G/H=2.4$ and SFV20, $G/H=1.25$) or that (\ref{eq:TpPi}) does not hold
(as is the case for the $r_1$ dependence at SFV10 of the $G/H=2.4$
wake where two-point
turbulence production rate is negligible).

Incidentally, it can be inferred from
Fig. \ref{fig:G_H_1.25_SFV20_sum_transfer_rates} and
Fig. \ref{fig:6mps_G_H_1.25_SFV20_sum_transfer_rates} that the
validity of (\ref{eq:TpPi}) can be very sensitive to the global
Reynolds number as these figures suggest that (\ref{eq:TpPi}) more or
less holds versus both $r_1$ and $r_2$ for $Re_H = 1.0\times 10^4$ but
not versus $r_2$ for $Re_H = 1.2\times 10^4$ at SFV20 of the
$G/H=1.25$ wake.

\section{Concluding discussion}

In all the centreline stations in the turbulence decay region or
towards the downstream edge of the turbulence buildup region of all
three qualitatively turbulent wakes examined here, the space-time
average inter-space turbulence transfer rate is very considerable and
positive down to the smallest turbulence length scales even though the
local $Re_{\lambda}$ reaches values of up to nearly $500$. There is no
tendency whatsoever towards local homogeneity, even at the smallest
(viscosity affected) turbulent length scales.

In all these stations, the space-time average inter-scale turbulence
transfer rate is negative and therefore has the opposite sign to the
average inter-space turbulence transfer rate. Turbulence cascade and
two-point turbulence diffusion act, on average, against each other at
dissipative and inertial scales and even at scales above the integral
scale $\langle\mathcal{L}_v\rangle$. In fact the average inter-scale
transfer rate is comparable to and often considerably faster than the
turbulence dissipation rate. Provided that the non-homogeneity is not
a turbulence producing one, i.e. that two-point turbulence production
rate is negligible, and that the two-point pressure-velocity term does
not violate the self-similar balance (\ref{eq:TpPi}), the sum of the
inter-space and inter-scale turbulence transfer rates is about
constant (equation (\ref{eq:space-scaleEQ})) in the inertial range of
length scales. This constant is between -0.6 and -1.0 depending on
wake
and streamwise centreline location within the wake.  The simple
self-similar balance (\ref{eq:space-scaleEQ}) is in fact obeyed versus
both $r_1$ and $r_2$ at any station sampled in the turbulence decay
region of our turbulent wakes.

Our results for the $G/H =3.5$ and $G/H=2.4$ turbulent wakes also show
that the normalised inter-scale and inter-space transfer rates
$\langle \Pi_r \rangle/\langle \varepsilon\rangle$ and $\langle T_X
\rangle/\langle \varepsilon\rangle$ are constant with a slight,
perhaps linear, trend with $r_1$, $r_2$ in some cases. Similar
observations were made by \cite{beaumard2024scale} under the rotating
blades in a baffled water tank, who also found the same signs of
$\langle \Pi_r \rangle/\langle \varepsilon\rangle$ and $\langle T_X
\rangle/\langle \varepsilon\rangle$ as we do here. The theory of
non-homogeneous turbulence of \cite{chen2022scalings} and
\cite{beaumard2024scale} predicts constant $\langle \Pi_r
\rangle/\langle \varepsilon\rangle$ and $\langle T_X \rangle/\langle
\varepsilon\rangle$ in the inertial range under the hypothesis of
similar two-point physics at different locations of the
non-homogeneous turbulence and under the assumption that two-point
turbulence production is negligible. This theory therefore also
predicts the self-similar balance (\ref{eq:space-scaleEQ}) in the
inertial range. However, the departure from constancy evidenced by the
slight, perhaps linear, trend mentioned at the start of this paragraph
may suggest the need of a sub-leading-order Reynolds number correction
to the theory. This is one of the three open questions which now need
to be addressed. The other two questions are (i) what sets the
turbulence dissipation rate and (ii) what physical mechanism
determines whether the inter-scale and the inter-space turbulence
transfer rates cooperate or counteract each other.

Future research can begin to tackle these open questions
and help test the bounds of the proposed theory by
investigating off-centreline regions across a broader range of wake
types and by extending the analysis
to diverse types of non-homogeneous turbulent
  flows. For example, shearless mixing layers introduced by \citet{gilbert1980diffusion} and \citet{veeravalli1989shearless} are also cases of non-producing non-homogeneity where
  two adjacent regions of shearless turbulence with different
  turbulence levels and/or length scales progressively mix. On the
  other hand, the layer sometimes referred to as a log-layer in
  turbulent channel flows and zero pressure gradient boundary layers
  are a case of non-transporting non-homogeneity where two-point
  turbulence production is present down to very small scales but
  average two-point inter-space transport vanishes. In the production
  region of a temporally evolving turbulent jet, \citet{cimarelli2021spatially} report an intermediate range of length scales where
  two-point inter-space turbulence transfer and inter-scale turbulence
  transfer have opposite signs (see their figure 10). However,
  \citet{cimarelli2024spatially} find these two types of
  two-point transfer to have the same sign above the, so-called
  log-layer in wall turbulence (see their figure 7). This
  variability of behaviour needs to be addressed and explained perhaps
  in terms of a classification of turbulence non-homogeneities into a
  small number of universality classes each one governed by its own
  comprehensive theory. Ultimately, the study of turbulent
diffusion–cascade interactions offers a promising entry point for
advancing our understanding of the physics governing non-homogeneous
and unsteady turbulent flows which are well beyond the assumptions of
Kolmogorov's equilibrium framework.

\backsection[Acknowledgements]{We are thankful to the authors of 
	\cite{chen2021turbulence} for granting us access to the data
	analysed in the present work. We thank C. Cuvier for the helpful discussions.}

\backsection[Funding]{This work was funded by the European Union 
	(ERC, NoStaHo, 101054117). Views and opinions expressed are, however, those of the authors only and do not necessarily reflect those of the European
	Union or the European Research Council. Neither the European Union 
	nor the granting authority can be held responsible for them.}

\backsection[Declaration of interests]{The authors report
  no conflict of interest.}

\backsection[Data availability statement]{The data that support the
  findings of this study are available upon request.}

\backsection[Author ORCIDs]{E. Fuentes-Noriega, https://orcid.org/0000-0002-9931-7061;
  J.C. Vassilicos, https://orcid.org/0000-0003-1828-6628}

  \bibliographystyle{jfm}
  
  \bibliography{jfm}

\appendix

\section{Scale-by-scale behaviour of $\overline{\delta K_h}$}\label{appA:deltaKh_r_profiles}

\begin{figure}
        \centering
        \begin{subfigure}{0.95\textwidth}
                \includegraphics[width=0.97 \textwidth]{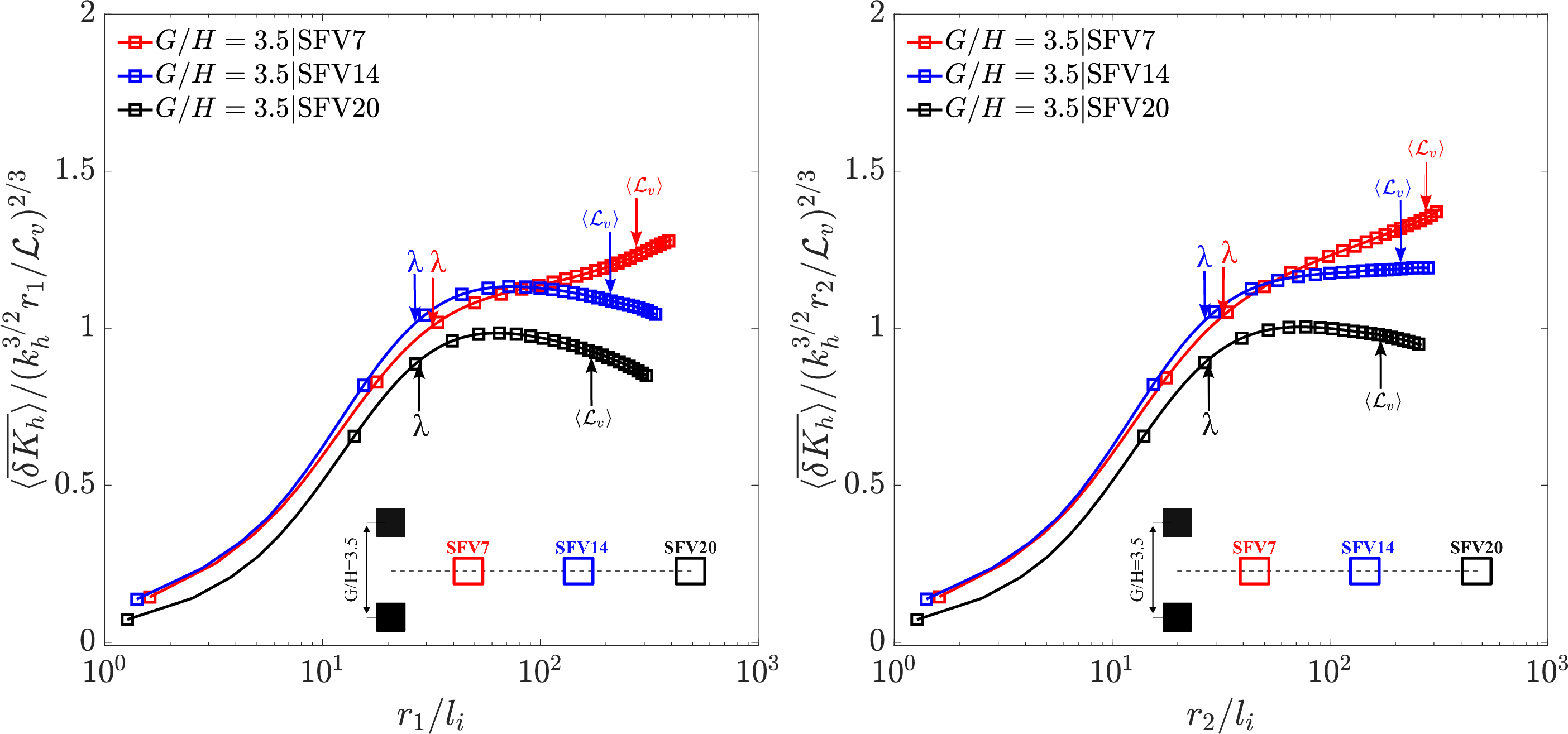}
                \caption{$G/H = 3.5$}
        \end{subfigure}
       
        \centering
        \begin{subfigure}{0.95\textwidth}
                \includegraphics[width=0.97 \textwidth]{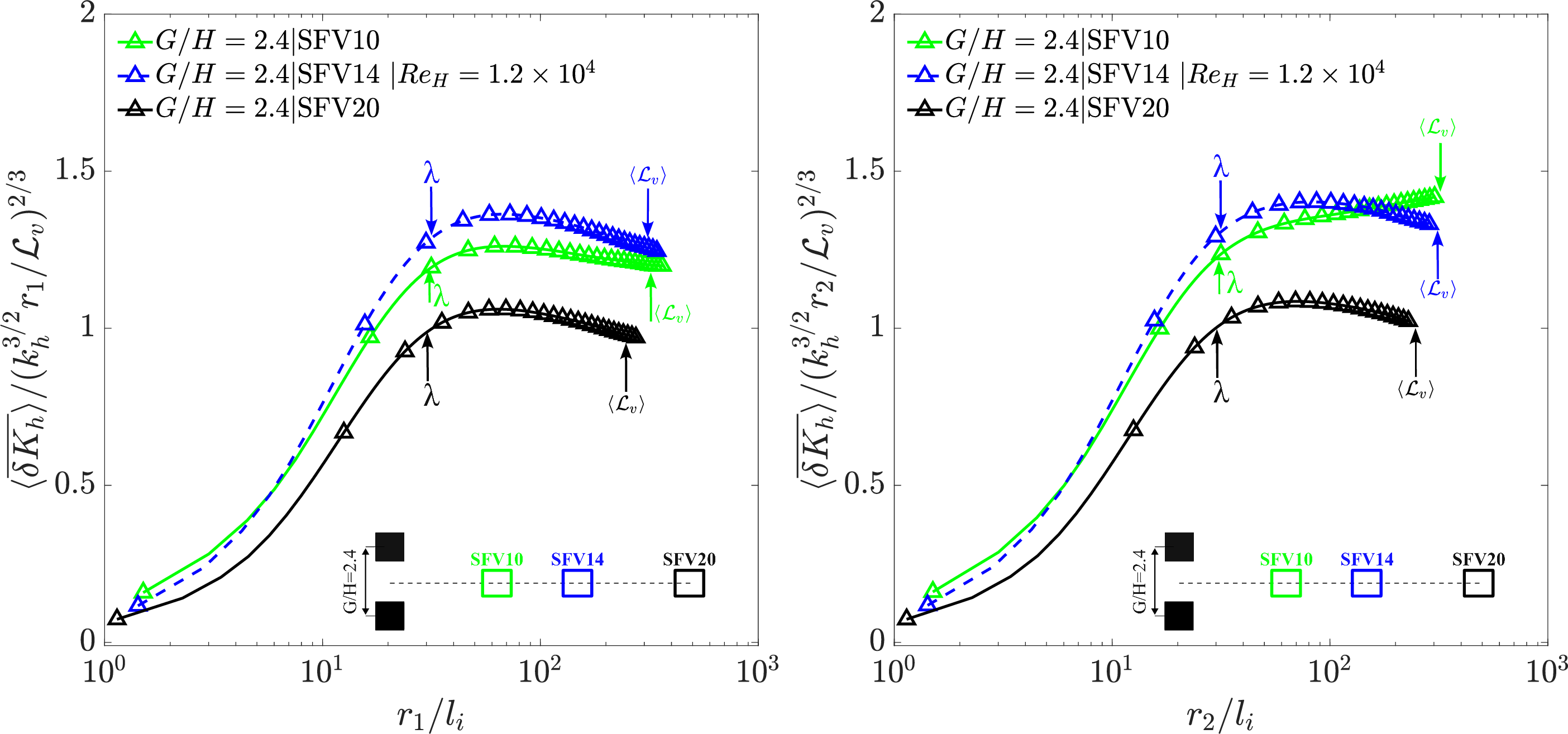}
                \caption{$G/H = 2.4$}
        \end{subfigure}
       
        \centering
        \begin{subfigure}{0.95\textwidth}
                \includegraphics[width=0.97 \textwidth]{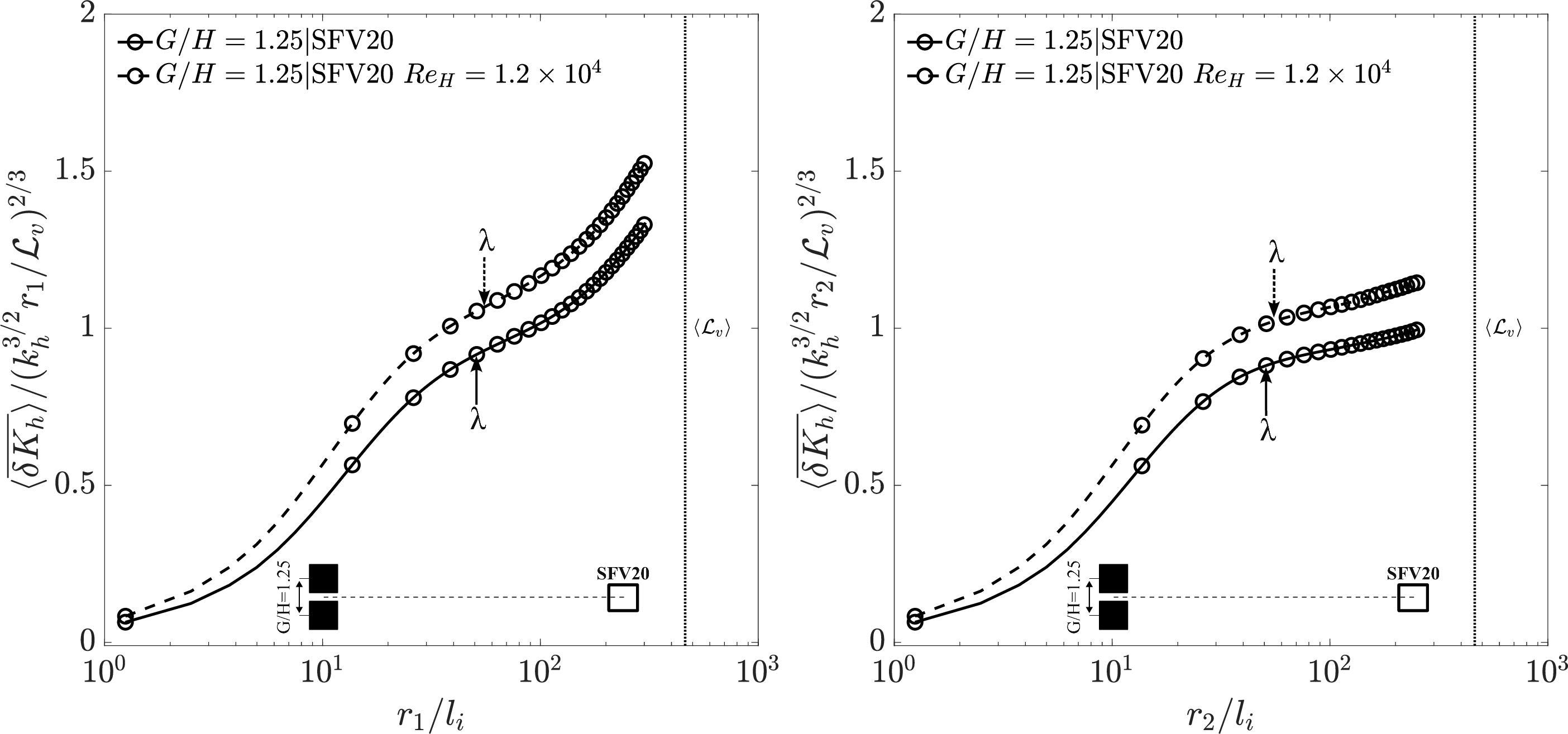}
                \caption{$G/H = 1.25$}
        \end{subfigure}
       
        \caption{Normalised space-averaged two-point horizontal
          turbulent kinetic energy $\langle \overline{\delta K_h}
          \rangle$ as a function of $r_1/l_i$ and $r_2/l_i$, where
          $l_i=\langle \mathcal{L}_v \rangle R^{-3/4}$ with
          $R=\sqrt{k_h}\langle \mathcal{L}_v \rangle /\nu$ for each
          wake and SFV. The Taylor length $\lambda$ and the integral
          length scale $\mathcal{L}_v$ are indicated in the plots for
          each SFV considered in each one of the three wakes.}
        \label{fig:deltaKh_profiles}
\end{figure}

  In this Appendix we report the profiles of $\langle
  \overline{\delta K_h} \rangle$ as function of the separation scales
  $r_1$ and $r_2$ normalised using the non-equilibrium scaling
  proposed in \citet{chen2022scalings}. These profiles are plotted in
  lin-log axes in Figs. \ref{fig:deltaKh_profiles}(a,b,c) for the
  $G/H=3.5$, $G/H=2.4$ and $G/H=1.25$ wakes, respectively. Unless
  specified otherwise, the profiles correspond to the $Re_H=1.0 \times
  10^4$ global Reynolds. Near-identical results are obtained for the
  other $Re_H$. In \citet{chen2022scalings}, similar plots where shown
  but for the longitudinal and transverse structure functions
  individually as a function of $r_1$ at different $X_2$
  locations. Here, we report the behaviour of $\overline{\delta K_h}$
  averaged over SFV area as a function of both separation scale
  directions $r_1$ and $r_2$. Same conclusions are drawn as in
  \citet{chen2022scalings}. As can be seen in the lin-log plots of
  Figs. \ref{fig:deltaKh_profiles}a and \ref{fig:deltaKh_profiles}b,
  an approximate plateau indicating approximate $2/3$ power law
  dependence on length-scale appears to be forming over nearly one
  decade between Taylor length and integral scale in the analysed SFVs
  for $G_H=3.5$ and $G_H=2.4$ with perhaps a slight deviation for $G/H=3.5$ SFV7. There is a
  significant departure from $2/3$ power law at the SFV20 station of
  the $G/H=1.25$ wake.

\section{Inter-space energy transfer rate: dependence on averaging procedure and individual terms}\label{appA}

\subsection{The spatial averaging dependence}

\begin{figure}
	\centering
	\begin{subfigure}{0.48\textwidth}
		\includegraphics[width=\textwidth]{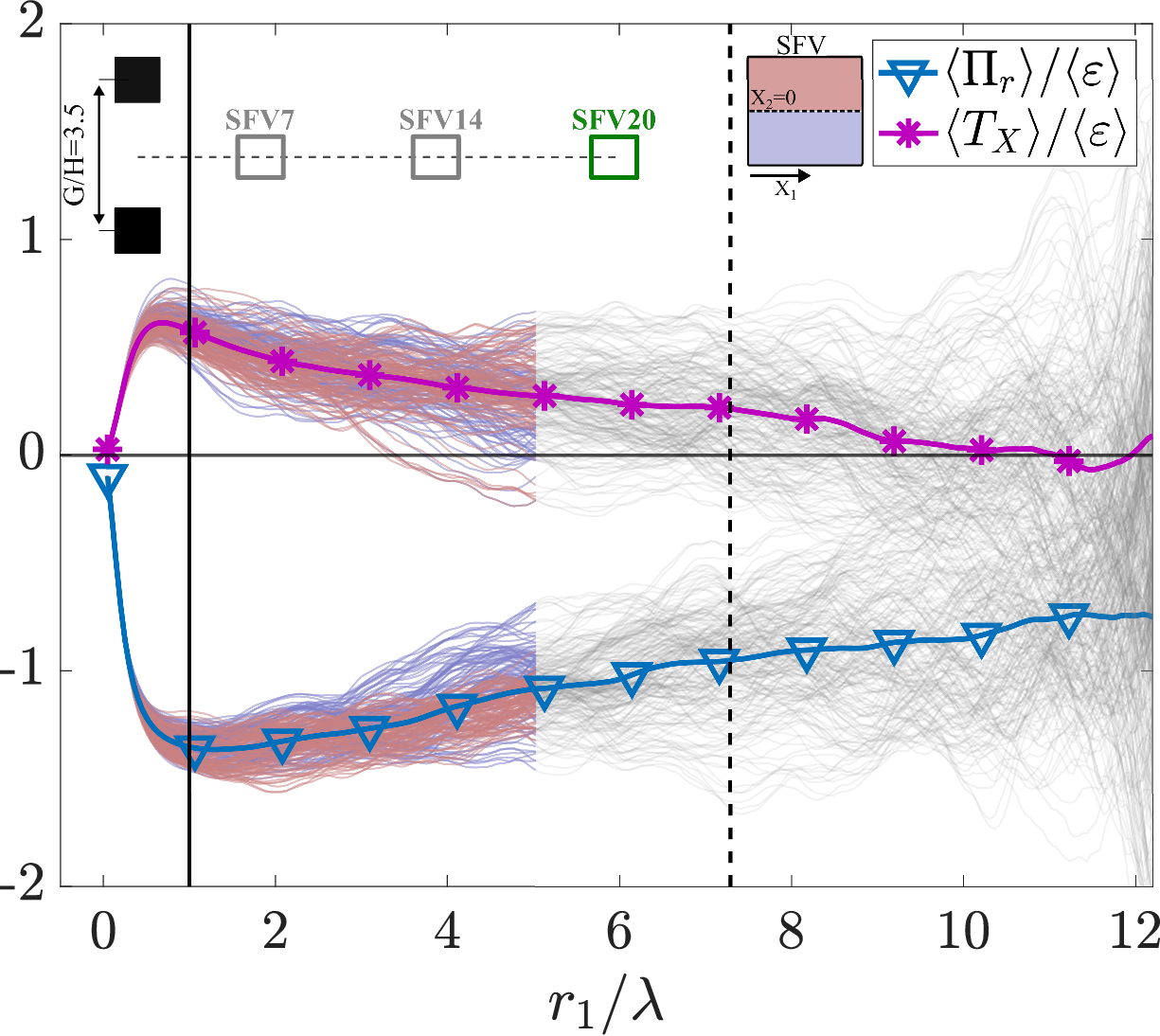}
		\caption{}
	\end{subfigure}
	\begin{subfigure}{0.48\textwidth}
		\includegraphics[width=\textwidth]{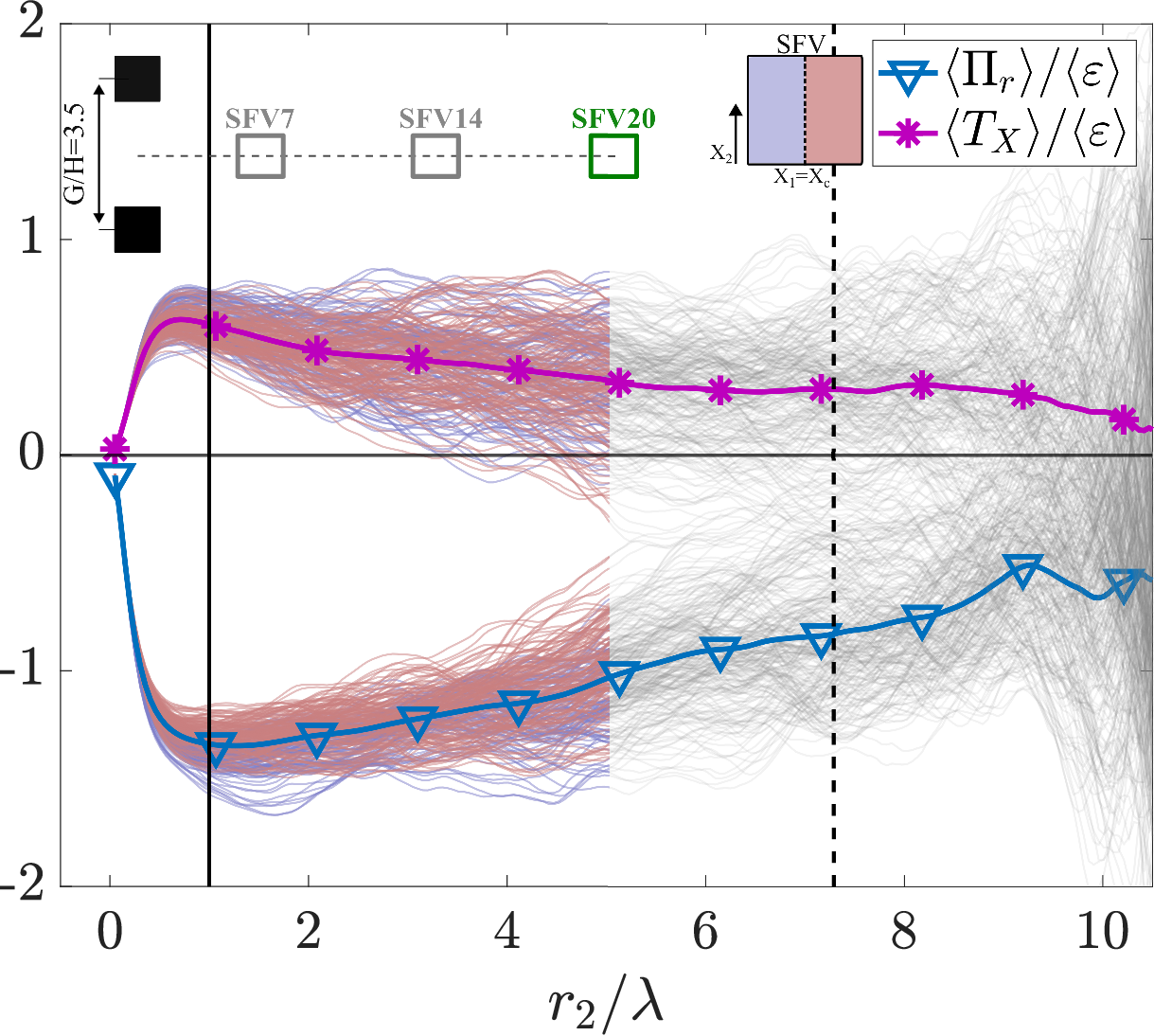}
		\caption{}
	\end{subfigure}
	
\caption{Analysis of the spatial averaging of inter-scale and inter-space energy transfer rates for case $G/H=3.5$ SFV20 at $Re_H=1.5\times 10^4$. Plots are shown for (a) quantities line averaged over $X_1$ at a given $X_2$ as a function of $r_1$ and (b) quantities line averaged over $X_2$ at a given $X_1$ as a function of $r_2$. Each thin curve represents a different position in the SFV and the line averaged profiles are coloured depending on their location with respect to the center of the SFV. The thick magenta and blue lines represent the fully averaged (over $X_1$ and $X_2$) quantities. The vertical dashed line locates $\langle \mathcal{L}_v \rangle/\lambda$.}
	\label{fig:7.35mps_G_H_3.5_SFV20_avg_impact}
\end{figure}

\begin{figure}
	\centering
	\begin{subfigure}{0.48\textwidth}
		\includegraphics[width=\textwidth]{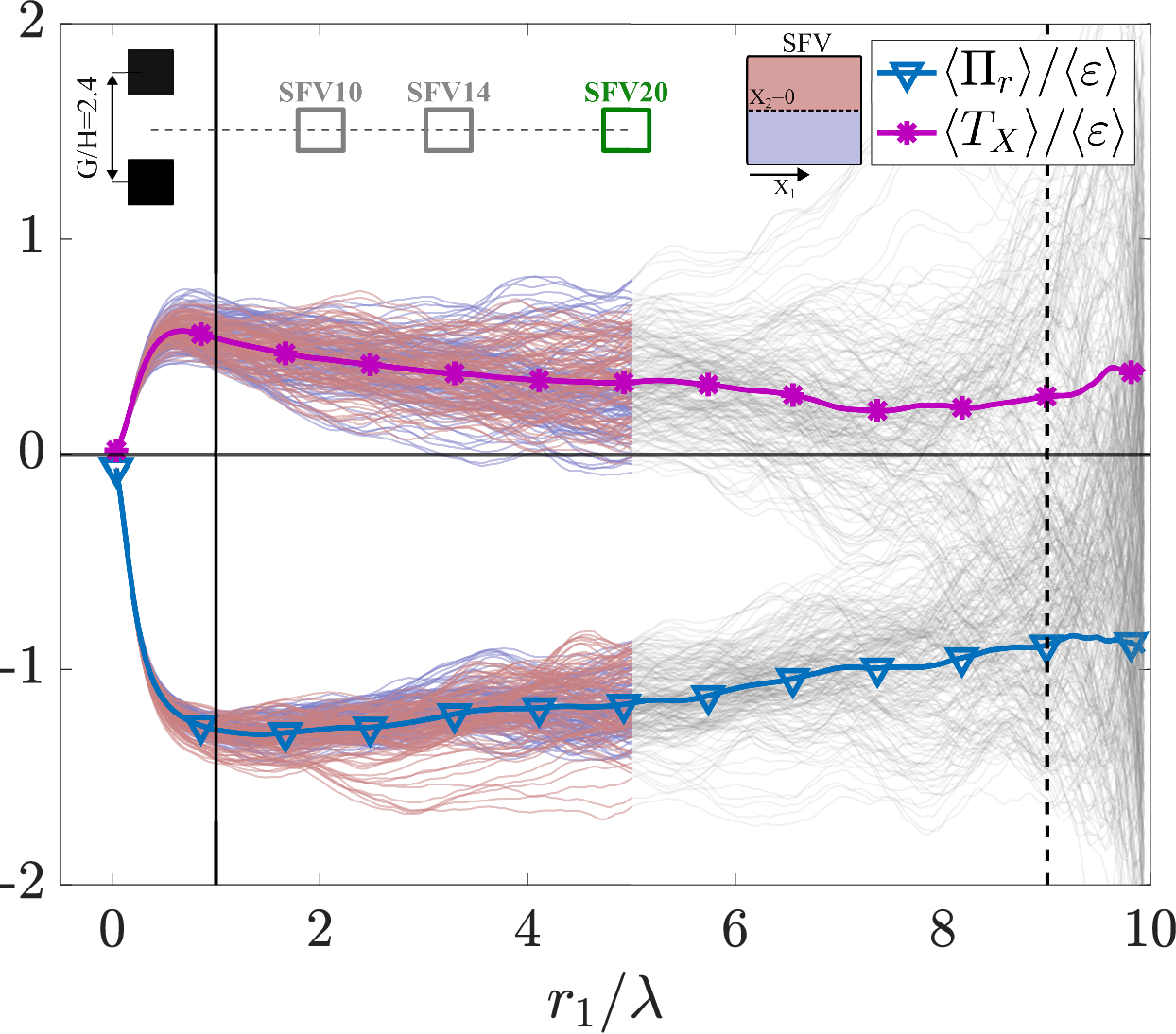}
		\caption{}
	\end{subfigure}
	\begin{subfigure}{0.48\textwidth}
		\includegraphics[width=\textwidth]{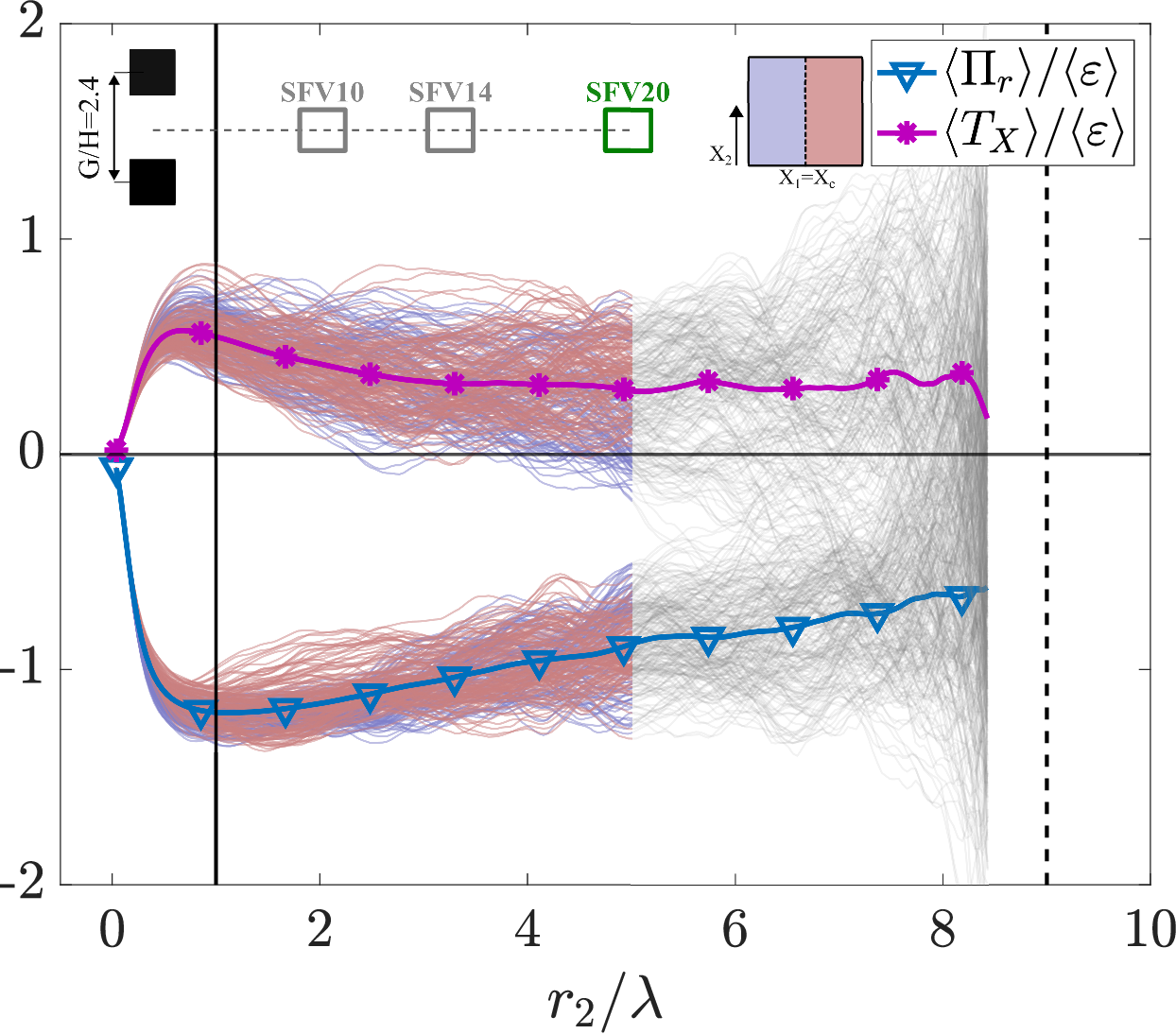}
		\caption{}
	\end{subfigure}

\caption{Analysis of the spatial averaging of inter-scale and inter-space energy transfer rates for case $G/H=2.4$ SFV20 at $Re_H=1.2\times 10^4$. Plots are shown for (a) quantities line averaged over $X_1$ at a given $X_2$ as a function of $r_1$ and (b) quantities line averaged over $X_2$ at a given $X_1$ as a function of $r_2$. Each thin curve represents a different position in the SFV and the line averaged profiles are coloured depending on their location with respect to the center of the SFV. The thick magenta and blue lines represent the fully averaged (over $X_1$ and $X_2$) quantities. The vertical dashed line locates $\langle \mathcal{L}_v \rangle/\lambda$.}
\label{fig:6mps_G_H_2.4_SFV20_avg_impact}
\end{figure}

The data reported in this experimental investigation was averaged over time (20 000 uncorrelated samples) and over the entire field of view (space averaging). The latter was done to yield converged third order moments, i.e converged averages of $T_X$ and $\Pi_r$ given that convergence of two-point statistics gradually weakens as the separation scale $r_1$ or $r_2$ increases. In the following, the effect of the space-average operation is assessed by replacing it with a straight line average over $X_1$ within the field of view's bounds for a given $X_2$ or over $X_2$ within the field of view's bounds for a given $X_1$. Figures \ref{fig:7.35mps_G_H_3.5_SFV20_avg_impact} and \ref{fig:6mps_G_H_2.4_SFV20_avg_impact} show scale-by-scale plots of averaged $T_X$ and $\Pi_r$ normalised by $\avepsilonxt$ comparing the full space average (over $X_1$ and $X_2$) to line averaged quantities (over either $X_1$ or $X_2$) for $G/H=3.5$ and $G/H=2.4$. Only results for the maximum available global Reynolds number and SFV20 are shown as they are representative of all other configurations. At small scales $r_1$ or $r_2$ where convergence is best, it is clear that the line averages of $T_X$ do not vanish and are in fact positive for any location $X_1$ or $X_2$ within the SFV in both wakes. Convergence obviously weakens with increasing $r_1$ and $r_2$ and line-averages increasingly fluctuate as a result. Curves are blue or red if the line-average they represent is over a straight line on one or the other side of the center of the field of view. The distribution of red and blue profiles demonstrates that there is no preferential departure from the fully averaged quantities above or below the centreline or upstream versus downstream locations in the SFV. This result legitimises the use of the spatial averaging operator over the entire SFV and reinforces one of the key results of this investigation which is the absence of local homogeneity over a range of length scales reaching down to sizes as small  as the Taylor length or even a fraction of it.

\subsection{Individual inter-space transfer terms}

 The inter-space transfer rate $T_X$ defined in
  eq. (\ref{eq:TX2D2C}) comprises two terms: 

\begin{equation}
  T_{X} \equiv T_{X_1} + T_{X_2}
\end{equation}
  where $T_{X_1} \equiv {\partial \over \partial X_{1}}
  \overline{u'_{X1} \delta K_h}$ characterises non-homogeneity in the
  streamwise direction and $T_{X_2} \equiv {\partial \over \partial
    X_{2}} \overline{u'_{X2} \delta K_h}$ characterises
  non-homogeneity in the cross-stream direction. In this Appendix, the
  individual contributions $T_{X_1}$ and $T_{X_2}$ are reported for
  all three wakes in each one of the SFV locations studied in the
  paper. In the same manner as the total inter-space transfer rate,
  each term is averaged over SFV space for converged results.

\subsubsection{Coupled-street $G/H=3.5$ wake}

\begin{figure}
	\centering
	\begin{subfigure}{0.49\textwidth}
		\includegraphics[width=\textwidth]{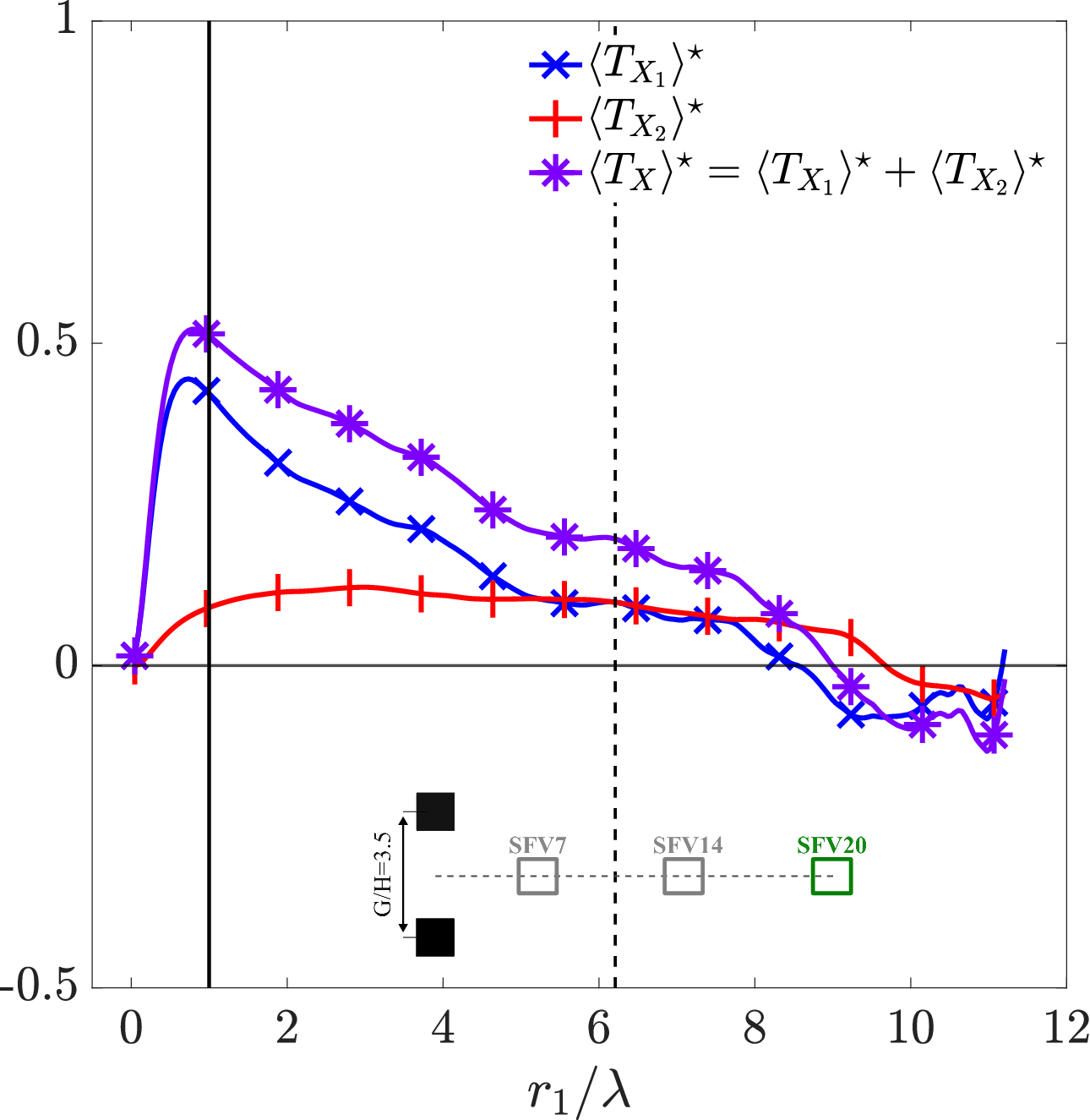}
		\caption{}
	\end{subfigure}
	\begin{subfigure}{0.49\textwidth}
		\includegraphics[width=\textwidth]{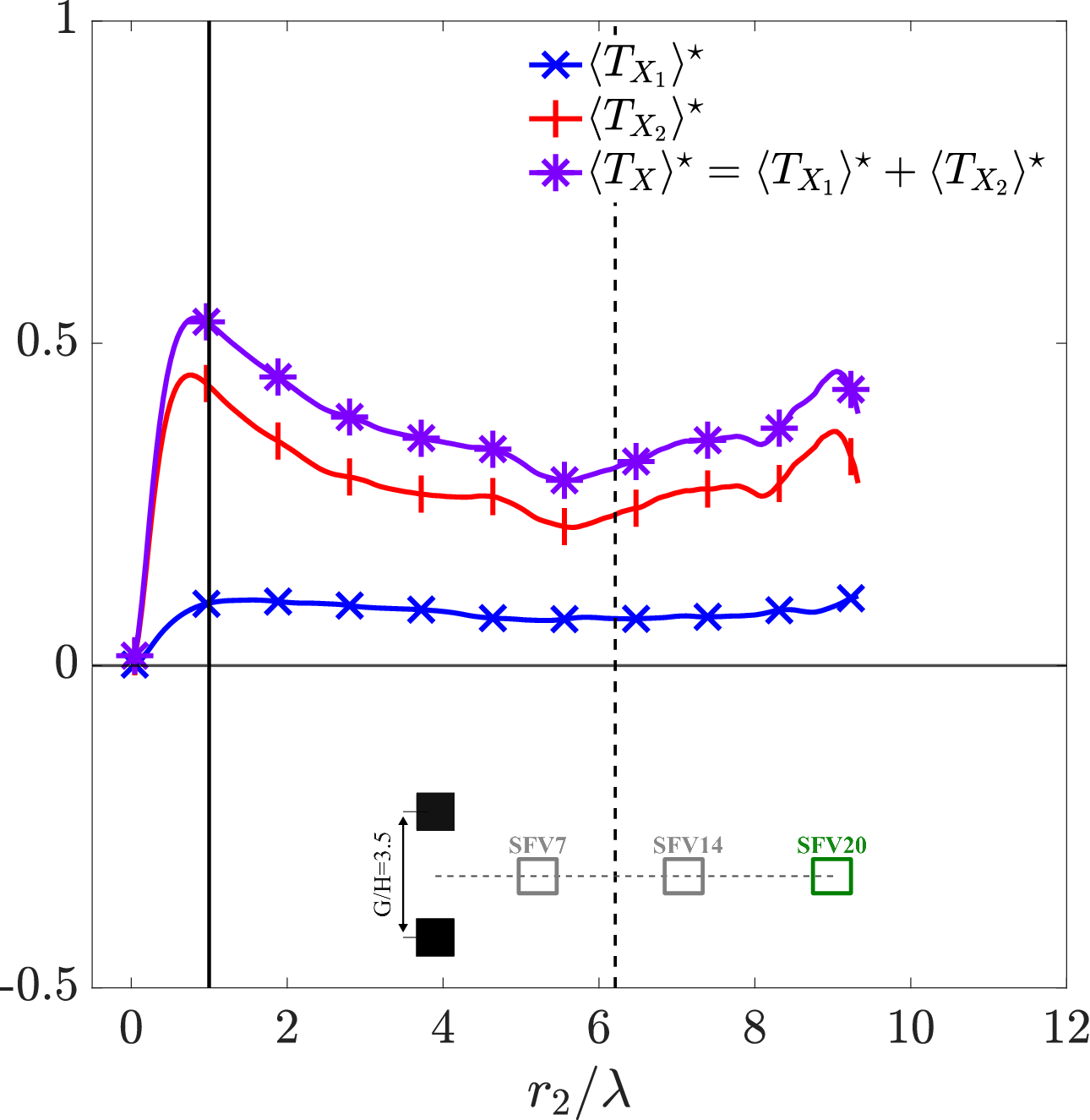}
		\caption{}
	\end{subfigure}
	\caption{Normalised average scale-by-scale inter-space energy transfer rate terms $T_{X_1}$ and $T_{X_2}$ as a function of $r_1$ (a) and $r_2$ (b) for case $G/H=3.5$ SFV20 at $Re_H=1.0 \times 10^4$. The vertical dashed line corresponds to the integral length scale location.}
	\label{fig:TX_terms_G_H_3.5_SFV20_5mps}
\end{figure}

\begin{figure}
	\centering
	\begin{subfigure}{0.49\textwidth}
		\includegraphics[width=\textwidth]{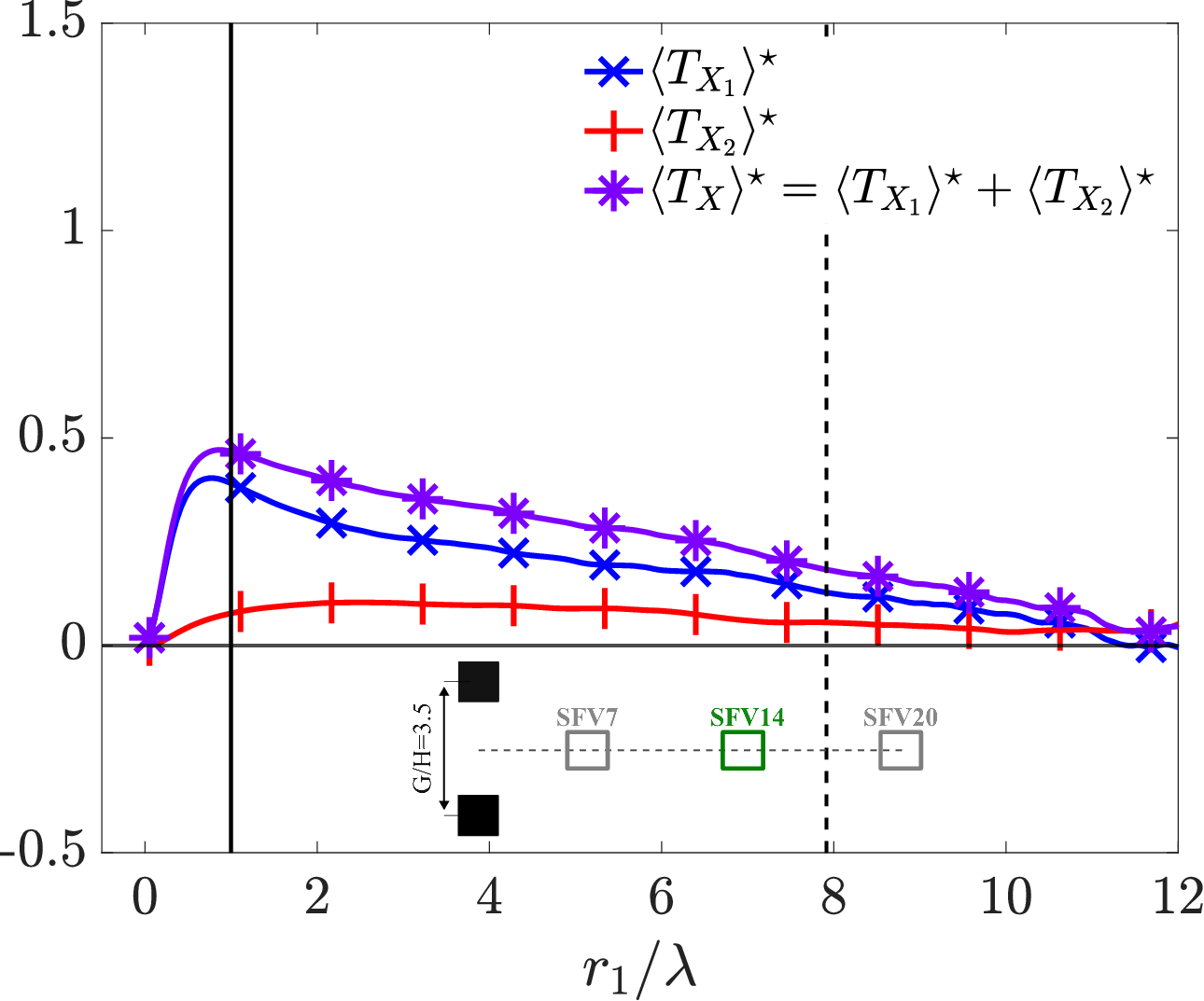}
		\caption{}
	\end{subfigure}
	\begin{subfigure}{0.49\textwidth}
		\includegraphics[width=\textwidth]{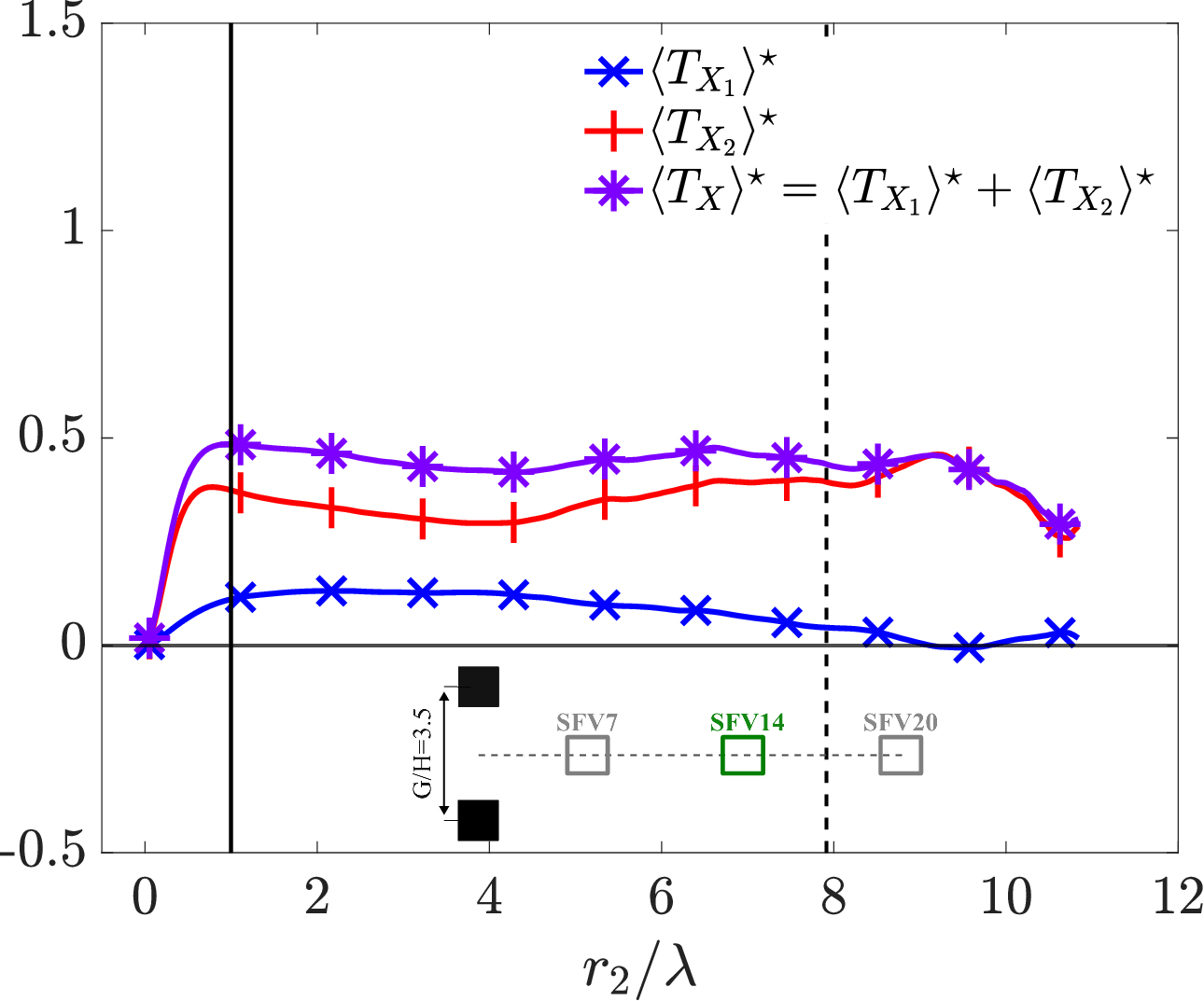}
		\caption{}
	\end{subfigure}
	\caption{Normalised average scale-by-scale inter-space energy transfer rate terms $T_{X_1}$ and $T_{X_2}$ as a function of $r_1$ (a) and $r_2$ (b) for case $G/H=3.5$ SFV14 at $Re_H=1.0 \times 10^4$. The vertical dashed line corresponds to the integral length scale location.}
	\label{fig:TX_terms_G_H_3.5_SFV14_5mps}
\end{figure}

\begin{figure}
	\centering
	\begin{subfigure}{0.49\textwidth}
		\includegraphics[width=\textwidth]{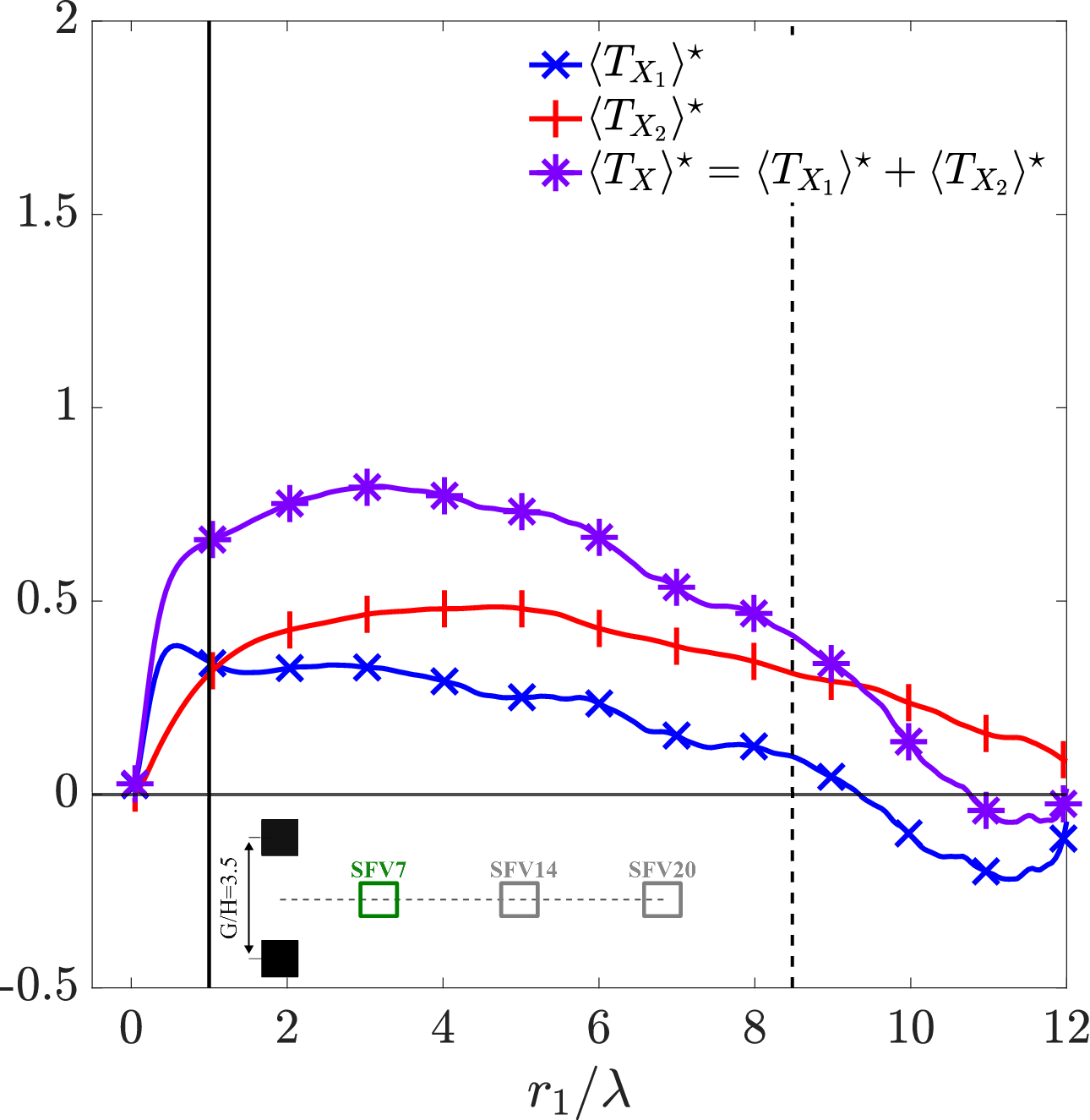}
		\caption{}
	\end{subfigure}
	\begin{subfigure}{0.49\textwidth}
		\includegraphics[width=\textwidth]{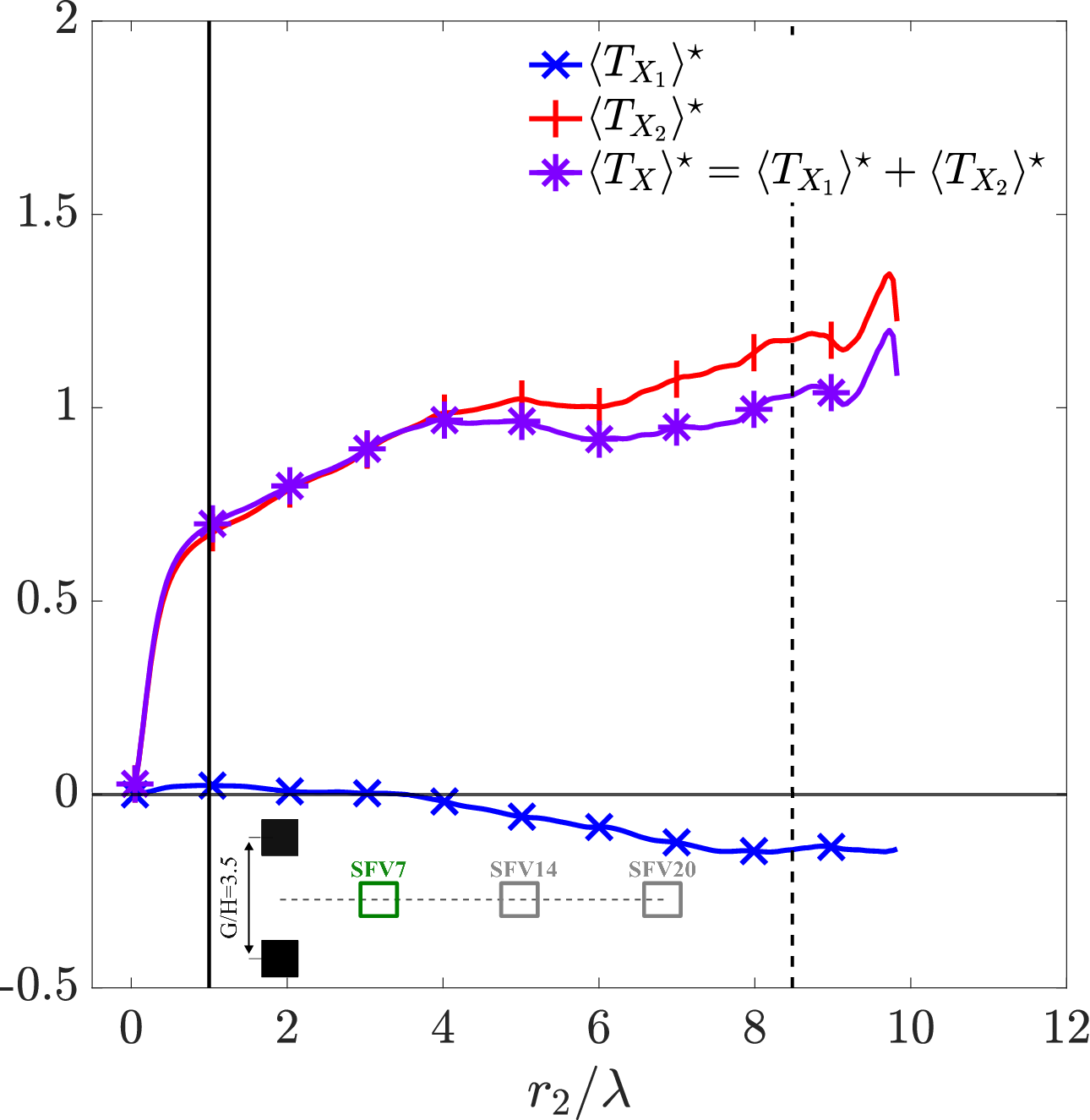}
		\caption{}
	\end{subfigure}
	\caption{Normalised average scale-by-scale inter-space energy transfer rate terms $T_{X_1}$ and $T_{X_2}$ as a function of $r_1$ (a) and $r_2$ (b) for case $G/H=3.5$ SFV7 at $Re_H=1.0 \times 10^4$. The vertical dashed line corresponds to the integral length scale location.}
	\label{fig:TX_terms_G_H_3.5_SFV7_5mps}
\end{figure}

  Figures \ref{fig:TX_terms_G_H_3.5_SFV20_5mps},
  \ref{fig:TX_terms_G_H_3.5_SFV14_5mps} and
  \ref{fig:TX_terms_G_H_3.5_SFV7_5mps} show scale-by-scale plots of
  normalised inter-space energy transfer rates $\langle
  T_{X_1}\rangle^\star = \langle T_{X_1} \rangle / \avepsilonxt$ and
  $\langle T_{X_2} \rangle^\star = \langle T_{X_2} \rangle /
  \avepsilonxt$ for $G/H=3.5$ at SFV20, SFV14 and SFV7,
  respectively. The results are shown for $Re_H=1.0\times 10^4$. It is
  found that, in most situations, both terms are non-zero and of same
  (positive) sign, indicating that both directions exhibit
  non-homogeneity even at small scales and that the transfer of
  two-point turbulent kinetic energy away from the SFV operates in
  both directions. For the two farthest locations SFV14 and SFV20
  (Figs. \ref{fig:TX_terms_G_H_3.5_SFV20_5mps} and
  \ref{fig:TX_terms_G_H_3.5_SFV14_5mps}), $\langle
  T_{X_1}\rangle^\star$ is generally dominant for scales $r_1$ (with
  $r_2=0$) up to around $\langle \mathcal{L}_v \rangle$, while
  $\langle T_{X_2} \rangle^\star$ is the dominant term for all sampled
  scales $r_2$ (with $r_1=0$). At SFV7, however, $\langle
  T_{X_2}\rangle^\star$ is dominant for both $r_1$ and (especially)
  $r_2$ (Fig. \ref{fig:TX_terms_G_H_3.5_SFV7_5mps}).

  In this case, $\langle T_{X_1}\rangle^\star$ shows
  small negative values for $r_2>5\lambda$.

\subsubsection{Coupled street bi-stable $G/H=2.4$ wake}

\begin{figure}
	\centering
	\begin{subfigure}{0.49\textwidth}
		\includegraphics[width=\textwidth]{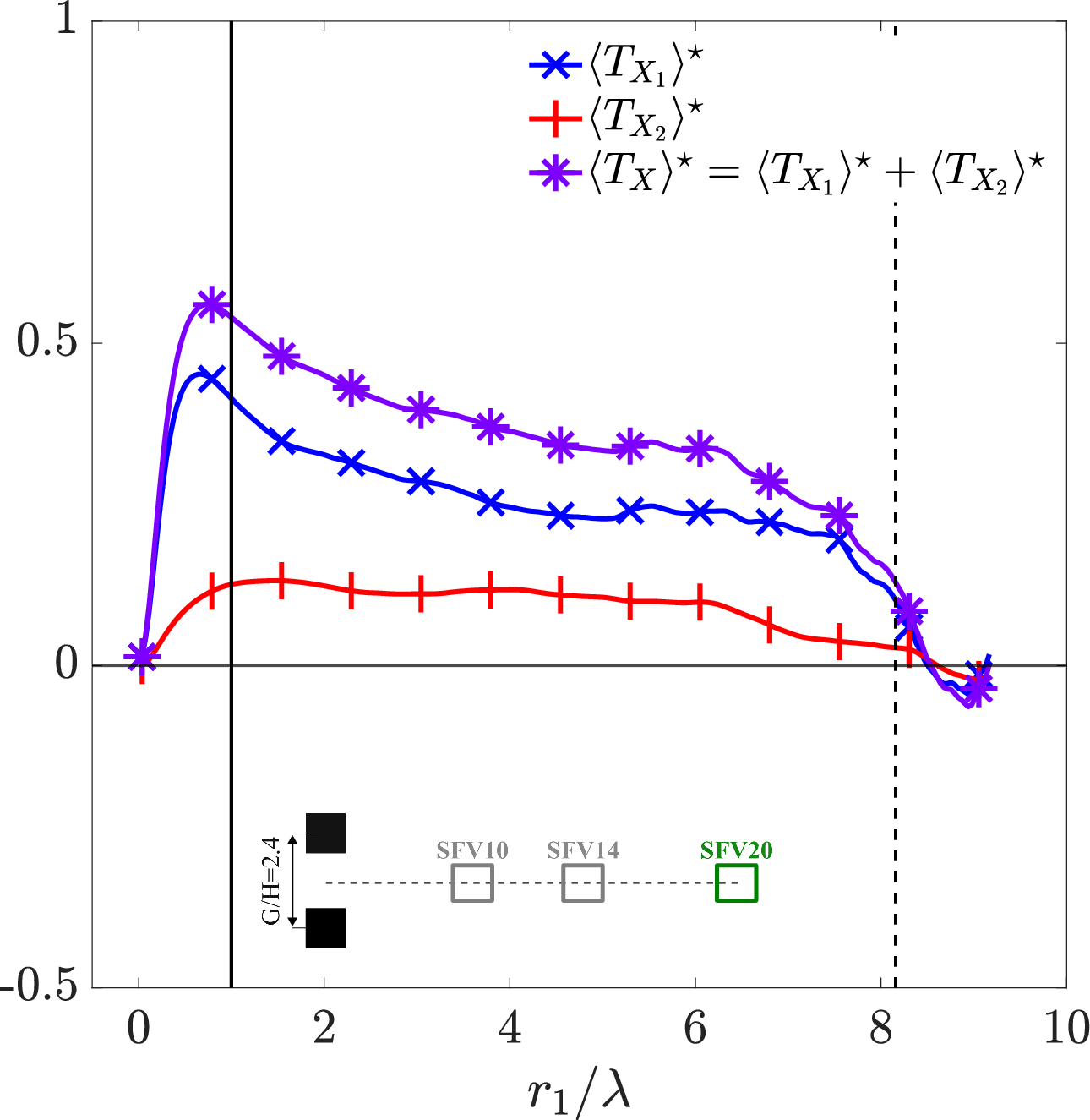}
		\caption{}
	\end{subfigure}
	\begin{subfigure}{0.49\textwidth}
		\includegraphics[width=\textwidth]{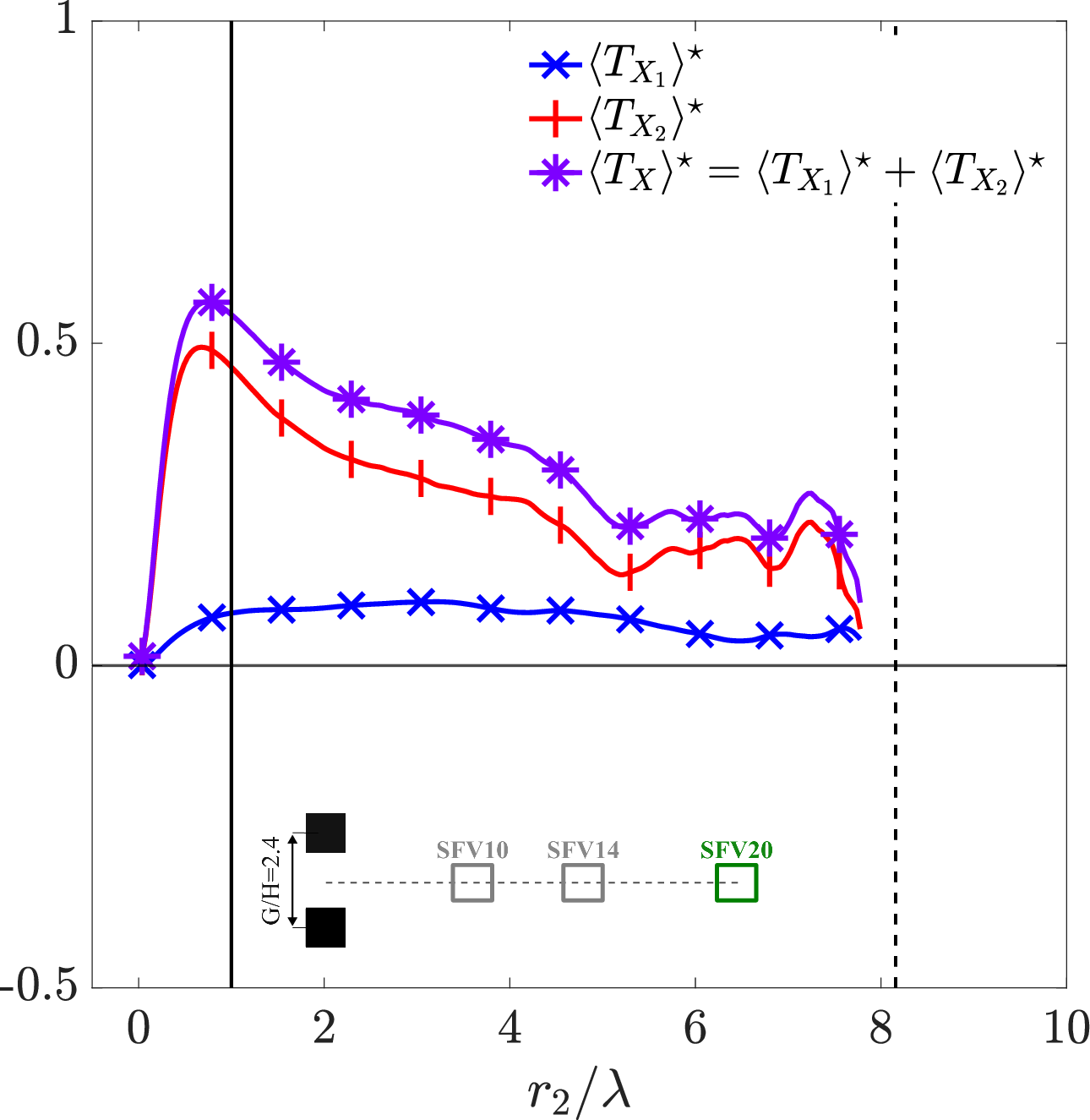}
		\caption{}
	\end{subfigure}
	\caption{Normalised average scale-by-scale inter-space energy
          transfer rate terms $T_{X_1}$ and $T_{X_2}$ as a function of
          $r_1$ (a) and $r_2$ (b) for case $G/H=2.4$ SFV20 at
          $Re_H=1.0 \times 10^4$. The vertical dashed line corresponds
          to the integral length scale location.}
	\label{fig:TX_terms_G_H_2.4_SFV20_5mps}
\end{figure}

\begin{figure}
	\centering
	\begin{subfigure}{0.49\textwidth}
		\includegraphics[width=\textwidth]{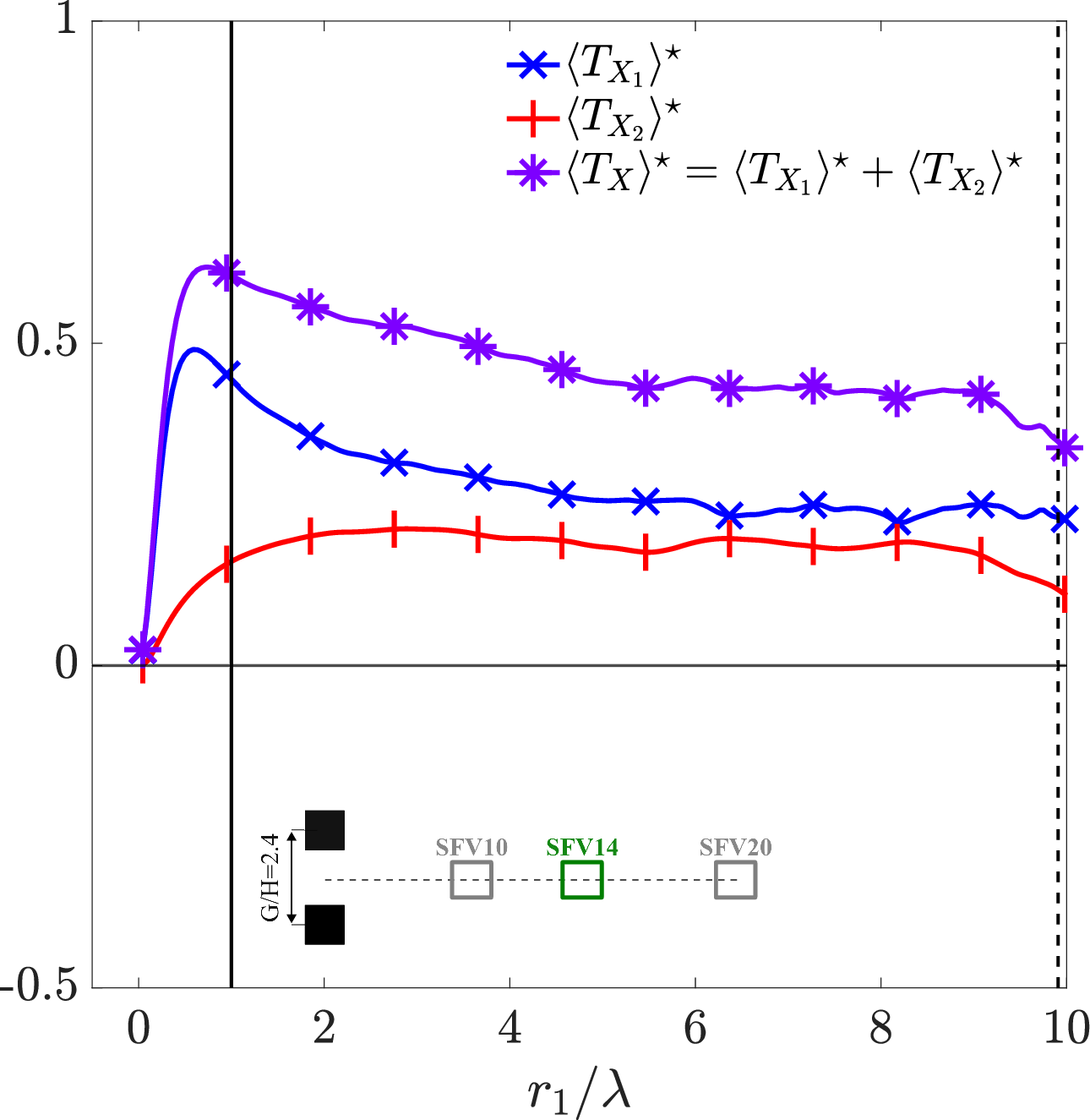}
		\caption{}
	\end{subfigure}
	\begin{subfigure}{0.49\textwidth}
		\includegraphics[width=\textwidth]{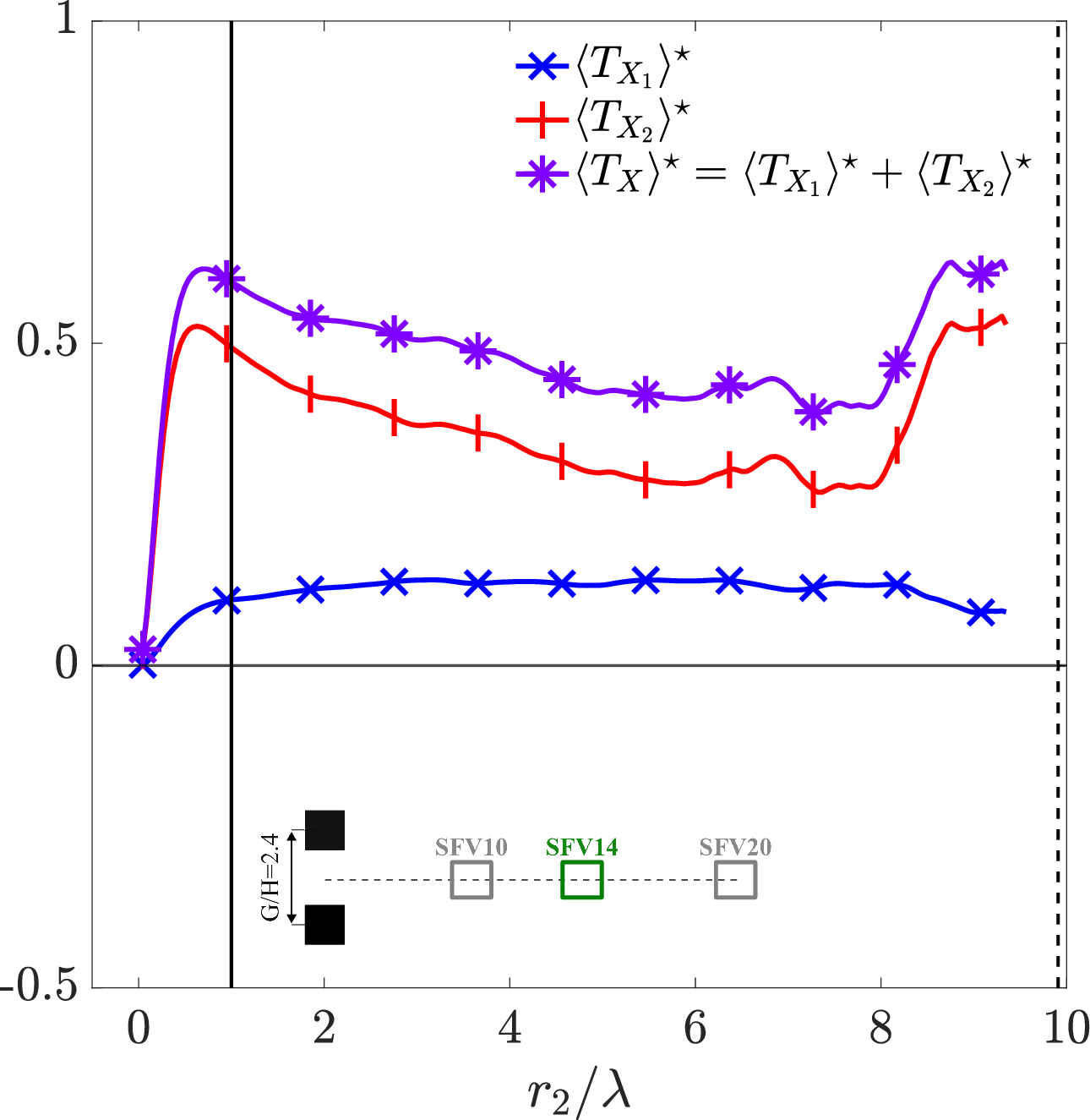}
		\caption{}
	\end{subfigure}
	\caption{Normalised average scale-by-scale inter-space energy
          transfer rate terms $T_{X_1}$ and $T_{X_2}$ as a function of
          $r_1$ (a) and $r_2$ (b) for case $G/H=2.4$ SFV14 at
          $Re_H=1.2 \times 10^4$. The vertical dashed line corresponds
          to the integral length scale location.}
	\label{fig:TX_terms_G_H_2.4_SFV14_6mps}
\end{figure}

\begin{figure}
	\centering
	\begin{subfigure}{0.49\textwidth}
		\includegraphics[width=\textwidth]{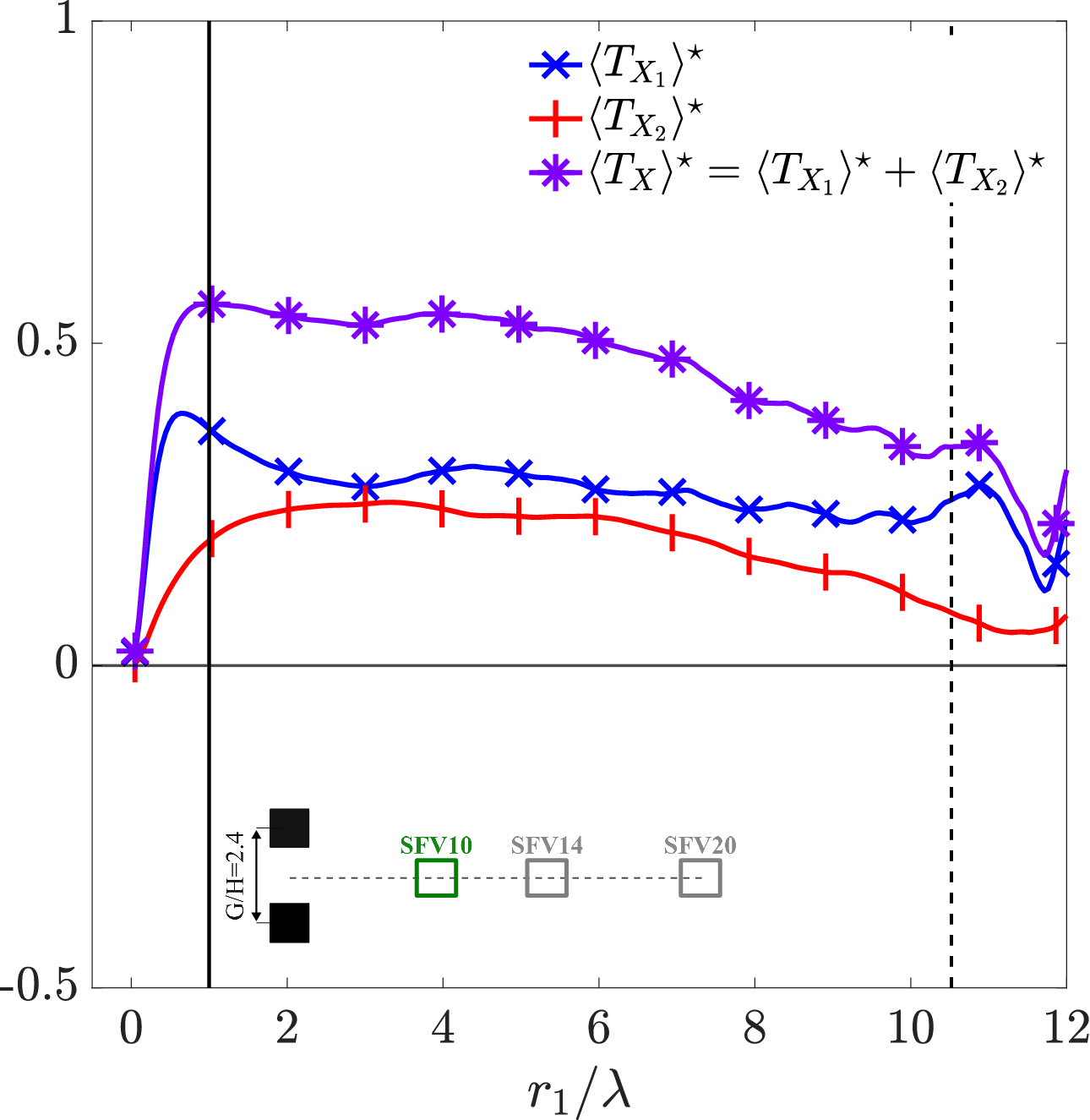}
		\caption{}
	\end{subfigure}
	\begin{subfigure}{0.49\textwidth}
		\includegraphics[width=\textwidth]{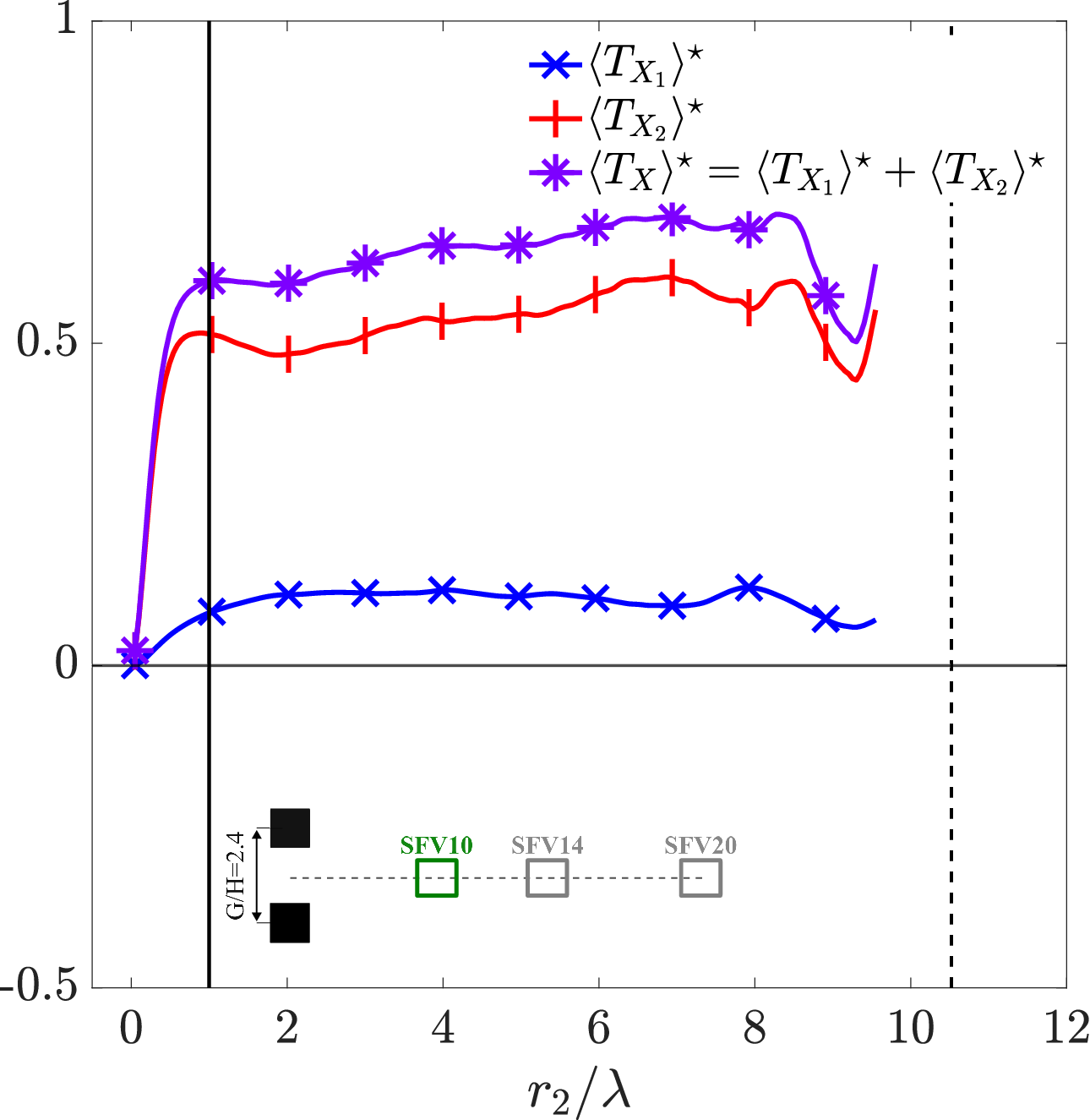}
		\caption{}
	\end{subfigure}
	\caption{Normalised average scale-by-scale inter-space energy
          transfer rate terms $T_{X_1}$ and $T_{X_2}$ as a function of
          $r_1$ (a) and $r_2$ (b) for case $G/H=2.4$ SFV10 at
          $Re_H=1.0 \times 10^4$. The vertical dashed line corresponds
          to the integral length scale location.}
	\label{fig:TX_terms_G_H_2.4_SFV10_5mps}
\end{figure}

  Figures \ref{fig:TX_terms_G_H_2.4_SFV20_5mps},
  \ref{fig:TX_terms_G_H_2.4_SFV14_6mps} and
  \ref{fig:TX_terms_G_H_2.4_SFV10_5mps} show the scale-by-scale plots
  of normalised inter-space energy transfer rates $\langle
  T_{X_1}\rangle^\star = \langle T_{X_1} \rangle / \avepsilonxt$ and
  $\langle T_{X_2} \rangle^\star = \langle T_{X_2} \rangle /
  \avepsilonxt$ for $G/H=3.5$ at SFV20, SFV14 and SFV7, respectively.
  Results are shown for $Re_H=1.0\times 10^4$ at SFV20 and SFV10 and
  for $Re_H=1.2\times 10^4$ at SFV14. In all cases, $\langle T_{X_1}
  \rangle^\star$ and $\langle T_{X_1} \rangle^\star$ exhibit the same
  positive sign as the total inter-space transfer rate and for all
  accessible scales. For the farthest location SFV20, the observations
  are very similar to that of $G/H=3.5$. Closer to the square prisms,
  at SFV14 and SFV10, both $\langle T_{X_1} \rangle^\star$ and
  $\langle T_{X_2} \rangle^\star$ show similar magnitudes for $r_1 >
  3\lambda$ (Figs. \ref{fig:TX_terms_G_H_2.4_SFV14_6mps} and
  \ref{fig:TX_terms_G_H_2.4_SFV10_5mps}) . However, $\langle T_{X_2}
  \rangle^\star$ still dominates for all $r_2$ scales. Once again,
  non-homogeneity is clearly present in both directions and such that
  the turbulence transfers two-point turbulent energy away from the
  SFV in both directions too.

\subsubsection{Close-to-single bluff body bistable $G/H=1.25$ wake}

\begin{figure}
	\centering
	\begin{subfigure}{0.49\textwidth}
		\includegraphics[width=\textwidth]{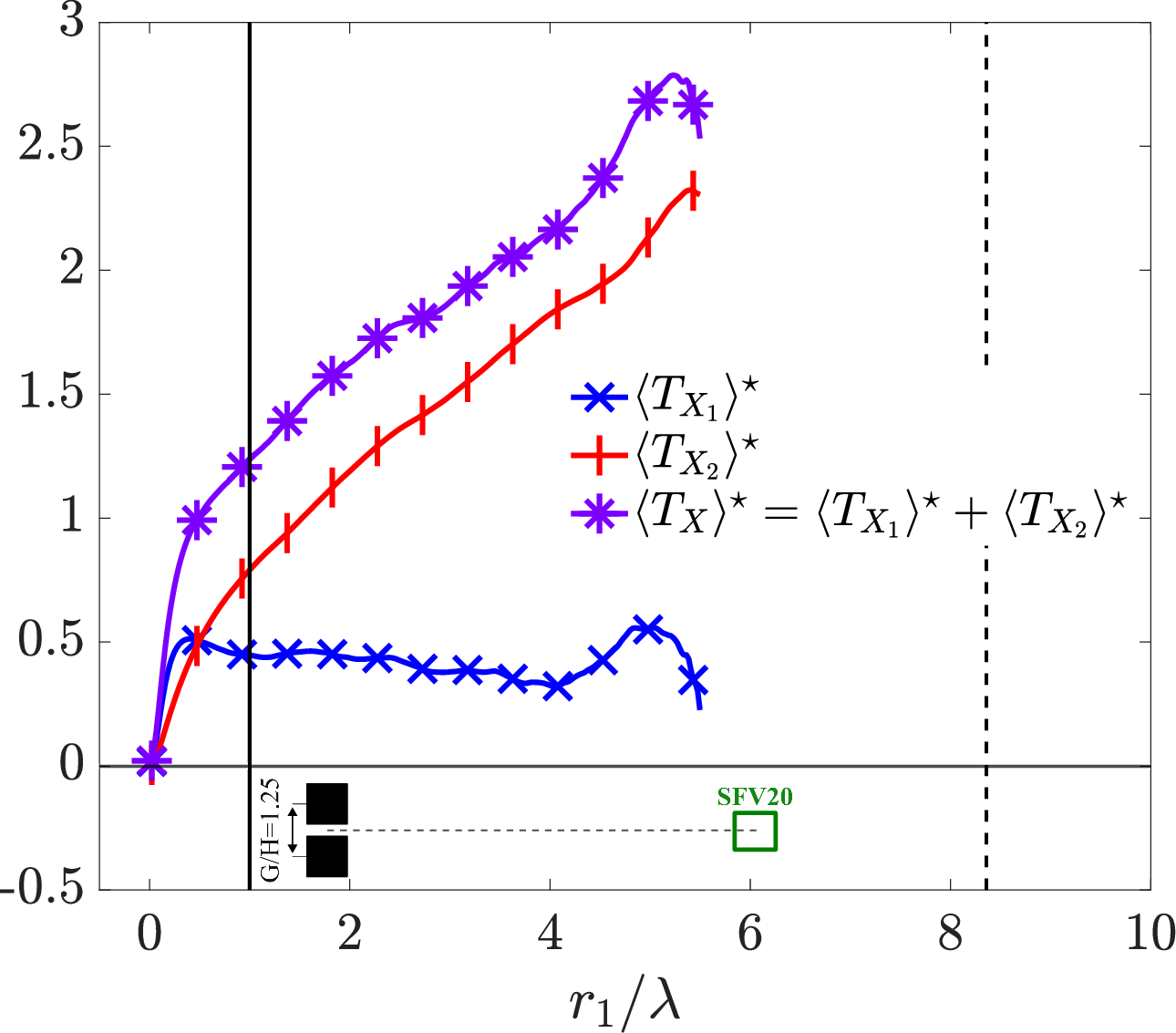}
		\caption{}
	\end{subfigure}
	\begin{subfigure}{0.49\textwidth}
		\includegraphics[width=\textwidth]{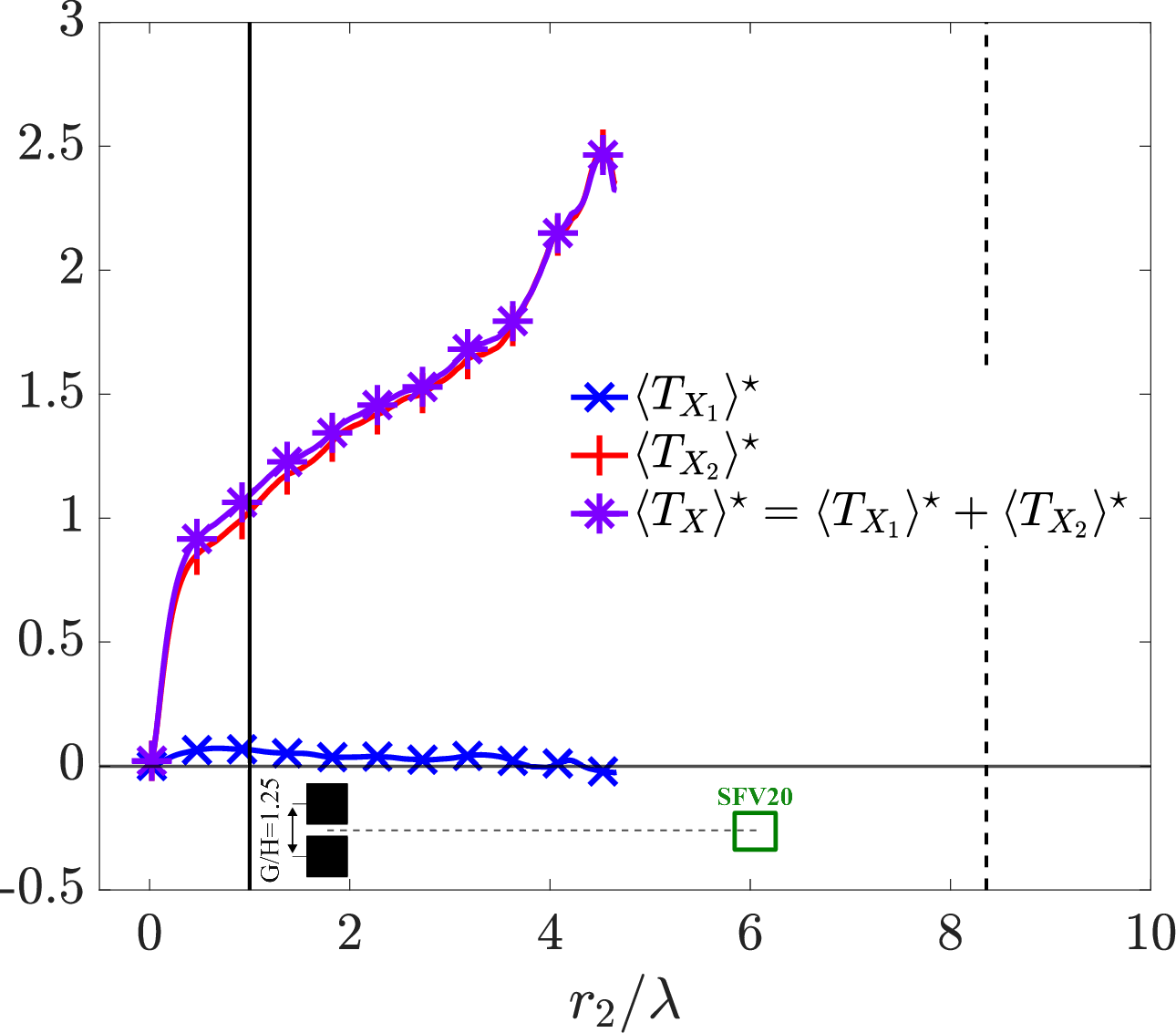}
		\caption{}
	\end{subfigure}
	
	\begin{subfigure}{0.49\textwidth}
		\includegraphics[width=\textwidth]{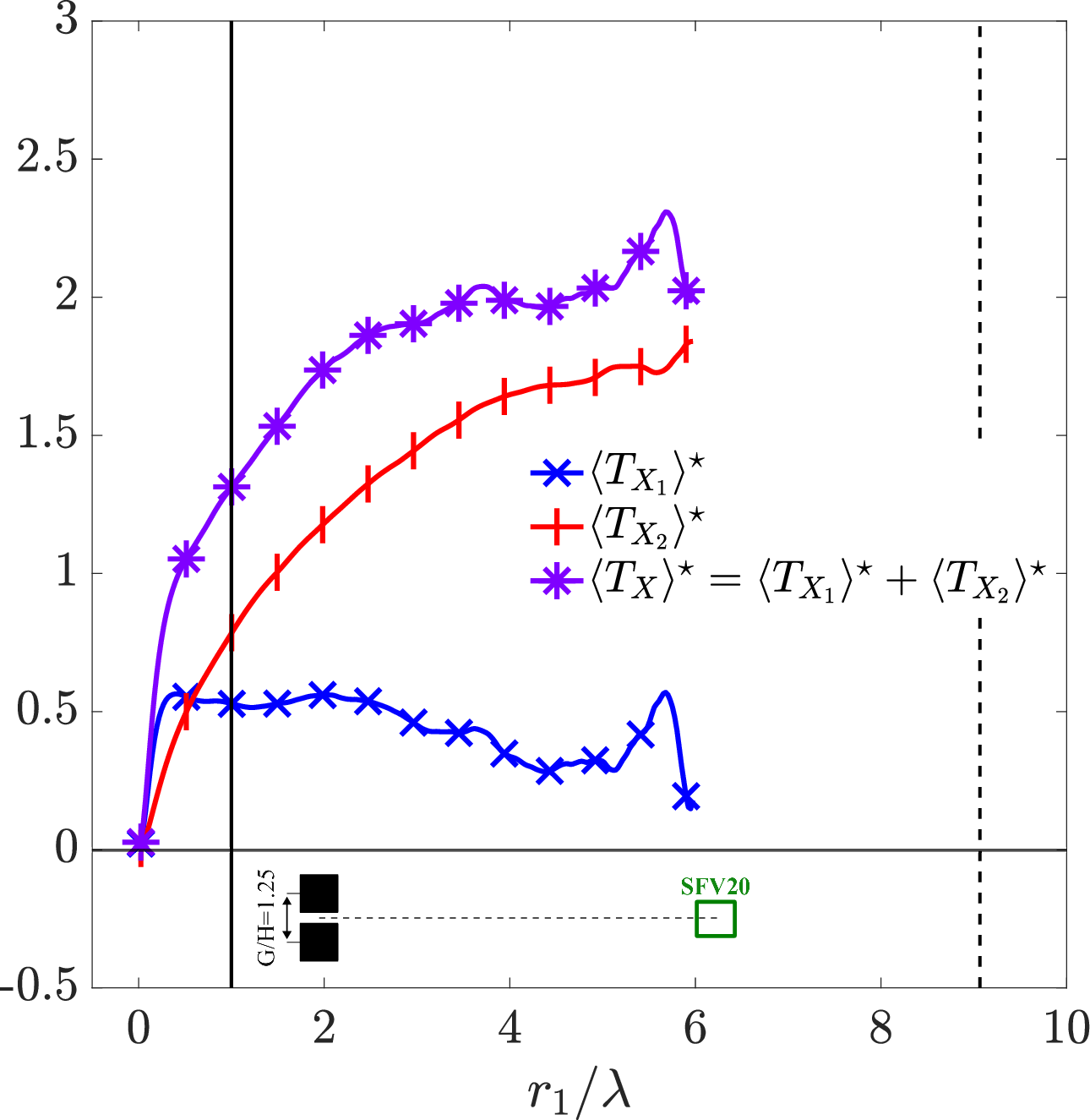}
		\caption{}
	\end{subfigure}
	\begin{subfigure}{0.49\textwidth}
		\includegraphics[width=\textwidth]{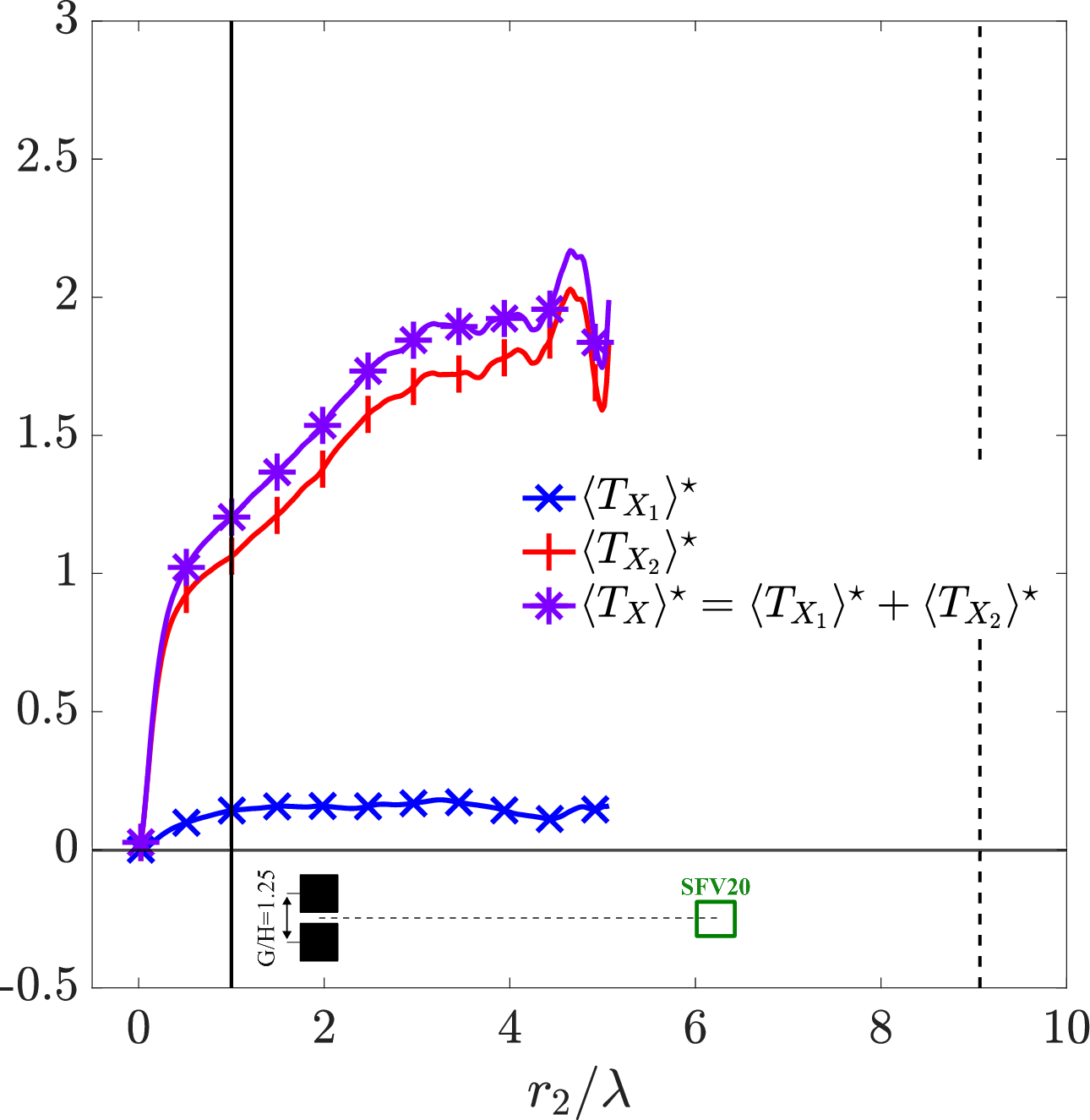}
		\caption{}
	\end{subfigure}
	
	\caption{Normalised average scale-by-scale inter-space energy transfer rate terms $T_{X_1}$ and $T_{X_2}$ as a function of $r_1$ (a) and $r_2$ (b) for case $G/H=1.25$ SFV20 at $Re_H=1.0 \times 10^4$ and at $Re_H=1.2 \times 10^4$ (c) \& (d). The vertical dashed line corresponds to the integral length scale location.}
	\label{fig:TX_terms_G_H_2.4_SFV20_5and6mps}
\end{figure}

  Figure \ref{fig:TX_terms_G_H_2.4_SFV20_5and6mps}
  shows the scale-by-scale plots of normalised inter-space energy
  transfer rates $\langle T_{X_1}\rangle^\star = \langle T_{X_1}
  \rangle / \avepsilonxt$ and $\langle T_{X_2} \rangle^\star = \langle
  T_{X_2} \rangle / \avepsilonxt$ for $G/H=1.25$ at SFV20 for two
  global Reynolds number values $Re_H=1.0\times 10^4$ and
  $Re_H=1.2\times 10^4$. As explained at the end of section 3 and in
  sub-section 4.4, the SFV20 location in the $G/H=1.25$ wake is
  qualitatively different from the SFV locations analysed in the other
  two wakes. Interestingly, it is the $\langle T_{X_2} \rangle$
  contribution to the inter-space transfer rate which dominates and
  grows sharply with both $r_1$ and $r_2$ while $\langle T_{X_1}
  \rangle$ reaches a plateau beyond $r_1>0.5\lambda$ and
  $r_2>\lambda$. This indicates that the non-homogeneity is very much
  stronger if not mainly in the cross-stream direction in this case
  unlike the SFV locations we analysed in the other two wakes.

\section{Two-point correlation part of $\langle T_X \rangle$}\label{sec:appC_corr}

\begin{figure}
	\centering
	\begin{subfigure}{0.48\textwidth}
		\includegraphics[width=\textwidth]{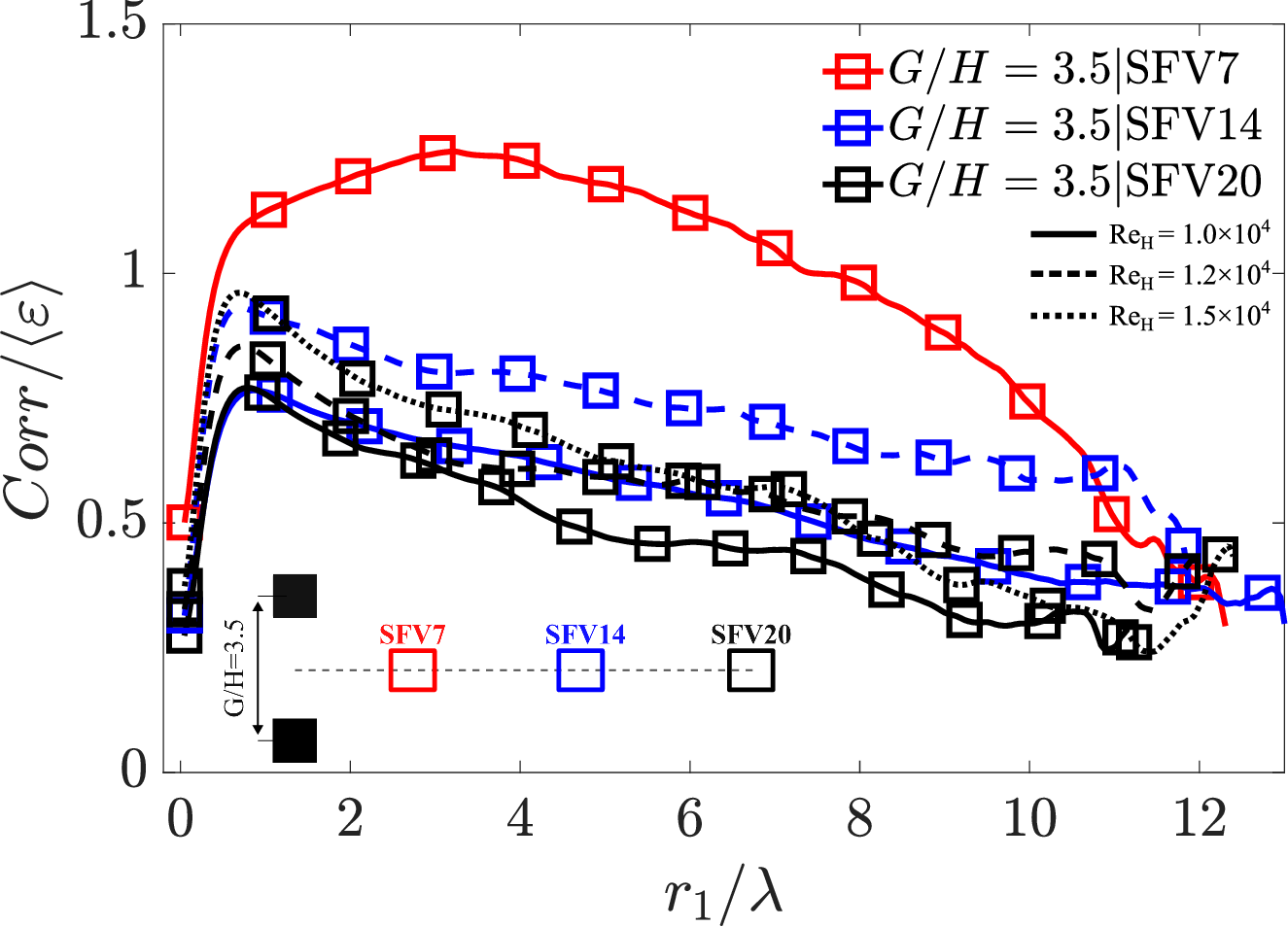}
		\caption{}
	\end{subfigure}
	\begin{subfigure}{0.48\textwidth}
		\includegraphics[width=\textwidth]{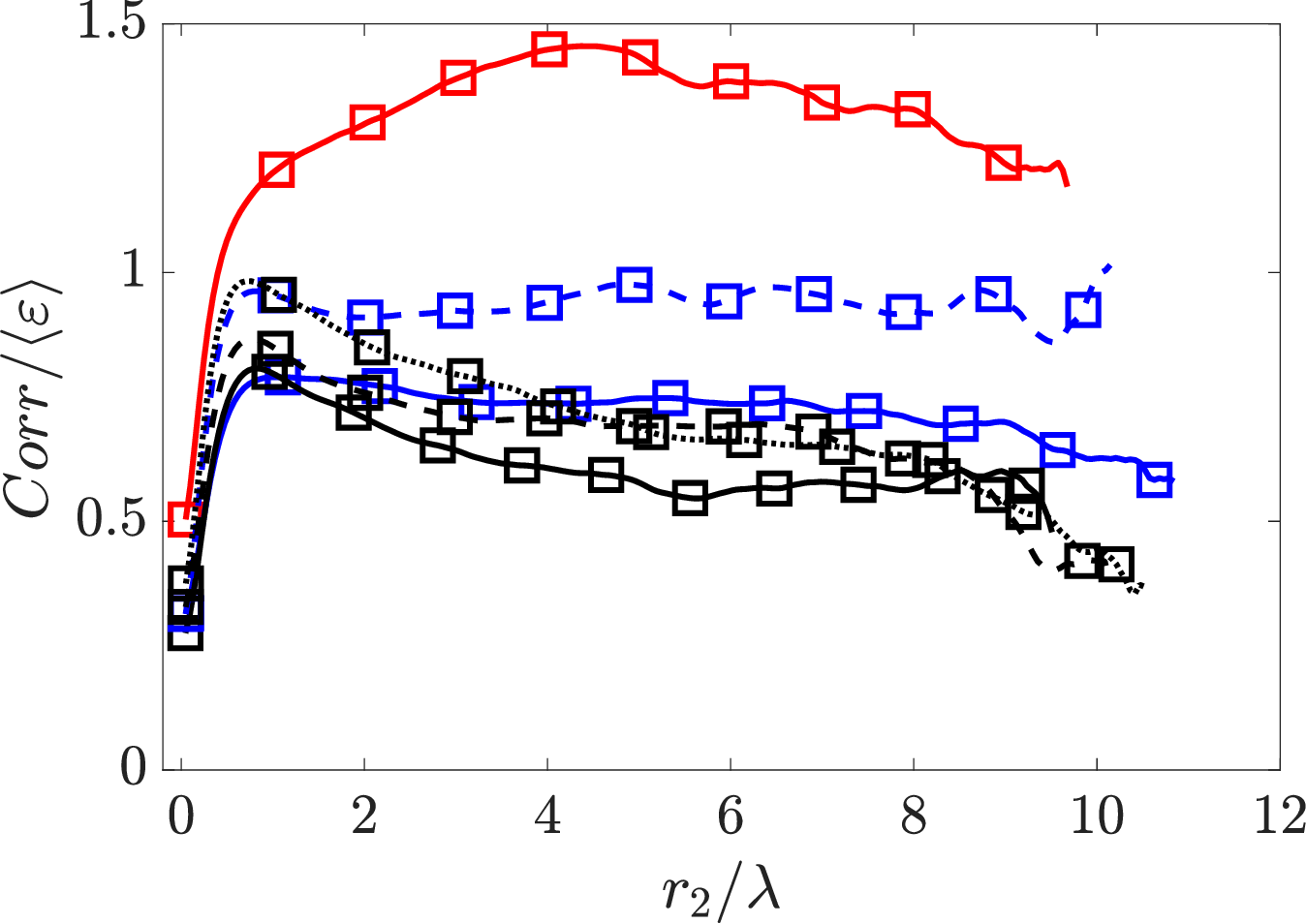}
		\caption{}
	\end{subfigure}
	
	\centering
	\begin{subfigure}{0.48\textwidth}
		\includegraphics[width=\textwidth]{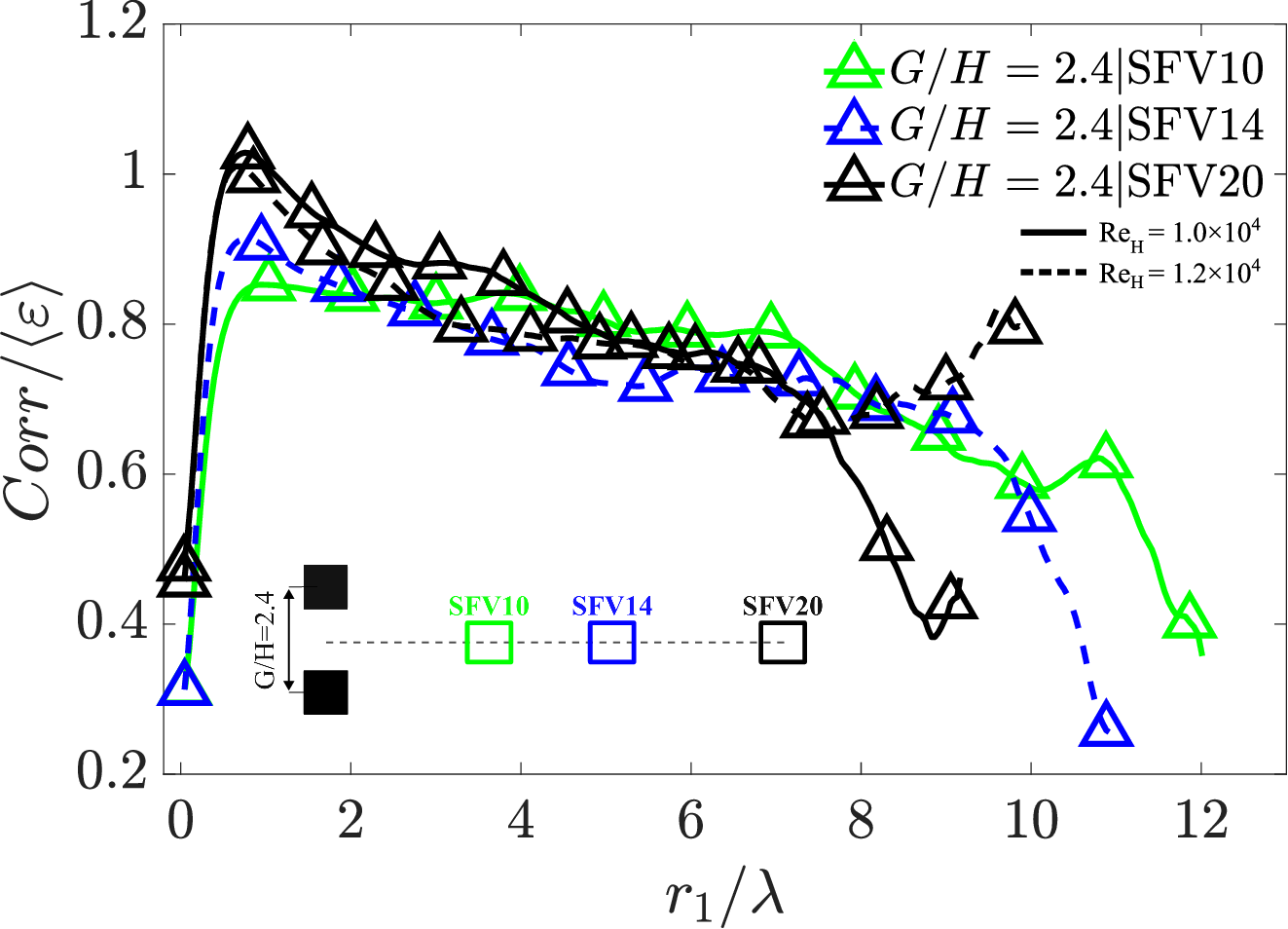}
		\caption{}
	\end{subfigure}
	\begin{subfigure}{0.48\textwidth}
		\includegraphics[width=\textwidth]{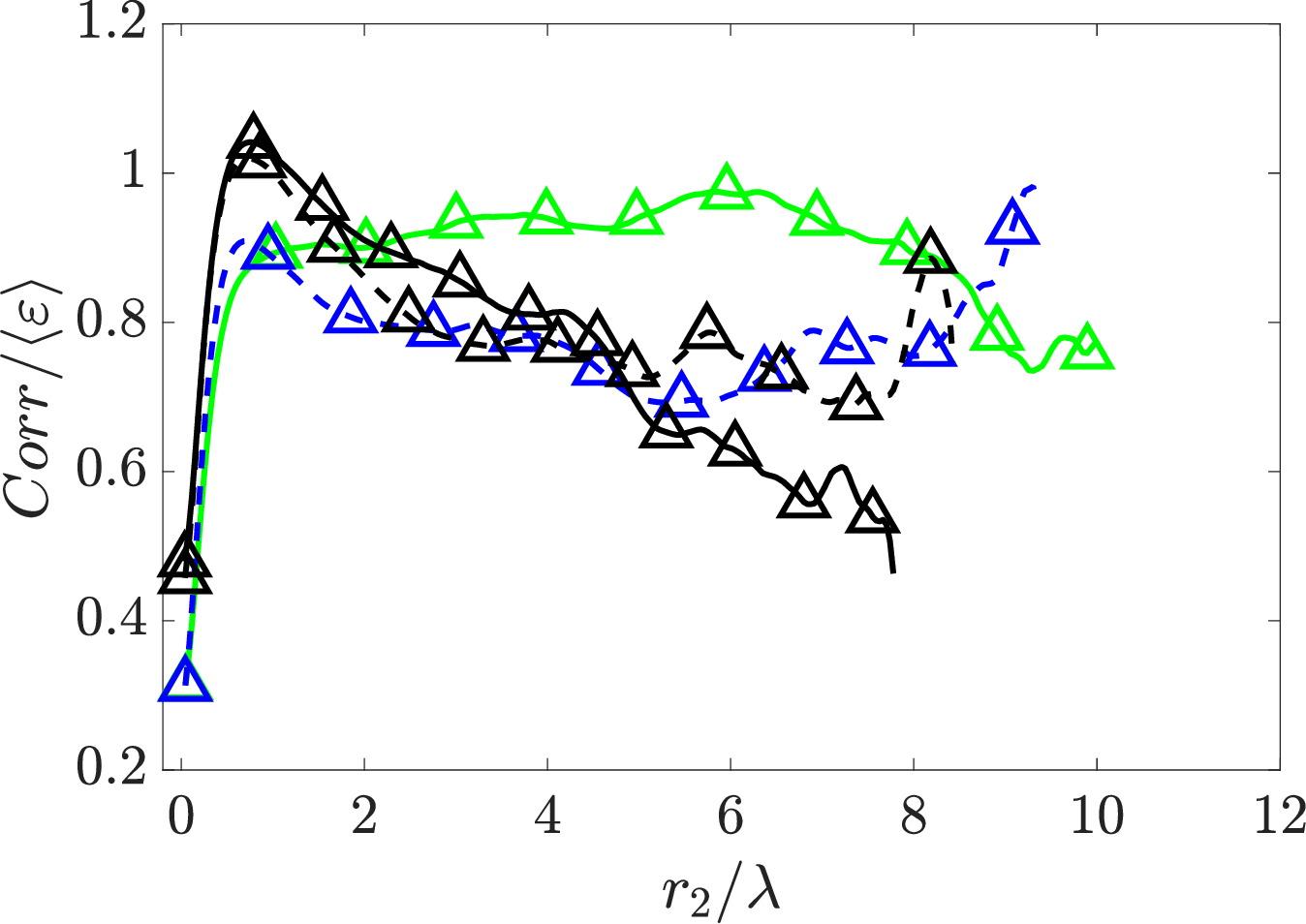}
		\caption{}
	\end{subfigure}
	\centering
	\begin{subfigure}{0.48\textwidth}
		\includegraphics[width=\textwidth]{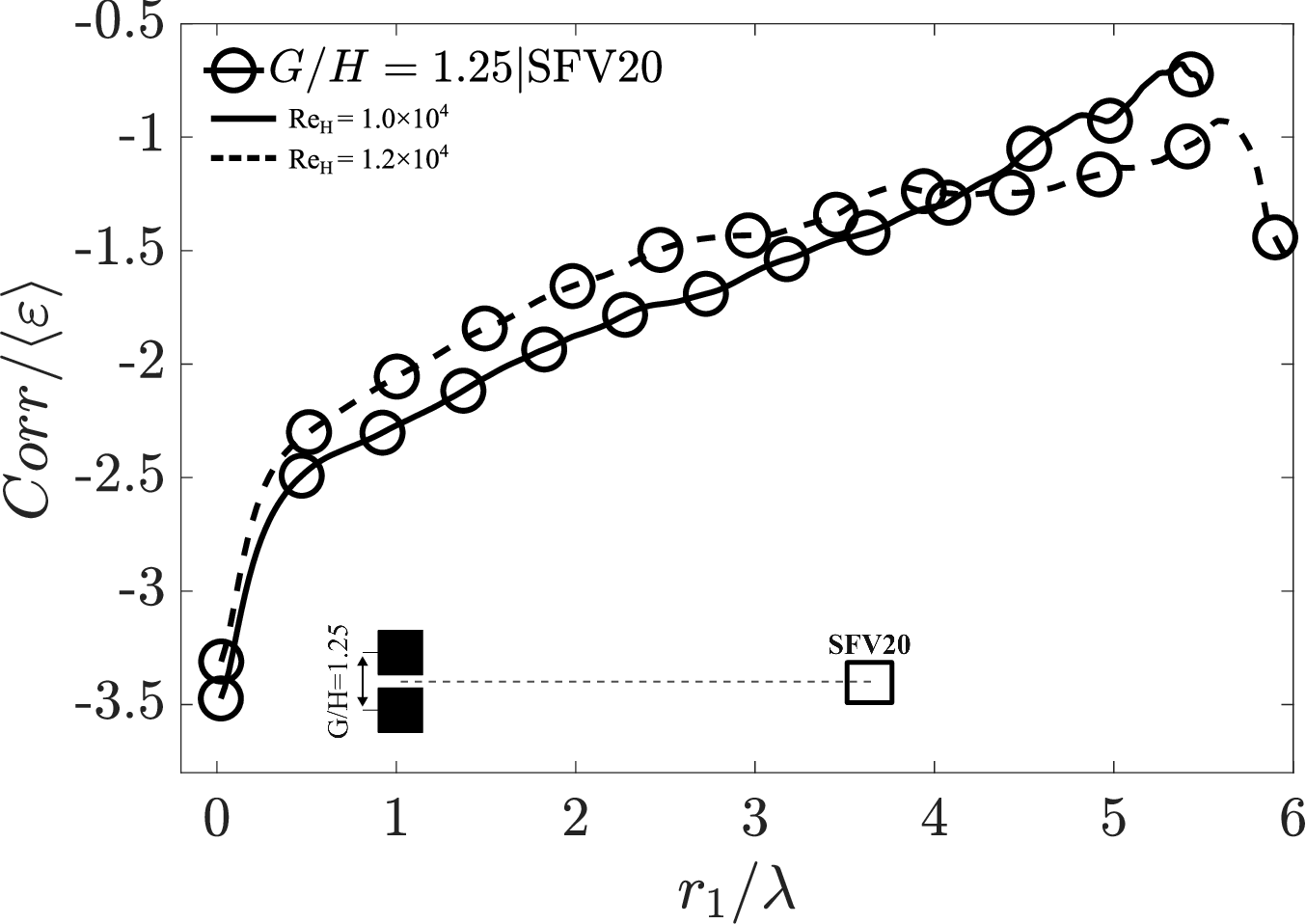}
		\caption{}
	\end{subfigure}
	\begin{subfigure}{0.48\textwidth}
		\includegraphics[width=\textwidth]{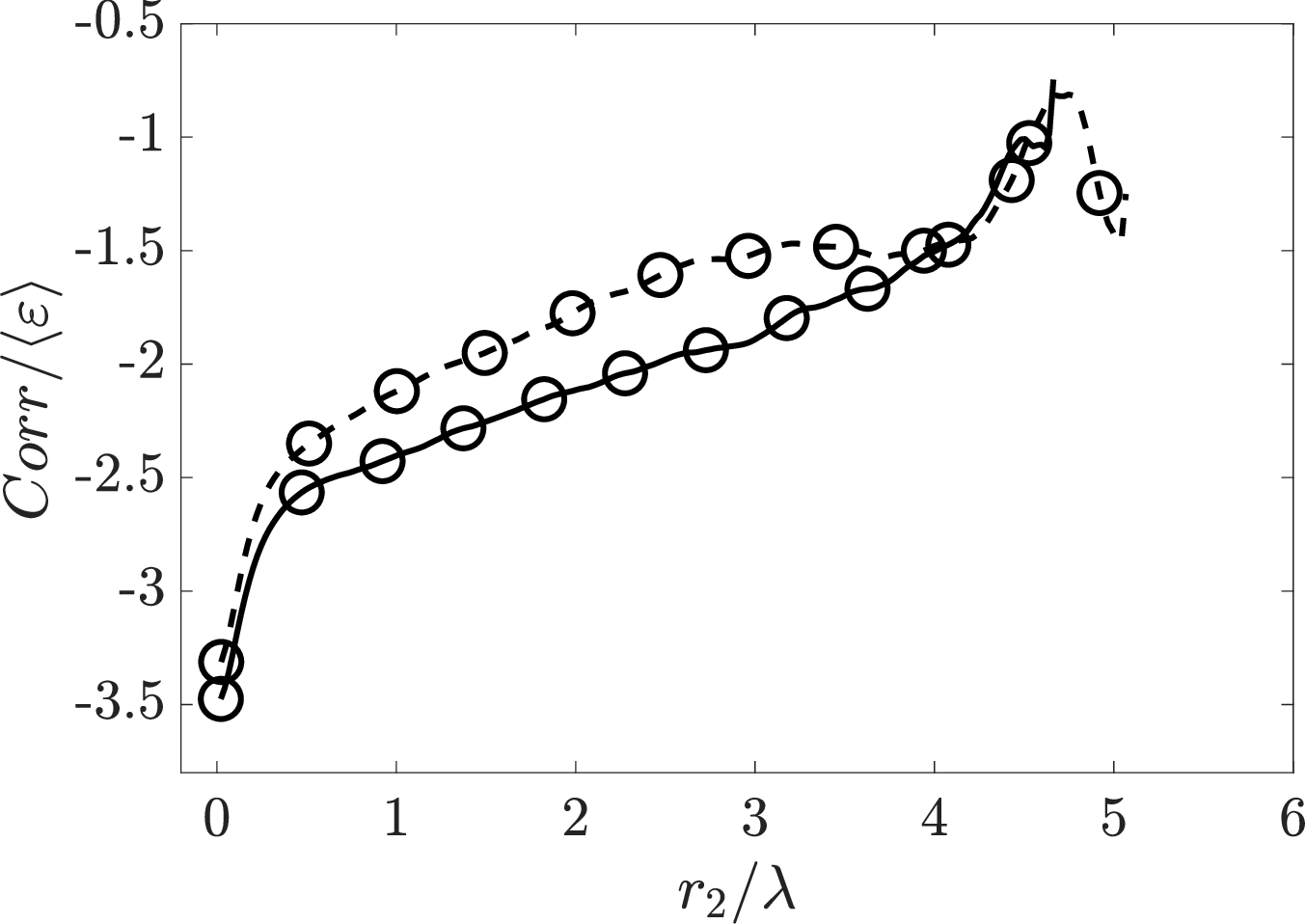}
		\caption{}
	\end{subfigure}
	\caption{Average two-point correlation contribution to the inter-space transport term $Corr=\langle T_X \rangle - \langle \mathcal{T}^+ \rangle - \langle \mathcal{T}^- \rangle  $ normalised by $\avepsilonxt$ for (a-b) $G/H=3.5$, (c-d) $G/H=2.4$ and (e-f) $G/H=1.25$ in the addressed SFVs and for all available $Re_H$.}
	\label{fig:Tcorr}
\end{figure}

Figure \ref{fig:Tcorr}, shows the average contribution of the correlation term in the two-point inter-space transfer rate $\langle T_X \rangle$ as described in $\S$\ref{sec:results} for all wakes and SFVs addressed in the present investigation. For all non-producing regions, i.e Figs. \ref{fig:Tcorr}a-d $Corr$ is positive and decreasing or constant in magnitude for separation scales larger than $\lambda$. An exception to this rule is the $G/H=3.5$ wake at SFV7 (Figs. \ref{fig:Tcorr}a-b) where $Corr$ increases with $r_1$ and $r_2$ from $\lambda$ to 3$\lambda$ and 4$\lambda$ respectively but is still positive. On the other hand, as shown in Figs. \ref{fig:Tcorr}(e-f), $Corr$ is negative and decreasing in magnitude with increasing $r_1$ and $r_2$ for both values of $Re_H$.

\end{document}